\documentclass[aps,prl,twocolumn,longbibliography,superscriptaddress]{revtex4-2}
\usepackage{bbm}
\usepackage{multirow}
\usepackage{graphicx}
\usepackage{dcolumn}
\usepackage{bm}
\usepackage{subfigure}
\usepackage{amsmath}
\usepackage{feynmf}
\usepackage{hyperref}
\usepackage{CJK}
\usepackage{amssymb}
\usepackage{attachfile}
\newcommand{\bk}{\boldsymbol k}

\usepackage{braket}
\usepackage{esint}
\usepackage{times}

\begin{document}

\title{Light-Induced  Even-Parity Unidirectionally Polarized Spin Splitting in Coplanar Antiferromagnets}

\author{Di Zhu}
\email{ zhud33@mail2.sysu.edu.cn}
\affiliation{Guangdong Provincial Key Laboratory of Magnetoelectric Physics and Devices,
State Key Laboratory of Optoelectronic Materials and Technologies,
School of Physics, Sun Yat-sen University, Guangzhou 510275, China}

\author{Dongling Liu}
\affiliation{Guangdong Provincial Key Laboratory of Magnetoelectric Physics and Devices,
State Key Laboratory of Optoelectronic Materials and Technologies,
School of Physics, Sun Yat-sen University, Guangzhou 510275, China}

\author{Zheng-Yang Zhuang}
\affiliation{Guangdong Provincial Key Laboratory of Magnetoelectric Physics and Devices,
State Key Laboratory of Optoelectronic Materials and Technologies,
School of Physics, Sun Yat-sen University, Guangzhou 510275, China}

\author{Zhigang Wu}
\affiliation{Quantum Science Center of Guangdong-Hong Kong-Macao Greater Bay Area (Guangdong), Shenzhen 508045, China}

\author{Zhongbo Yan}
\email{yanzhb5@mail.sysu.edu.cn}
\affiliation{Guangdong Provincial Key Laboratory of Magnetoelectric Physics and Devices,
State Key Laboratory of Optoelectronic Materials and Technologies,
School of Physics, Sun Yat-sen University, Guangzhou 510275, China}

\date{\today}

\begin{abstract}
When a coplanar antiferromagnet (AFM) with $xy$-plane magnetic moments 
exhibits a spin-split band structure and unidirectional spin polarization along $z$, 
the spin polarization is forced to be an odd function of momentum by the fundamental 
symmetry $[\bar{C}_{2z}\|\mathcal{T}]$. Coplanar AFMs displaying such odd-parity, unidirectionally polarized spin splittings 
are known as odd-parity magnets. In this work, we propose the realization of their 
missing even-parity counterparts.  We begin by deriving the symmetry conditions required 
for an even-parity, out-of-plane spin splitting. We then show that irradiating a spin-degenerate 
coplanar AFM with circularly polarized light lifts the $[\bar{C}_{2z}\|\mathcal{T}]$ constraint, 
dynamically generating this even-parity state. 
Specifically,  the light-induced unidirectionally polarized spin splitting exhibits a $d$-wave texture in momentum space, 
akin to that of a $d$-wave altermagnet.  
We prove this texture's robustness against spin canting and  propose time-, spin- and angle-resolved photoemission photocurrent
as a direct probe. Our work demonstrates 
that optical driving can generate novel spin-split phases in coplanar AFMs, 
thereby diversifying the landscape of materials exhibiting distinct spin splittings.
\end{abstract}

\maketitle

Antiferromagnets (AFMs) have recently gained renewed research interest due to 
a refined classification based on spin groups~\cite{Liu2022AM,Xiao2024SSG,Yi2024SSG,Chen2023AM}. This framework reveals 
that momentum-dependent spin splitting (MDSS) can arise even in AFMs with zero net magnetization 
and vanishing spin-orbit coupling~\cite{Hayami2019AM,LDYuan2020,LDYuan2021,Ma2021AM,Shao2021NC,
libor2022AMa,Libor2022AMb,Libor2022AMc}.  This has led to the prediction and experimental confirmation of spin-split AFMs~\cite{Mazin2021,Hu2025CAM,Zhu2024observation,Lee2024AM,Osumi2024MnTe,Reimers2024,Krempasky2024,Yang2024CrSb,Zeng2024CrSb,Jiang2024KV2Se2O,zhangAM2025,
Ding2024AM}, a class of materials with properties fundamentally distinct from conventional spin-degenerate AFMs~\cite{Bai2024AM,Song2025AMreview,Liu2025AMreview,Fukaya2025}.
Importantly, the symmetry-dictated MDSS pattern---a defining property of these materials---provides 
a natural taxonomy, enabling their classification broadly by parity (even or odd) and more 
specifically by wave type ($p$, $d$, $f$, {\it etc.})~\cite{libor2022AMa,Jungwirth2025AMreview,Luo2025pwave}, with each category leading to distinct physical phenomena~\cite{Brekke2024pwave,Ezawa2024pwave,Hedayati2025,Soori2025pwave,Nagae2025,Fukaya2025AM,
Ouassou2023AM,Cheng2024AM,Lu2024AM,Lin2024AM12,
Zhang2024AM,Zhu2023TSC,Zhu2024dislocation,Li2023AMHOTSC,Li2024AMHOTI,Ghorashi2024AM,Antonenko2025AM,
Parshukov2025AM7,Yang2025AM2,Qu2025AM,Li2025AM1,Rao2024AM7,Ma2024AM8,Fernandes2024AM,Zhuang2025SNL,Liu2025AM,
Chakraborty2024AM8,Brekke2023AM,Carvalho2024AM12,Maeda2025AM4,
Liu2025AM8,Bose2024AM11,Hong2025AM4,Parshukov2025,Fang2023NHE,Hu2025NLME,
Ezawa2025magnet,Ezawa2025nonlinear,Zarzuela2025transport,Zhu2025AMTopo,Zhu2025AMFET,Fu2025AM}.

Both collinear and coplanar AFMs share the fundamental symmetry $[\bar{C}_{2\alpha}\|\mathcal{T}]$
if  the spin-independent part of their Hamiltonians respects time-reversal symmetry (TRS).  
Here, $\mathcal{T}$ is the antiunitary time-reversal operator, and $\bar{C}_{2\alpha}$ 
denotes a $180^\circ$ spin-space rotation about the $\alpha$ axis---perpendicular to the magnetic moments---combined 
with a reversal of spin enforced by time reversal (overbar notation). The notation for composite symmetries 
is defined such that operators to the left of the double vertical bar act exclusively in spin space, 
while those to the right act in real space.~\cite{Livtin1974,Litvin1977}.
The symmetry $[\bar{C}_{2\alpha}\|\mathcal{T}]$ imposes a strict constraint on Bloch band spin polarization:
$\langle s_{\alpha}(\bk)\rangle=-\langle s_{\alpha}(-\bk)\rangle$ and $\langle s_{\alpha_{\perp}}(\bk)\rangle=\langle s_{\alpha_{\perp}}(-\bk)\rangle$, 
where $\alpha_{\perp}$ denotes directions orthogonal to $\alpha$. In a collinear 
spin-split AFM (altermagnet) with $z$-directed magnetic moments, 
the spin-rotation axis $\alpha$ has two choices,  $\alpha=\{x,y\}$. These collectively force $\langle s_{x,y}(\bk)\rangle=0$ and $\langle s_{z}(\bk)\rangle=\langle s_{z}(-\bk)\rangle$. 
Consequently, the MDSS is constrained to be even-parity. In a coplanar AFM with the magnetic moments
confined to the $xy$ plane, the only choice for the spin-rotation axis is $\alpha=z$. 
This mandates $\langle s_{x,y}(\bk)\rangle=\langle s_{x,y}(-\bk)\rangle$ and $\langle s_{z}(\bk)\rangle=-\langle s_{z}(-\bk)\rangle$, but the 
spin polarization is not unidirectional in general~\cite{Hayami2020AM1,Hayami2020AM2,Lee2024dwaveNAFM}.
Unidirectional polarization can occur when an extra symmetry is present. This is exemplified by 
$p$-wave magnets~\cite{Birk2023}, a recently identified class of coplanar AFMs featuring an effective TRS~\cite{Yamada2025}. 
This supplemental symmetry, in concert with $[\bar{C}_{2z}\|\mathcal{T}]$, suppresses all in-plane polarization, leading to purely {unidirectionally polarized} (perpendicular to the moments) 
and odd-parity MDSS~\cite{Yu2025pwave}. Given the fundamental importance of discovering new spin-split phases, a central 
question arises: Can breaking the $[\bar{C}_{2\alpha}\|\mathcal{T}]$ symmetry lead to phases 
with spin-splitting parity opposite to that of the $p$-wave magnet, thereby producing an entirely new type of
coplanar AFM?

Some insight into this problem has been provided by recent developments in collinear systems. Several studies demonstrate that 
breaking the $[\bar{C}_{2\alpha}\|\mathcal{T}]$ symmetry via mechanisms such as sublattice currents~\cite{Lin2025OAM,Zeng2025OPAMa,Zeng2025OPAMb}, 
orbital order~\cite{zhuang2025AM9}, or irradiation with circularly polarized light (CPL)~\cite{Huang2025AM7,Li2025AM7,Zhu2025AM8,liu2025AM10,Pan2025OPAM,Li2025FloAM} can lift the spin degeneracy of a $\mathcal{PT}$-symmetric collinear AFM, yielding an altermagnetic 
state with odd-parity MDSS. However, whether this mechanism also 
applies to coplanar AFMs remains an open question. 

Here, we answer this fundamental question affirmatively by studying a 
bilayer coplanar AFM under CPL irradiation. 
While the pristine system exhibits spin-degenerate bands, 
the CPL simultaneously breaks both the 
$[\bar{C}_{2z}|\mathcal{T}]$ symmetry and the symmetry enforcing spin degeneracy. This results in 
a $d$-wave, unidirectionally polarized spin-splitting texture on the Fermi surface, analogous to that of a $d$-wave altermagnet. 
Crucially, this induced unidirectional polarization is even-parity and oriented out-of-plane, making it as the direct 
even-parity counterpart to odd-parity magnets~\cite{Birk2023}. We show that time-, spin- and angle-resolved photoemission spectroscopy (tr-SARPES)
provides a natural probe of this Floquet-engineered even-parity spin-split magnet.

{\it Symmetry-guided route.---}Before investigating specific models, we outline the 
general route to realize the target phase. We begin with a pre-driven coplanar 
AFM possessing three fundamental symmetries:  $[\bar{C}_{2z}\|\mathcal{T}]$, 
$[\bar{E}\|\mathcal{T}U_{I}]$, and $[\bar{E}\|\mathcal{T}|\bm{\tau}]$, where $\bar{E}$ is the identity operator
combined with time-reversal, $\bm{\tau}$ denotes a fractional translation, and $U_{I}$ denotes an operation reversing the momentum. 
In three-dimensional (3D) bulk systems, $U_{I}=\mathcal{P}$ (inversion), while in two-dimensional (2D)
layer systems, $U_{I}$ can be either $\mathcal{P}$ 
or $C_{2z}$. The coexistence of $[\bar{C}_{2z}\|\mathcal{T}]$ and $[\bar{E}\|\mathcal{T}U_{I}]$ yields a composite symmetry 
$[C_{2z}\|U_{I}]$. Similarly, the coexistence of $[\bar{C}_{2z}\|\mathcal{T}]$ and $[\bar{E}\|\mathcal{T}|\bm{\tau}]$ 
yields another composite symmetry $[C_{2z}\|E|\bm{\tau}]$. The $[\bar{E}\|\mathcal{T}U_{I}]$  symmetry  (commonly termed 
$\mathcal{PT}$ symmetry when $U_{I}=\mathcal{P}$) enforces a spin-degenerate band structure prior to driving. 
We then apply CPL to the system. The optical field dynamically 
breaks all three fundamental symmetries simultaneously, while preserving the two composite symmetries, 
$[C_{2z}\|U_{I}]$ and $[C_{2z}\|E|\bm{\tau}]$. 
These preserved symmetries impose distinct constraints on the spin polarization. 
The $[C_{2z}\|U_{I}]$ symmetry forces 
 $\langle s_{x,y}(\bm{k})\rangle=-\langle s_{x,y}(-\bm{k})\rangle$ and 
$\langle s_{z}(\bm{k})\rangle=\langle s_{z}(-\bm{k})\rangle$, while the $[C_{2z}\|E|\bm{\tau}]$ 
symmetry mandates  $\langle s_{x,y}(\bm{k})\rangle=-\langle s_{x,y}(\bm{k})\rangle=0$.
Consequently, any nonzero spin polarization generated by the drive must be even in parity 
and oriented exclusively along the $z$-direction. 

Although the two symmetries $[C_{2z}\|U_{I}]$ and $[C_{2z}\|E|\bm{\tau}]$ 
ensure the MDSS is even-parity and unidirectionally polarized (with only $\langle s_{z}(\bm{k})\rangle$ being nonzero), 
engineering a specific wave type (e.g., $s$-wave or $d$-wave) requires additional 
symmetries to constrain the pattern.
A natural candidate for imposing such a constraint is a symmetry
$[C_{2\parallel}\|U_{nz}|\bm{\tau}]$ (or $[C_{2\parallel}\|U_{nz}]$), with $n=\{4, 6\}$. Here,  $C_{2\parallel}$ denotes a $180^{\circ}$ rotation
about an axis lying within the plane of the magnetic moments. In 3D bulk systems,  
$U_{nz}=C_{nz}$,  while in 2D layer systems, $U_{nz}$ can be either $C_{nz}$ or $C_{nz}\mathcal{M}_{z}$, 
where $C_{nz}$ denotes a $360^{\circ}/n$ rotation about the $z$ axis, and $\mathcal{M}_{z}$ denotes 
mirror reflection about the midplane. 
When the AFM possesses $[C_{2\parallel}\|U_{4z}|\bm{\tau}]$ symmetry, the out-of-plane spin polarization obeys
$\langle s_{z}(k_{x},k_{y})\rangle=-\langle s_{z}(k_{y},-k_{x})\rangle$, which enforces a 
$d$-wave pattern for the MDSS. Similarly, the presence of $[C_{2\parallel}\|U_{6z}|\bm{\tau}]$
symmetry leads to a $g$-wave pattern.

\begin{figure}[t]
\centering
\includegraphics[width=0.45\textwidth]{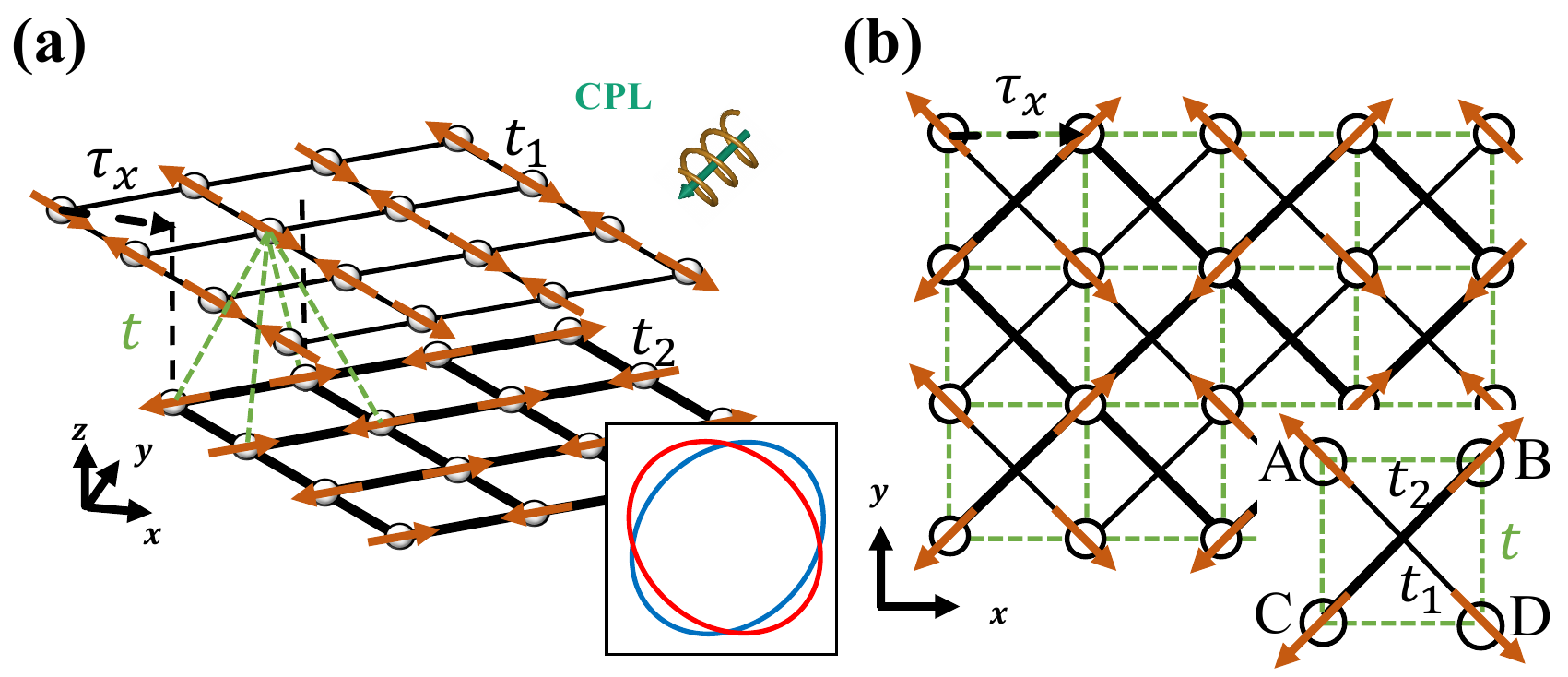}
\caption{(a) Schematic of a bilayer coplanar AFM under CPL irradiation. 
The primitive lattice vectors of the top layer are $a(1,1)$ and $a(-1,1)$, 
and the two layers are shifted relative to each other by $\bm{\tau}_x=a(1,0)$, 
illustrated by the  black dashed line with arrow.
Orange arrows on the lattice represent the magnetic order. Thin (thick) 
black solid lines denote the nearest-neighbor hoppings $t_1$ ($t_2$) in 
the top (bottom) layer, 
and green dashed lines denote the interlayer hoppings $t$. The inset illustrates the CPL-induced $d_{xy}$-wave out-of-plane spin splitting.
(b) Top view of the bilayer system. The system can be regarded as a coplanar AFM model with all-out spin 
configuration ($M_x=M_y=M$). The inset shows the schematic of a unit cell, where sublattices and hoppings are labelled.
}\label{fig1}
\end{figure}

{\it 2D bilayer coplanar AFM.---}Having established the necessary symmetries, we now construct an 
explicit bilayer model to show how to achieve the proposed spin-split phases. As shown in Fig.~\ref{fig1}(a), the system 
consists of two square-lattice monolayers of collinear AFMs, shifted relative to each other by the vector 
$\bm{\tau}_{x}=a(1,0)$. Both monolayers share an identical lattice constant 
$\sqrt{2}a$. Crucially, their Néel vectors are oriented perpendicularly to each other, yielding a resultant 
coplanar all-out magnetic configuration when viewed from above [Fig.~\ref{fig1}(b)]. 
It is evident that the hopping pattern and moment configuration respect the following symmetries: 
$[\bar{C}_{2z}\|\mathcal{T}]$, $[\bar{E}\|\mathcal{T}C_{2z}]$, $[\bar{E}\|\mathcal{T}|\bm{\tau}_{d}]$, 
$[C_{2z}\|C_{2z}]$, $[C_{2z}\|E|\bm{\tau}_{d}]$, $[\bar{E}\|\mathcal{T}C_{2x}\mathcal{M}_{z}|\bm{\tau}_{x}]$, 
$[\bar{E}\|\mathcal{T}C_{2y}\mathcal{M}_{z}|\bm{\tau}_{y}]$, and $[C_{2(x-y)}\|C_{4z}|\bm{\tau}_{x}]$, 
where $\bm{\tau}_{y}=a(0,1)$ and $\bm{\tau}_{d}=a(1,-1)$. 

The single-particle tight-binding Hamiltonian for this system is
\begin{align}
H=&\sum_{\langle i,j\rangle,m\neq n,\alpha}tc^\dagger_{i,m,\alpha}c_{j,n,\alpha}+\sum_{\langle i,j\rangle,m,\alpha}t_{m}c^\dagger_{i,m,\alpha}c_{j,m,\alpha}\nonumber\\
&+\sum_{i,m,\alpha,\beta}c^\dagger_{i,m,\alpha}(\bm{M}_{i,m}\cdot \bm{s})_{\alpha,\beta}c_{i,m,\beta},
\end{align}
where $c_{i,m,\alpha} (c_{i,m,\alpha}^\dagger)$ denotes the annihilation (creation) operator for an electron at site $i$, in layer $m$, with spin $\alpha$. The first term describes interlayer nearest-neighbor hopping with amplitude $t$ [Fig.~\ref{fig1}(a)]. The second term corresponds to intralayer hopping with a layer-dependent amplitude $t_m$ ($m=1,2$) [Fig.~\ref{fig1}(b)]. The final term is the exchange field due to the local magnetic moments, with $\bm{M}_{i,m}=(\pm M_x,\pm M_y)$ specifying the moment orientation and  $\bm{s}=(s_x,s_y)$ 
denoting the vector of Pauli matrices in spin space.

In the Fourier-transformed basis $\psi_{\bm{k}}=(c_{\bm{k},\uparrow},c_{\bm{k},\downarrow})^{T}$ with
b $c_{\bm{k},\alpha}=(c_{A,\bm{k},\alpha},c_{B,\bm{k},\alpha},c_{C,\bm{k},\alpha},c_{D,\bm{k},\alpha})$, the momentum-space Hamiltonian is given by 
\begin{align}
\mathcal{H}(\bm{k})=&2t h^{c}_{x}\sigma_x+2t h^{c}_{y}\tau_x+4t_{s}h^{c}_{x}h^{c}_{y}\tau_x\sigma_x\nonumber\\
&+4t_{a}h^{c}_{x}h^{c}_{y}\tau_y\sigma_y-M_x\sigma_zs_x+M_y\tau_zs_y.
\end{align}
Here, we introduce a compact notation by defining $h^{c}_{j}\equiv\cos k_{j}$ for $j=x,y$
and omitting the identity matrices. 
The Pauli matrices $\tau_a$ and $\sigma_b$ act on the four sublattice degrees of freedom, and $s_c$ acts on spin, 
with products in the Hamiltonian following the order $\tau_{a}\sigma_{b} s_{c}$. 
The intralayer nearest-neighbor hoppings are parameterized by symmetric and antisymmetric parts, 
$t_{s}=(t_1+t_2)/2$ and $t_{a}=(t_2-t_1)/2$. Throughout this work, we adopt $a \equiv 1$ as the unit of length and 
take all hopping parameters ($t$, $t_s$, $t_a$) to be positive. 

When the magnetic order is absent, i.e., $M_{x}=M_{y}=0$, the spectrum of the Hamiltonian is 
\begin{eqnarray}
E_{\beta,\pm}(\bk)=\beta4t_{s}h^{c}_{x}h^{c}_{y}\pm\sqrt{(4t_{a}h^{c}_{x}h^{c}_{y})^{2}+4t^{2}(h^{c}_{x}+\beta h^{c}_{y})^{2}},\quad
\end{eqnarray}
where $\beta=\pm$. A key feature of the doubly-degenerate spectrum is the emergence of Dirac points for non-zero $t_{a}$. 
These Dirac points are essential for generating spin splitting and a topological band structure under CPL~\cite{Huang2025AM7,Li2025AM7,Zhu2025AM8,liu2025AM10}, as CPL breaks
$\mathcal{PT}$ symmetry by inducing a nontrivial Dirac mass term when coupled to the Dirac fermions~\cite{Oka2009}.

When magnetic order is present, the spectrum remains spin-degenerate due to the Kramers degeneracy 
dictated by the $\mathcal{PT}$-like $[\bar{E}\|\mathcal{T}C_{2z}]$ symmetry. This spin degeneracy leads to an exact cancellation 
of spin polarization for all bands at every momentum.

{\it Light-induced $d$-wave MDSS.---}Light driving provides a powerful method for 
tuning electronic band structures~\cite{Oka2009,lindner2011floquet,yan2016tunable,Chan2016type,Zhou2023Floquet}. Recent research has shown that by modifying the band structure and MDSS, 
it can induce a wide range of intriguing phenomena in altermagnets~\cite{Pal2025FloAM,Fu2025FloAM,Fu2025FloAM2,Ghorashi2025FloAM,
Yokoyama2025FloAM,Ganguli2025FloAM,Liu2025FWSM,Yarmohammadi2025FloAM,Yarmohammadi2026FloAM}.

We now investigate the influence of CPL on the coplanar AFM. The light is incident perpendicular to the plane, with a vector 
potential $\bm{A}(t)=A_0(\cos{\omega t},\sin{\omega t})$. 
The effect of CPL is incorporated into the Hamiltonian via the Peierls substitution, $\bm{k}\rightarrow \bm{k}+\bm{A}(t)$ (we set $e=\hbar=1$ for notational simplicity). Since the Hamiltonian is time-periodic, it can be expanded by Fourier transformation as $\mathcal{H}(\bm{k}+\bm{A}(t))=\sum_n\mathcal{H}_ne^{i\omega t}$, with $n\in\mathbb{Z}$. 
To obtain an analytical description of the key physics, we focus on the high-frequency off-resonant regime
where the driven system is described by an effective static Hamiltonian given by~\cite{Kitagawa2011Floquet,Goldman2014} 
\begin{align}
\mathcal{H}_{{\rm eff}}(\bm{k})=\mathcal{H}_0(\bm{k})+\sum_{n\ge 1}\frac{[\mathcal{H}_{n},\mathcal{H}_{-n}]}{n\omega}+O(\omega^{-2}).
\end{align}
Keeping only the leading-order contributions from the $n=1$ commutator (one-photon processes), 
we obtain (further details can be found 
in Section I of the Supplemental Material (SM)~\cite{supplemental})
\begin{align}
\mathcal{H}_{{\rm eff}}(\bm{k})&=2J_0(A_0)t(h^{c}_{x}\sigma_x
+h^{c}_{y}\tau_x)-M\sigma_zs_x+M\tau_zs_y\nonumber\\
&+4J_0(\sqrt{2}A_0)h^{c}_{x}h^{c}_{y}(t_{s}\tau_x\sigma_x+t_{a}\tau_y\sigma_y)\nonumber\\
&-F(A_0,\omega)h^{s}_{x}h^{s}_{y}(h^{c}_{x}\tau_y\sigma_z-h^{c}_{y}\tau_z\sigma_y),
\label{Heff}
\end{align}
where $h^{s}_{j}\equiv\sin k_{j}$ and $F(A_0,\omega)={16\sqrt{2}tt_{a}J_1(A_0)J_1(\sqrt{2}A_0)}/{\omega}$ with $J_n(x)$ denoting the $n$th order Bessel function of the first kind. 
Since the moments are aligned along the $x=\pm y$ directions, the magnetic exchange field is set to satisfy $M_x=M_y=M$.
Compared to the original Hamiltonian, the primary modifications are contained in the last two terms. 
These CPL-induced terms break three fundamental symmetries: $[\bar{C}_{2z}\|\mathcal{T}]$, $[\bar{E}\|\mathcal{T}C_{2z}]$, and $[\bar{E}\|\mathcal{T}|\bm{\tau}_{d}]$, giving rise to band spin splitting. 
Nevertheless, three other symmetries---$[\bar{E}\|\mathcal{T}C_{2x}\mathcal{M}_{z}|\bm{\tau}_{x}]$, $[\bar{E}\|\mathcal{T}C_{2y}\mathcal{M}_{z}|\bm{\tau}_{y}]$, 
and $[C_{2(x-y)}\|C_{4z}|\bm{\tau}_{x}]$---remain intact, 
along with the two composite symmetries 
$[C_{2z}\|C_{2z}]$ and $[C_{2z}\|E|\bm{\tau}_{d}]$ (a more detailed discussion is provided in the SM Section II~\cite{supplemental}). 
As noted earlier, these preserved symmetries enforce the spin 
polarization that is strictly unidirectional along $z$, and which exhibits a $d$-wave symmetry. 

\begin{figure}[t]
\centering
\includegraphics[width=0.45\textwidth]{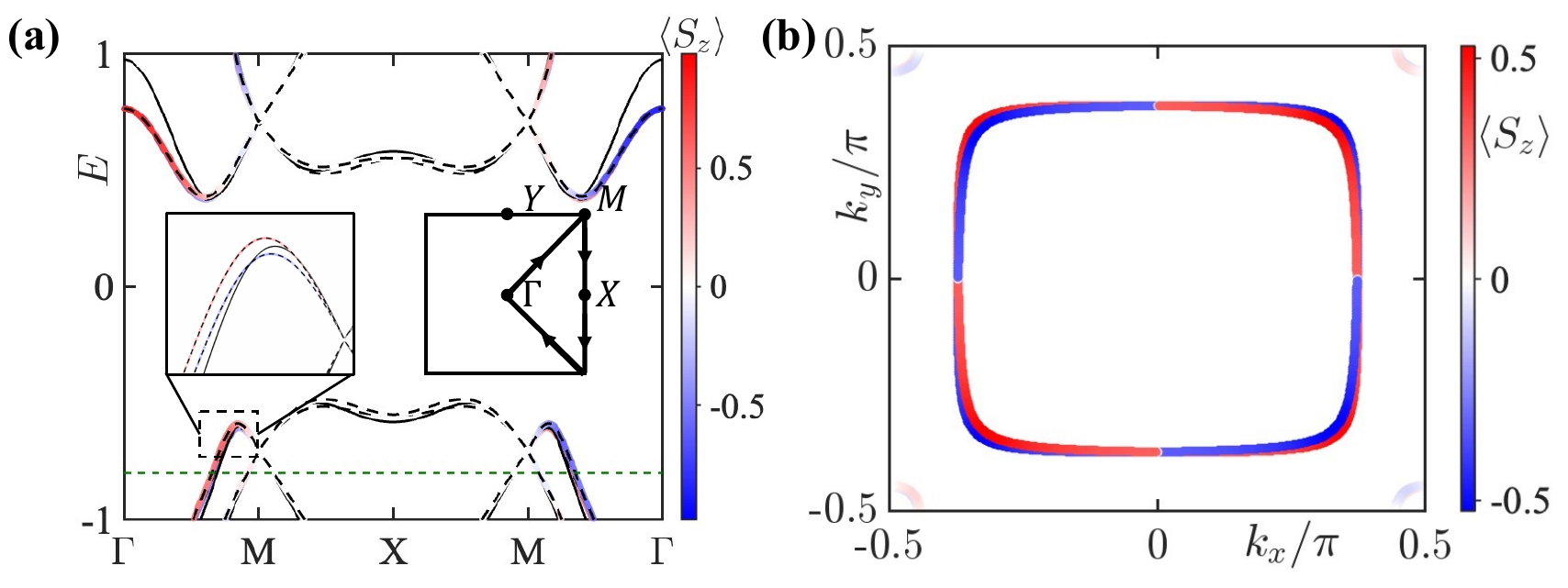}
\caption{(a) Energy bands of the static system (solid black lines, spin-degenerate) and of the system driven by CPL (dashed red and blue lines, spin-split). The left inset shows a detailed view of the spin-split band structure near the M point; the right inset shows the Brillouin zone with the high-symmetry paths used in the plot. The green dash line is located at energy $E=-0.8$. (b) The corresponding CPL-induced $d$-wave spin splitting on the equal-energy surface at $E=-0.8$. The parameters 
are $t=0.4$, $t_s=0.7$, $t_a=0.3$, $M=0.5$, $A_0=0.6$, and $\omega=10$.
}
\label{fig2}
\end{figure}

To verify the light-induced emergence of an out-of-plane 
$d$-wave MDSS, we calculate the quantity~\cite{Hayami2020AM1,Hayami2020AM2}, 
\begin{align}
{\rm Tr}[e^{-\beta \mathcal{H}_{\rm eff}(\bm{k})}s_z]=\sum_s\frac{(-\beta)^s}{s!}g_s^z(\bm{k}),
\end{align}
where $\beta$ represents the inverse temperature. Physically, this trace represents the unnormalized 
expectation value of the spin polarization along the $z$-axis for a given momentum $\bm{k}$. 
Performing a high-temperature expansion in powers of $\beta$, the $s$th-order coefficient $g_s^z(\bm{k})$
determines an effective spin-splitting field whose momentum dependence will be manifested as the spin texture. 
The leading, non-vanishing contribution to the spin polarization is given by the term of lowest order 
$s$ for which $g_s^z(\bm{k})\neq0$. We find that this leading coefficient is (details are provided in the 
SM Section III~\cite{supplemental})
\begin{align}
g_5^z(\bm{k})=&{\rm Tr}[(\mathcal{H}_{\rm eff}(\bm{k}))^5s_z],\nonumber\\
=&-64J_0(A_0)J_0(\sqrt{2}A_0)tt_sF(A_0,\omega)M^2\nonumber\\
&[2+\cos{(2k_x)}+\cos{(2k_y)}]\sin{(2k_x)}\sin{(2k_y)}.
\end{align}
Near the $\Gamma$ point with $\bm{k}\rightarrow\bm{0}$, the coefficient can be approximated as 
\begin{align}
g_5^z(\bm{k})&\simeq-1024J_0(A_0)J_0(\sqrt{2}A_0)tt_sF(A_0,\omega)M^2k_xk_y.
\end{align}
The momentum dependence, $\propto k_xk_y$, exhibits a clear $d_{xy}$-wave symmetry. 
Crucially, the spin polarization is explicitly light-induced, as $g_5^z(\bm{k})$ 
vanishes if the light-dependent factor $F(A_0,\omega)$ is zero. Furthermore, 
$F(A_0,\omega)$ depends linearly on the antisymmetric intralayer hopping
$t_{a}$. Since a nonzero $t_{a}$ is essential for the formation 
of Dirac points, this linear dependence directly demonstrates that the 
underlying Dirac band structure is indispensable for generating the light-induced
$d$-wave spin polarization.
 
To unambiguously demonstrate the generation of $d$-wave MDSS, we directly calculate the band spin polarization. 
As shown in Fig.~\ref{fig2}(a), the bands exhibit a MDSS under CPL, where
$\langle s_{x,y}(\bk)\rangle$ vanishes (thereby not shown in the figure) but $\langle s_{z}(\bk)\rangle$ is nonzero. 
The corresponding Fermi surface, plotted for a given Fermi energy, reveals a spin texture with clear
$d_{xy}$-wave symmetry, confirming our analytical prediction. 
In the high-frequency off-resonant regime, the MDSS is generically small because
$F(A_0,\omega)\propto 1/\omega$. Following the off-resonant high-frequency condition, we present the spin splitting at $\omega=10$ in Fig.~\ref{fig2} based on the effective Hamiltonian, where the $d$-wave texture is clearly visible. However, we emphasize that the emergence of $d$-wave spin splitting is symmetry-dictated and thus insensitive to the frequency condition. In other frequency regimes, the band structure can be calculated using the full Floquet Hamiltonian with a proper cutoff, and the $d$-wave spin splitting also exists. When the frequency becomes sufficiently large, the spectrum obtained from the full Floquet Hamiltonian becomes consistent with that obtained from the effective Hamiltonian (see SM Section IV~\cite{supplemental}). 

To underscore the decisive role of symmetry in determining light-induced spin textures, 
we contrast our results with a prior study of a coplanar chiral AFM on 
the kagomé lattice~\cite{Hu2025NAFM2}. There, adjacent moments form a  $120^{\circ}$ angle. 
In the undriven system, the lack of $\mathcal{PT}$ symmetry (while $[\bar{C}_{2z}\|\mathcal{T}]$ is preserved) 
leads to an even-parity, in-plane spin polarization with a winding texture. Under CPL, 
this system develops a nonzero $\langle s_{z}(\bm{k})\rangle$ with 
$s$-wave symmetry, producing a finite net spin magnetization. In stark contrast, 
our model produces a  $d$-wave MDSS that inherently yields zero net magnetization. 
This zero magnetization is enforced not only by the $[C_{2(x-y)}\|C_{4z}|\bm{\tau}_{x}]$ symmetry, but also by the 
symmetries $[\bar{E}\|\mathcal{T}C_{2x}\mathcal{M}_{z}|\bm{\tau}_{x}]$ and $[\bar{E}\|\mathcal{T}C_{2y}\mathcal{M}_{z}|\bm{\tau}_{y}]$. 
The latter two symmetries  impose the sign reversals
$\langle s_{z}(k_{x},k_{y})\rangle=-\langle s_{z}(k_{x},-k_{y})\rangle$ and 
 $\langle s_{z}(k_{x},k_{y})\rangle=-\langle s_{z}(-k_{x},k_{y})\rangle$, respectively. 
In magnetic space group terminology, their action is equivalent to that of the vertical 
 mirror symmetries $\mathcal{M}_x$ and $\mathcal{M}_y$.
Consequently, $\langle s_{z}(\bm{k})\rangle$ must vanish along 
$k_x=0$ and $k_y=0$, forming the characteristic nodal lines of a $d_{xy}$-wave symmetry. 
As we will show, this robust nodal structure is key to stabilizing the
$d$-wave MDSS even when the $[C_{2(x-y)}\|C_{4z}|\bm{\tau}_{x}]$ symmetry is broken by spin canting.

\begin{figure}[t]
\centering
\includegraphics[width=0.45\textwidth]{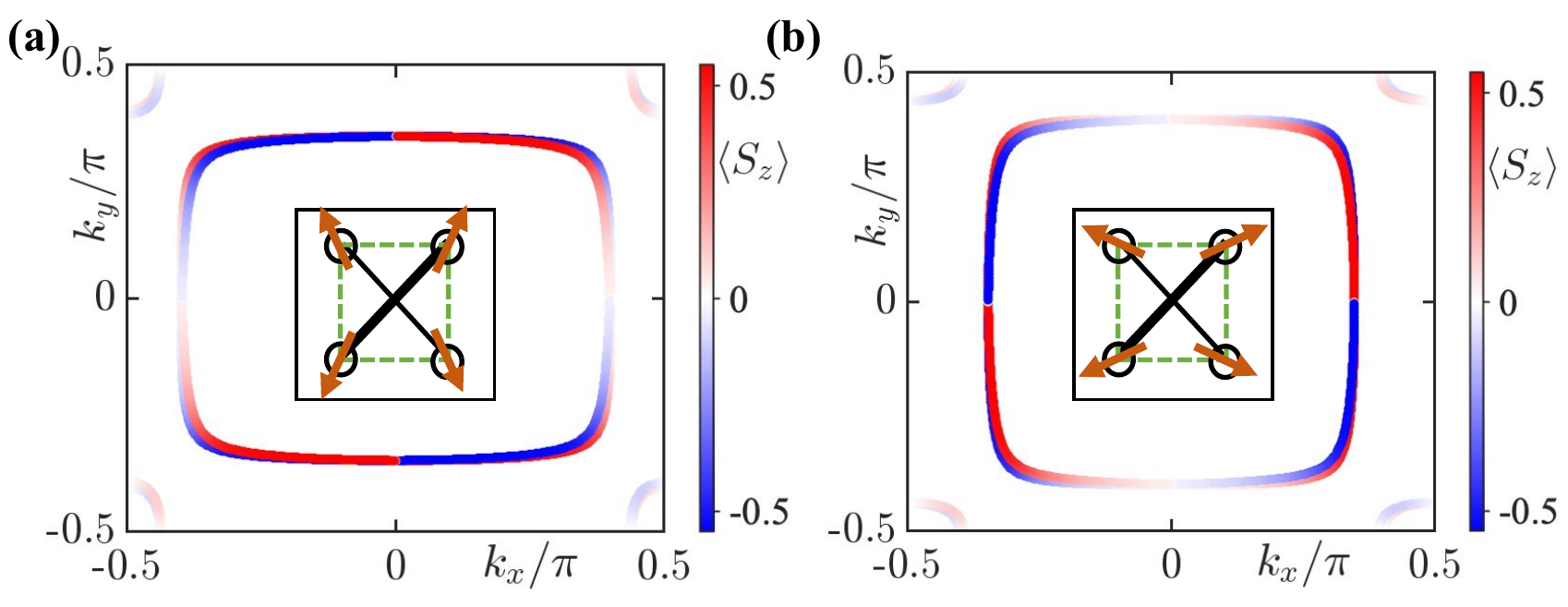}
\caption {Spin polarization on the equal-energy surface at $E=-0.8$. The corresponding
 real-space canted magnetic moment configurations are shown in the insets. 
(a) Canting along the $y$ direction ($M_x=0.3$, $M_y=0.6$). (b) Canting along the $x$ direction ($M_x=0.6$, $M_y=0.3$). 
Other parameters are $t=0.4$, $t_s=0.7$, $t_a=0.3$, $A_0=0.6$, and $\omega=10$.
}\label{fig3}
\end{figure}

{\it Robustness of the $d$-wave MDSS against canting.---}The Floquet Hamiltonian in Eq.~(\ref{Heff}) employs a 
symmetric exchange field with $M_x=M_y=M$ to describe the high-symmetry moment configuration, which preserves 
$[\bar{E}\|\mathcal{T}C_{2x}\mathcal{M}_{z}|\bm{\tau}_{x}]$, $[\bar{E}\|\mathcal{T}C_{2y}\mathcal{M}_{z}|\bm{\tau}_{y}]$, 
and $[C_{2(x-y)}\|C_{4z}|\bm{\tau}_{x}]$ symmetries. 
Introducing spin canting via $M_x \neq M_y$ explicitly breaks the $[C_{2(x-y)}\|C_{4z}|\bm{\tau}_{x}]$ symmetry while preserving $[\bar{E}\|\mathcal{T}C_{2x}\mathcal{M}_{z}|\bm{\tau}_{x}]$ and $[\bar{E}\|\mathcal{T}C_{2y}\mathcal{M}_{z}|\bm{\tau}_{y}]$. 
This symmetry reduction induces a structural 
transition of the Fermi surface from a square [Fig.~\ref{fig2}(b)] to a rectangular geometry (Fig.~\ref{fig3}). 
As shown in Figs.~\ref{fig3}(a) and \ref{fig3}(b), the $d$-wave spin texture on the Fermi surface persists 
for both canting cases. As discussed, this stability stems from the nodal lines of $\langle s_{z}(\bm{k})\rangle$ enforced by $[\bar{E}\|\mathcal{T}C_{2x}\mathcal{M}_{z}|\bm{\tau}_{x}]$ and $[\bar{E}\|\mathcal{T}C_{2y}\mathcal{M}_{z}|\bm{\tau}_{y}]$. 
These nodal lines partition the Brillouin zone into quadrants, where the spin 
polarization reverses sign under reflection across $k_{x}=0$ or  $k_{y}=0$, thereby sustaining a deformed 
$d$-wave texture despite the symmetry reduction. This robustness against spin canting is a notable feature 
that could facilitate the experimental realization of this spin-split phase.

\begin{figure}[t]
\centering
\includegraphics[width=0.45\textwidth]{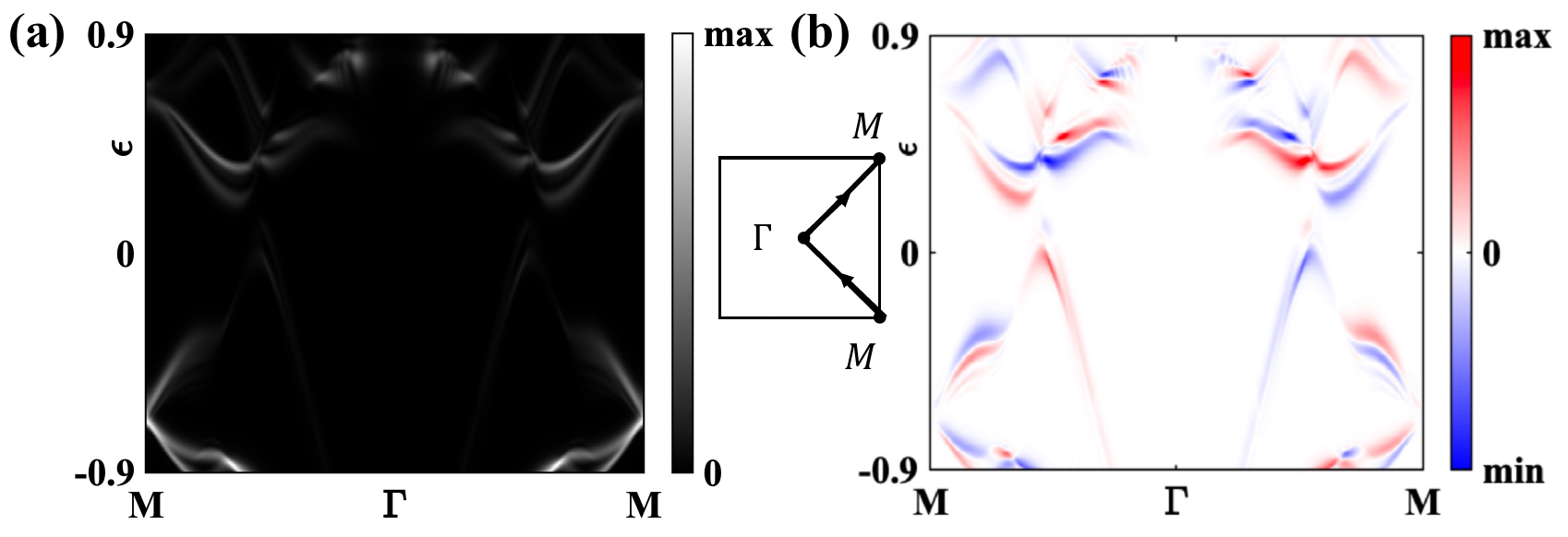}
\caption{(Color online) Simulated (a) time-resolved photocurrent and (b) spin-resolved photocurrent along the selected high-symmetry path $M-\Gamma-M$. The middle inset shows this path in the Brillouin zone. The parameters used in the calculation are $T_{\rm probe}=80$, $T_{\rm pump}=100$, $\omega=2$, $A_0=0.6$, $t_d=20$, $\mu=0.5$, $T=0.1$, $t=0.4$, $t_s=0.7$, $t_a=0.3$, and $M=0.5$. }\label{fig4}
\end{figure}

{\it Time-, spin- and angle-resolved photoemission photocurrent as an experimental fingerprint.---}The applying of CPL gives rise to a distinctive $d$-wave spin splitting. To detect the light-induced spin splitting in our coplanar AFM system, pump-probe experiments with tr-SARPES provide a viable approach\cite{Zhu2025AM8,SentefTrARPES,FreeeicksTrARPES,FarrellTrARPES}. Based on our model, we simulate the time- and spin-resolved energy spectra in a pump-probe setup. Specifically, the pump laser vector potential and the probe pulse are assumed to take the Gaussian function form of
\begin{align}
 \bm{A}_{\rm pump}(t)=e^{-\frac{(t-t_0)^2}{2T_{\rm pump}^2}}\bm{A}(t),\quad s(t,t_d)=e^{-\frac{(t-t_d)^2}{2T_{\rm probe}^2}},
\end{align}
where $\bm{A}(t)=A_0(\cos{\omega t},\sin{\omega t})$,  $T_{\rm pump}$ and $T_{\rm probe}$ denotes the pump and probe pulse width, 
and $t_0$ and $t_d$ are their amplitude-peak times.   We set $t_0=0$ without loss of generality; $t_{d}$ then characterizes the delay between the pump and probe peaks. The time-resolved photocurrent (tr-ARPES intensity) at momenta $\bm{k}$, binding energy $\epsilon$, and pump-probe delay time $t_{d}$ is obtained as
\begin{align}
I(\bm{k},\epsilon, t_{d})=&{\rm Im}\int dt_1\int dt_2s(t_1, t_{d}) \nonumber\\
&s(t_2, t_{d})e^{i\epsilon (t_1-t_2)} \text{Tr}[G^<(\bm{k},t_1,t_2)],
\end{align}
where $G^<(\bm{k},t_1,t_2)$ represents the lesser Green's function. To obtain the spin polarization information, 
one replaces the trace of lesser Green's function by ${\rm Tr}[s_zG^<(\bm{k},t_1,t_2)]$ (details are provided in the 
SM Section VI~\cite{supplemental}). The detection can be regarded as a quench process, in which the lesser Green's function characterizes the corresponding time evolution from the original equilibrium state.  

In Fig.~\ref{fig4}, we present the time-resolved photocurrent and spin-resolved photocurrent along the selected high-symmetry path $M-\Gamma-M$
in the Brillouin zone. Our calculations are performed with the following parameters: pump pulse duration  $T_{\rm pump}=100$, probe pulse duration $T_{\rm probe}=80$,  CPL of frequency $\omega=2$ and amplitude $A_0=0.6$, delay time $t_d=20$, initial-state chemical potential $\mu=0.5$, temperature $T=0.1$, and other model parameters $t=0.4$, $t_s=0.7$, $t_a=0.3$, and $M=0.5$. In the photocurrent spectrum shown in Fig.~\ref{fig4}(a), the light-induced spin-split band structure is clearly visible. More importantly, the $d$-wave spin splitting is manifest in the spin-resolved photocurrent spectrum shown in Fig.~\ref{fig4}(b), where the spin polarization is inverted after a $90^\circ$ rotation in the Brillouin zone. 

{\it Discussions and conclusions.---}In this work, we have established the general 
symmetry conditions for realizing even-parity,unidirectionally polarized MDSS in 
coplanar AFMs. As a proof of principle, we constructed a bilayer model with 
in-plane magnetic moments that satisfies these symmetry requirements. The application 
of CPL selectively breaks certain symmetries, particularly $[\bar{C}_{2\alpha}\|\mathcal{T}]$, inducing an 
out-of-plane spin splitting whose momentum-space pattern exhibits 
a $d$-wave symmetry. More broadly, our results reveal that the MDSS patterns in $[\bar{C}_{2\alpha}\|\mathcal{T}]$-preserved and $[\bar{C}_{2\alpha}\|\mathcal{T}]$-broken systems are like two sides of the same coin. Each type of MDSS in $[\bar{C}_{2\alpha}\|\mathcal{T}]$-preserved systems can have a counterpart with the opposite parity in $[\bar{C}_{2\alpha}\|\mathcal{T}]$-broken systems. Previous studies mainly focused on one side. Combined with previous discoveries of odd-parity 
unidirectionally polarized MDSS in coplanar AFMs and of both parities in collinear AFMs, 
our finding not only provides a foundational demonstration of the existence and experimental accessibility of 
even-parity unidirectionally polarized MDSS in coplanar AFM systems, but also gives a complete account of the symmetry-allowed spin-split phases with unidirectional spin polarization and well-defined parity in these two broad families of AFMs.

 Regarding experimental realization (detailed discussions are provided in SM Sections V–VIII~\cite{supplemental}), 
our symmetry analysis suggests two routes: 
(i) screening intrinsic coplanar AFMs for the key symmetries, 
validated by first-principles calculations; or (ii) synthetically constructing 
the required state in van der Waals heterostructures by stacking two collinear 
antiferromagnetic monolayers into a bilayer with the appropriate coplanar 
magnetic order.  We emphasize that the even-parity unidirectionally polarized spin-splitting phase
is generic rather than model-specific in $[\bar{C}_{2\alpha}\|\mathcal{T}]$-broken AFMs.
 Its realization can be readily generalized to other coplanar systems with different lattice structures, spin configurations, and symmetries.  Experimental verification is directly accessible through imaging of the spin-splitting texture via tr-SARPES, or through detection of its distinctive spin-transport signatures. 

{\it Note added.---}During the preparation of this work, we became aware of a related 
preprint (arXiv:2512.08901) that also discusses symmetry conditions for even-parity 
unidirectional spin polarization in coplanar AFMs~\cite{Song2025coplanar}. The coplanar AFMs studied 
therein respect the fundamental symmetry 
$[\bar{C}_{2z}\|\mathcal{T}]$, confining the even-parity spin polarization within the moment plane. 
In contrast, our work focuses on spin polarization along the axis perpendicular to the moment plane. 
These two studies are complementary, together providing a complete picture of the conditions for generating 
even-parity unidirectionally polarized MDSS along a general direction in coplanar AFMs.

{\it Acknowledgements.---}This work is supported by the National Natural Science Foundation of China (Grant No. 12174455, No. 12474264),  Guangdong Basic and Applied Basic Research Foundation (Grant No. 2023B1515040023), Guangdong Provincial Quantum Science Strategic Initiative (Grant No. GDZX2404007) and
National Key R\&D Program of China (Grant No. 2022YFA1404103).

\bibliography{dirac.bib}

\begin{thebibliography}{126}%
\makeatletter
\providecommand \@ifxundefined [1]{%
 \@ifx{#1\undefined}
}%
\providecommand \@ifnum [1]{%
 \ifnum #1\expandafter \@firstoftwo
 \else \expandafter \@secondoftwo
 \fi
}%
\providecommand \@ifx [1]{%
 \ifx #1\expandafter \@firstoftwo
 \else \expandafter \@secondoftwo
 \fi
}%
\providecommand \natexlab [1]{#1}%
\providecommand \enquote  [1]{``#1''}%
\providecommand \bibnamefont  [1]{#1}%
\providecommand \bibfnamefont [1]{#1}%
\providecommand \citenamefont [1]{#1}%
\providecommand \href@noop [0]{\@secondoftwo}%
\providecommand \href [0]{\begingroup \@sanitize@url \@href}%
\providecommand \@href[1]{\@@startlink{#1}\@@href}%
\providecommand \@@href[1]{\endgroup#1\@@endlink}%
\providecommand \@sanitize@url [0]{\catcode `\\12\catcode `\$12\catcode
  `\&12\catcode `\#12\catcode `\^12\catcode `\_12\catcode `\%12\relax}%
\providecommand \@@startlink[1]{}%
\providecommand \@@endlink[0]{}%
\providecommand \url  [0]{\begingroup\@sanitize@url \@url }%
\providecommand \@url [1]{\endgroup\@href {#1}{\urlprefix }}%
\providecommand \urlprefix  [0]{URL }%
\providecommand \Eprint [0]{\href }%
\providecommand \doibase [0]{https://doi.org/}%
\providecommand \selectlanguage [0]{\@gobble}%
\providecommand \bibinfo  [0]{\@secondoftwo}%
\providecommand \bibfield  [0]{\@secondoftwo}%
\providecommand \translation [1]{[#1]}%
\providecommand \BibitemOpen [0]{}%
\providecommand \bibitemStop [0]{}%
\providecommand \bibitemNoStop [0]{.\EOS\space}%
\providecommand \EOS [0]{\spacefactor3000\relax}%
\providecommand \BibitemShut  [1]{\csname bibitem#1\endcsname}%
\let\auto@bib@innerbib\@empty
\bibitem [{\citenamefont {Liu}\ \emph {et~al.}(2022)\citenamefont {Liu},
  \citenamefont {Li}, \citenamefont {Han}, \citenamefont {Wan},\ and\
  \citenamefont {Liu}}]{Liu2022AM}%
  \BibitemOpen
  \bibfield  {author} {\bibinfo {author} {\bibfnamefont {P.}~\bibnamefont
  {Liu}}, \bibinfo {author} {\bibfnamefont {J.}~\bibnamefont {Li}}, \bibinfo
  {author} {\bibfnamefont {J.}~\bibnamefont {Han}}, \bibinfo {author}
  {\bibfnamefont {X.}~\bibnamefont {Wan}},\ and\ \bibinfo {author}
  {\bibfnamefont {Q.}~\bibnamefont {Liu}},\ }\bibfield  {title} {\bibinfo
  {title} {Spin-group symmetry in magnetic materials with negligible spin-orbit
  coupling},\ }\href {https://doi.org/10.1103/PhysRevX.12.021016} {\bibfield
  {journal} {\bibinfo  {journal} {Phys. Rev. X}\ }\textbf {\bibinfo {volume}
  {12}},\ \bibinfo {pages} {021016} (\bibinfo {year} {2022})}\BibitemShut
  {NoStop}%
\bibitem [{\citenamefont {Xiao}\ \emph {et~al.}(2024)\citenamefont {Xiao},
  \citenamefont {Zhao}, \citenamefont {Li}, \citenamefont {Shindou},\ and\
  \citenamefont {Song}}]{Xiao2024SSG}%
  \BibitemOpen
  \bibfield  {author} {\bibinfo {author} {\bibfnamefont {Z.}~\bibnamefont
  {Xiao}}, \bibinfo {author} {\bibfnamefont {J.}~\bibnamefont {Zhao}}, \bibinfo
  {author} {\bibfnamefont {Y.}~\bibnamefont {Li}}, \bibinfo {author}
  {\bibfnamefont {R.}~\bibnamefont {Shindou}},\ and\ \bibinfo {author}
  {\bibfnamefont {Z.-D.}\ \bibnamefont {Song}},\ }\bibfield  {title} {\bibinfo
  {title} {Spin space groups: Full classification and applications},\ }\href
  {https://doi.org/10.1103/PhysRevX.14.031037} {\bibfield  {journal} {\bibinfo
  {journal} {Phys. Rev. X}\ }\textbf {\bibinfo {volume} {14}},\ \bibinfo
  {pages} {031037} (\bibinfo {year} {2024})}\BibitemShut {NoStop}%
\bibitem [{\citenamefont {Jiang}\ \emph {et~al.}(2024)\citenamefont {Jiang},
  \citenamefont {Song}, \citenamefont {Zhu}, \citenamefont {Fang},
  \citenamefont {Weng}, \citenamefont {Liu}, \citenamefont {Yang},\ and\
  \citenamefont {Fang}}]{Yi2024SSG}%
  \BibitemOpen
  \bibfield  {author} {\bibinfo {author} {\bibfnamefont {Y.}~\bibnamefont
  {Jiang}}, \bibinfo {author} {\bibfnamefont {Z.}~\bibnamefont {Song}},
  \bibinfo {author} {\bibfnamefont {T.}~\bibnamefont {Zhu}}, \bibinfo {author}
  {\bibfnamefont {Z.}~\bibnamefont {Fang}}, \bibinfo {author} {\bibfnamefont
  {H.}~\bibnamefont {Weng}}, \bibinfo {author} {\bibfnamefont {Z.-X.}\
  \bibnamefont {Liu}}, \bibinfo {author} {\bibfnamefont {J.}~\bibnamefont
  {Yang}},\ and\ \bibinfo {author} {\bibfnamefont {C.}~\bibnamefont {Fang}},\
  }\bibfield  {title} {\bibinfo {title} {Enumeration of spin-space groups:
  Toward a complete description of symmetries of magnetic orders},\ }\href
  {https://doi.org/10.1103/PhysRevX.14.031039} {\bibfield  {journal} {\bibinfo
  {journal} {Phys. Rev. X}\ }\textbf {\bibinfo {volume} {14}},\ \bibinfo
  {pages} {031039} (\bibinfo {year} {2024})}\BibitemShut {NoStop}%
\bibitem [{\citenamefont {Chen}\ \emph {et~al.}(2024)\citenamefont {Chen},
  \citenamefont {Ren}, \citenamefont {Zhu}, \citenamefont {Yu}, \citenamefont
  {Zhang}, \citenamefont {Liu}, \citenamefont {Li}, \citenamefont {Liu},
  \citenamefont {Li},\ and\ \citenamefont {Liu}}]{Chen2023AM}%
  \BibitemOpen
  \bibfield  {author} {\bibinfo {author} {\bibfnamefont {X.}~\bibnamefont
  {Chen}}, \bibinfo {author} {\bibfnamefont {J.}~\bibnamefont {Ren}}, \bibinfo
  {author} {\bibfnamefont {Y.}~\bibnamefont {Zhu}}, \bibinfo {author}
  {\bibfnamefont {Y.}~\bibnamefont {Yu}}, \bibinfo {author} {\bibfnamefont
  {A.}~\bibnamefont {Zhang}}, \bibinfo {author} {\bibfnamefont
  {P.}~\bibnamefont {Liu}}, \bibinfo {author} {\bibfnamefont {J.}~\bibnamefont
  {Li}}, \bibinfo {author} {\bibfnamefont {Y.}~\bibnamefont {Liu}}, \bibinfo
  {author} {\bibfnamefont {C.}~\bibnamefont {Li}},\ and\ \bibinfo {author}
  {\bibfnamefont {Q.}~\bibnamefont {Liu}},\ }\bibfield  {title} {\bibinfo
  {title} {{Enumeration and Representation Theory of Spin Space Groups}},\
  }\href {https://doi.org/10.1103/PhysRevX.14.031038} {\bibfield  {journal}
  {\bibinfo  {journal} {Phys. Rev. X}\ }\textbf {\bibinfo {volume} {14}},\
  \bibinfo {pages} {031038} (\bibinfo {year} {2024})}\BibitemShut {NoStop}%
\bibitem [{\citenamefont {Hayami}\ \emph {et~al.}(2019)\citenamefont {Hayami},
  \citenamefont {Yanagi},\ and\ \citenamefont {Kusunose}}]{Hayami2019AM}%
  \BibitemOpen
  \bibfield  {author} {\bibinfo {author} {\bibfnamefont {S.}~\bibnamefont
  {Hayami}}, \bibinfo {author} {\bibfnamefont {Y.}~\bibnamefont {Yanagi}},\
  and\ \bibinfo {author} {\bibfnamefont {H.}~\bibnamefont {Kusunose}},\
  }\bibfield  {title} {\bibinfo {title} {Momentum-dependent spin splitting by
  collinear antiferromagnetic ordering},\ }\href
  {https://doi.org/10.7566/JPSJ.88.123702} {\bibfield  {journal} {\bibinfo
  {journal} {Journal of the Physical Society of Japan}\ }\textbf {\bibinfo
  {volume} {88}},\ \bibinfo {pages} {123702} (\bibinfo {year}
  {2019})}\BibitemShut {NoStop}%
\bibitem [{\citenamefont {Yuan}\ \emph {et~al.}(2020)\citenamefont {Yuan},
  \citenamefont {Wang}, \citenamefont {Luo}, \citenamefont {Rashba},\ and\
  \citenamefont {Zunger}}]{LDYuan2020}%
  \BibitemOpen
  \bibfield  {author} {\bibinfo {author} {\bibfnamefont {L.-D.}\ \bibnamefont
  {Yuan}}, \bibinfo {author} {\bibfnamefont {Z.}~\bibnamefont {Wang}}, \bibinfo
  {author} {\bibfnamefont {J.-W.}\ \bibnamefont {Luo}}, \bibinfo {author}
  {\bibfnamefont {E.~I.}\ \bibnamefont {Rashba}},\ and\ \bibinfo {author}
  {\bibfnamefont {A.}~\bibnamefont {Zunger}},\ }\bibfield  {title} {\bibinfo
  {title} {Giant momentum-dependent spin splitting in centrosymmetric low-$z$
  antiferromagnets},\ }\href {https://doi.org/10.1103/PhysRevB.102.014422}
  {\bibfield  {journal} {\bibinfo  {journal} {Phys. Rev. B}\ }\textbf {\bibinfo
  {volume} {102}},\ \bibinfo {pages} {014422} (\bibinfo {year}
  {2020})}\BibitemShut {NoStop}%
\bibitem [{\citenamefont {Yuan}\ \emph {et~al.}(2021)\citenamefont {Yuan},
  \citenamefont {Wang}, \citenamefont {Luo},\ and\ \citenamefont
  {Zunger}}]{LDYuan2021}%
  \BibitemOpen
  \bibfield  {author} {\bibinfo {author} {\bibfnamefont {L.-D.}\ \bibnamefont
  {Yuan}}, \bibinfo {author} {\bibfnamefont {Z.}~\bibnamefont {Wang}}, \bibinfo
  {author} {\bibfnamefont {J.-W.}\ \bibnamefont {Luo}},\ and\ \bibinfo {author}
  {\bibfnamefont {A.}~\bibnamefont {Zunger}},\ }\bibfield  {title} {\bibinfo
  {title} {Prediction of low-z collinear and noncollinear antiferromagnetic
  compounds having momentum-dependent spin splitting even without spin-orbit
  coupling},\ }\href {https://doi.org/10.1103/PhysRevMaterials.5.014409}
  {\bibfield  {journal} {\bibinfo  {journal} {Phys. Rev. Mater.}\ }\textbf
  {\bibinfo {volume} {5}},\ \bibinfo {pages} {014409} (\bibinfo {year}
  {2021})}\BibitemShut {NoStop}%
\bibitem [{\citenamefont {Ma}\ \emph {et~al.}(2021)\citenamefont {Ma},
  \citenamefont {Hu}, \citenamefont {Li}, \citenamefont {Liu}, \citenamefont
  {Yao}, \citenamefont {Jia},\ and\ \citenamefont {Liu}}]{Ma2021AM}%
  \BibitemOpen
  \bibfield  {author} {\bibinfo {author} {\bibfnamefont {H.-Y.}\ \bibnamefont
  {Ma}}, \bibinfo {author} {\bibfnamefont {M.}~\bibnamefont {Hu}}, \bibinfo
  {author} {\bibfnamefont {N.}~\bibnamefont {Li}}, \bibinfo {author}
  {\bibfnamefont {J.}~\bibnamefont {Liu}}, \bibinfo {author} {\bibfnamefont
  {W.}~\bibnamefont {Yao}}, \bibinfo {author} {\bibfnamefont {J.-F.}\
  \bibnamefont {Jia}},\ and\ \bibinfo {author} {\bibfnamefont {J.}~\bibnamefont
  {Liu}},\ }\bibfield  {title} {\bibinfo {title} {Multifunctional
  antiferromagnetic materials with giant piezomagnetism and noncollinear spin
  current},\ }\href {https://doi.org/10.1038/s41467-021-23127-7} {\bibfield
  {journal} {\bibinfo  {journal} {Nature Communications}\ }\textbf {\bibinfo
  {volume} {12}},\ \bibinfo {pages} {2846} (\bibinfo {year}
  {2021})}\BibitemShut {NoStop}%
\bibitem [{\citenamefont {Shao}\ \emph {et~al.}(2021)\citenamefont {Shao},
  \citenamefont {Zhang}, \citenamefont {Li}, \citenamefont {Eom},\ and\
  \citenamefont {Tsymbal}}]{Shao2021NC}%
  \BibitemOpen
  \bibfield  {author} {\bibinfo {author} {\bibfnamefont {D.-F.}\ \bibnamefont
  {Shao}}, \bibinfo {author} {\bibfnamefont {S.-H.}\ \bibnamefont {Zhang}},
  \bibinfo {author} {\bibfnamefont {M.}~\bibnamefont {Li}}, \bibinfo {author}
  {\bibfnamefont {C.-B.}\ \bibnamefont {Eom}},\ and\ \bibinfo {author}
  {\bibfnamefont {E.~Y.}\ \bibnamefont {Tsymbal}},\ }\bibfield  {title}
  {\bibinfo {title} {Spin-neutral currents for spintronics},\ }\href
  {https://doi.org/10.1038/s41467-021-26915-3} {\bibfield  {journal} {\bibinfo
  {journal} {Nature Communications}\ }\textbf {\bibinfo {volume} {12}},\
  \bibinfo {pages} {7061} (\bibinfo {year} {2021})}\BibitemShut {NoStop}%
\bibitem [{\citenamefont {\ifmmode~\check{S}\else \v{S}\fi{}mejkal}\ \emph
  {et~al.}(2022{\natexlab{a}})\citenamefont {\ifmmode~\check{S}\else
  \v{S}\fi{}mejkal}, \citenamefont {Sinova},\ and\ \citenamefont
  {Jungwirth}}]{libor2022AMa}%
  \BibitemOpen
  \bibfield  {author} {\bibinfo {author} {\bibfnamefont {L.}~\bibnamefont
  {\ifmmode~\check{S}\else \v{S}\fi{}mejkal}}, \bibinfo {author} {\bibfnamefont
  {J.}~\bibnamefont {Sinova}},\ and\ \bibinfo {author} {\bibfnamefont
  {T.}~\bibnamefont {Jungwirth}},\ }\bibfield  {title} {\bibinfo {title}
  {{Beyond Conventional Ferromagnetism and Antiferromagnetism: A Phase with
  Nonrelativistic Spin and Crystal Rotation Symmetry}},\ }\href
  {https://doi.org/10.1103/PhysRevX.12.031042} {\bibfield  {journal} {\bibinfo
  {journal} {Phys. Rev. X}\ }\textbf {\bibinfo {volume} {12}},\ \bibinfo
  {pages} {031042} (\bibinfo {year} {2022}{\natexlab{a}})}\BibitemShut
  {NoStop}%
\bibitem [{\citenamefont {\ifmmode~\check{S}\else \v{S}\fi{}mejkal}\ \emph
  {et~al.}(2022{\natexlab{b}})\citenamefont {\ifmmode~\check{S}\else
  \v{S}\fi{}mejkal}, \citenamefont {Sinova},\ and\ \citenamefont
  {Jungwirth}}]{Libor2022AMb}%
  \BibitemOpen
  \bibfield  {author} {\bibinfo {author} {\bibfnamefont {L.}~\bibnamefont
  {\ifmmode~\check{S}\else \v{S}\fi{}mejkal}}, \bibinfo {author} {\bibfnamefont
  {J.}~\bibnamefont {Sinova}},\ and\ \bibinfo {author} {\bibfnamefont
  {T.}~\bibnamefont {Jungwirth}},\ }\bibfield  {title} {\bibinfo {title}
  {{Emerging Research Landscape of Altermagnetism}},\ }\href
  {https://doi.org/10.1103/PhysRevX.12.040501} {\bibfield  {journal} {\bibinfo
  {journal} {Phys. Rev. X}\ }\textbf {\bibinfo {volume} {12}},\ \bibinfo
  {pages} {040501} (\bibinfo {year} {2022}{\natexlab{b}})}\BibitemShut
  {NoStop}%
\bibitem [{\citenamefont {\ifmmode~\check{S}\else \v{S}\fi{}mejkal}\ \emph
  {et~al.}(2022{\natexlab{c}})\citenamefont {\ifmmode~\check{S}\else
  \v{S}\fi{}mejkal}, \citenamefont {Hellenes}, \citenamefont
  {Gonz\'alez-Hern\'andez}, \citenamefont {Sinova},\ and\ \citenamefont
  {Jungwirth}}]{Libor2022AMc}%
  \BibitemOpen
  \bibfield  {author} {\bibinfo {author} {\bibfnamefont {L.}~\bibnamefont
  {\ifmmode~\check{S}\else \v{S}\fi{}mejkal}}, \bibinfo {author} {\bibfnamefont
  {A.~B.}\ \bibnamefont {Hellenes}}, \bibinfo {author} {\bibfnamefont
  {R.}~\bibnamefont {Gonz\'alez-Hern\'andez}}, \bibinfo {author} {\bibfnamefont
  {J.}~\bibnamefont {Sinova}},\ and\ \bibinfo {author} {\bibfnamefont
  {T.}~\bibnamefont {Jungwirth}},\ }\bibfield  {title} {\bibinfo {title}
  {{Giant and Tunneling Magnetoresistance in Unconventional Collinear
  Antiferromagnets with Nonrelativistic Spin-Momentum Coupling}},\ }\href
  {https://doi.org/10.1103/PhysRevX.12.011028} {\bibfield  {journal} {\bibinfo
  {journal} {Phys. Rev. X}\ }\textbf {\bibinfo {volume} {12}},\ \bibinfo
  {pages} {011028} (\bibinfo {year} {2022}{\natexlab{c}})}\BibitemShut
  {NoStop}%
\bibitem [{\citenamefont {Mazin}\ \emph {et~al.}(2021)\citenamefont {Mazin},
  \citenamefont {Koepernik}, \citenamefont {Johannes}, \citenamefont
  {González-Hernández},\ and\ \citenamefont {\ifmmode~\check{S}\else
  \v{S}\fi{}mejkal}}]{Mazin2021}%
  \BibitemOpen
  \bibfield  {author} {\bibinfo {author} {\bibfnamefont {I.~I.}\ \bibnamefont
  {Mazin}}, \bibinfo {author} {\bibfnamefont {K.}~\bibnamefont {Koepernik}},
  \bibinfo {author} {\bibfnamefont {M.~D.}\ \bibnamefont {Johannes}}, \bibinfo
  {author} {\bibfnamefont {R.}~\bibnamefont {González-Hernández}},\ and\
  \bibinfo {author} {\bibfnamefont {L.}~\bibnamefont {\ifmmode~\check{S}\else
  \v{S}\fi{}mejkal}},\ }\bibfield  {title} {\bibinfo {title} {Prediction of
  unconventional magnetism in doped {F}e{S}b$_{2}$},\ }\href
  {https://doi.org/10.1073/pnas.2108924118} {\bibfield  {journal} {\bibinfo
  {journal} {Proceedings of the National Academy of Sciences}\ }\textbf
  {\bibinfo {volume} {118}},\ \bibinfo {pages} {e2108924118} (\bibinfo {year}
  {2021})}\BibitemShut {NoStop}%
\bibitem [{\citenamefont {Hu}\ \emph {et~al.}(2025{\natexlab{a}})\citenamefont
  {Hu}, \citenamefont {Cheng}, \citenamefont {Huang},\ and\ \citenamefont
  {Liu}}]{Hu2025CAM}%
  \BibitemOpen
  \bibfield  {author} {\bibinfo {author} {\bibfnamefont {M.}~\bibnamefont
  {Hu}}, \bibinfo {author} {\bibfnamefont {X.}~\bibnamefont {Cheng}}, \bibinfo
  {author} {\bibfnamefont {Z.}~\bibnamefont {Huang}},\ and\ \bibinfo {author}
  {\bibfnamefont {J.}~\bibnamefont {Liu}},\ }\bibfield  {title} {\bibinfo
  {title} {Catalog of $c$-paired spin-momentum locking in antiferromagnetic
  systems},\ }\href {https://doi.org/10.1103/PhysRevX.15.021083} {\bibfield
  {journal} {\bibinfo  {journal} {Phys. Rev. X}\ }\textbf {\bibinfo {volume}
  {15}},\ \bibinfo {pages} {021083} (\bibinfo {year}
  {2025}{\natexlab{a}})}\BibitemShut {NoStop}%
\bibitem [{\citenamefont {Zhu}\ \emph {et~al.}(2024{\natexlab{a}})\citenamefont
  {Zhu}, \citenamefont {Chen}, \citenamefont {Liu}, \citenamefont {Liu},
  \citenamefont {Liu}, \citenamefont {Zha}, \citenamefont {Qu}, \citenamefont
  {Hong}, \citenamefont {Li}, \citenamefont {Jiang}, \citenamefont {Ma},
  \citenamefont {Hao}, \citenamefont {Zhu}, \citenamefont {Liu}, \citenamefont
  {Zeng}, \citenamefont {Jayaram}, \citenamefont {Lenger}, \citenamefont
  {Ding}, \citenamefont {Mo}, \citenamefont {Tanaka}, \citenamefont {Arita},
  \citenamefont {Liu}, \citenamefont {Ye}, \citenamefont {Shen}, \citenamefont
  {Wrachtrup}, \citenamefont {Huang}, \citenamefont {He}, \citenamefont {Qiao},
  \citenamefont {Liu},\ and\ \citenamefont {Liu}}]{Zhu2024observation}%
  \BibitemOpen
  \bibfield  {author} {\bibinfo {author} {\bibfnamefont {Y.~P.}\ \bibnamefont
  {Zhu}}, \bibinfo {author} {\bibfnamefont {X.}~\bibnamefont {Chen}}, \bibinfo
  {author} {\bibfnamefont {X.~R.}\ \bibnamefont {Liu}}, \bibinfo {author}
  {\bibfnamefont {Y.}~\bibnamefont {Liu}}, \bibinfo {author} {\bibfnamefont
  {P.}~\bibnamefont {Liu}}, \bibinfo {author} {\bibfnamefont {H.}~\bibnamefont
  {Zha}}, \bibinfo {author} {\bibfnamefont {G.}~\bibnamefont {Qu}}, \bibinfo
  {author} {\bibfnamefont {C.}~\bibnamefont {Hong}}, \bibinfo {author}
  {\bibfnamefont {J.}~\bibnamefont {Li}}, \bibinfo {author} {\bibfnamefont
  {Z.}~\bibnamefont {Jiang}}, \bibinfo {author} {\bibfnamefont {X.~M.}\
  \bibnamefont {Ma}}, \bibinfo {author} {\bibfnamefont {Y.~J.}\ \bibnamefont
  {Hao}}, \bibinfo {author} {\bibfnamefont {M.~Y.}\ \bibnamefont {Zhu}},
  \bibinfo {author} {\bibfnamefont {W.}~\bibnamefont {Liu}}, \bibinfo {author}
  {\bibfnamefont {M.}~\bibnamefont {Zeng}}, \bibinfo {author} {\bibfnamefont
  {S.}~\bibnamefont {Jayaram}}, \bibinfo {author} {\bibfnamefont
  {M.}~\bibnamefont {Lenger}}, \bibinfo {author} {\bibfnamefont
  {J.}~\bibnamefont {Ding}}, \bibinfo {author} {\bibfnamefont {S.}~\bibnamefont
  {Mo}}, \bibinfo {author} {\bibfnamefont {K.}~\bibnamefont {Tanaka}}, \bibinfo
  {author} {\bibfnamefont {M.}~\bibnamefont {Arita}}, \bibinfo {author}
  {\bibfnamefont {Z.}~\bibnamefont {Liu}}, \bibinfo {author} {\bibfnamefont
  {M.}~\bibnamefont {Ye}}, \bibinfo {author} {\bibfnamefont {D.}~\bibnamefont
  {Shen}}, \bibinfo {author} {\bibfnamefont {J.}~\bibnamefont {Wrachtrup}},
  \bibinfo {author} {\bibfnamefont {Y.}~\bibnamefont {Huang}}, \bibinfo
  {author} {\bibfnamefont {R.~H.}\ \bibnamefont {He}}, \bibinfo {author}
  {\bibfnamefont {S.}~\bibnamefont {Qiao}}, \bibinfo {author} {\bibfnamefont
  {Q.}~\bibnamefont {Liu}},\ and\ \bibinfo {author} {\bibfnamefont
  {C.}~\bibnamefont {Liu}},\ }\bibfield  {title} {\bibinfo {title}
  {{Observation of plaid-like spin splitting in a noncoplanar
  antiferromagnet}},\ }\href {https://doi.org/10.1038/s41586-024-07023-w}
  {\bibfield  {journal} {\bibinfo  {journal} {Nature}\ }\textbf {\bibinfo
  {volume} {626}},\ \bibinfo {pages} {523} (\bibinfo {year}
  {2024}{\natexlab{a}})}\BibitemShut {NoStop}%
\bibitem [{\citenamefont {Lee}\ \emph {et~al.}(2024{\natexlab{a}})\citenamefont
  {Lee}, \citenamefont {Lee}, \citenamefont {Jung}, \citenamefont {Jung},
  \citenamefont {Kim}, \citenamefont {Lee}, \citenamefont {Seok}, \citenamefont
  {Kim}, \citenamefont {Park}, \citenamefont {\ifmmode~\check{S}\else
  \v{S}\fi{}mejkal}, \citenamefont {Kang},\ and\ \citenamefont
  {Kim}}]{Lee2024AM}%
  \BibitemOpen
  \bibfield  {author} {\bibinfo {author} {\bibfnamefont {S.}~\bibnamefont
  {Lee}}, \bibinfo {author} {\bibfnamefont {S.}~\bibnamefont {Lee}}, \bibinfo
  {author} {\bibfnamefont {S.}~\bibnamefont {Jung}}, \bibinfo {author}
  {\bibfnamefont {J.}~\bibnamefont {Jung}}, \bibinfo {author} {\bibfnamefont
  {D.}~\bibnamefont {Kim}}, \bibinfo {author} {\bibfnamefont {Y.}~\bibnamefont
  {Lee}}, \bibinfo {author} {\bibfnamefont {B.}~\bibnamefont {Seok}}, \bibinfo
  {author} {\bibfnamefont {J.}~\bibnamefont {Kim}}, \bibinfo {author}
  {\bibfnamefont {B.~G.}\ \bibnamefont {Park}}, \bibinfo {author}
  {\bibfnamefont {L.}~\bibnamefont {\ifmmode~\check{S}\else \v{S}\fi{}mejkal}},
  \bibinfo {author} {\bibfnamefont {C.-J.}\ \bibnamefont {Kang}},\ and\
  \bibinfo {author} {\bibfnamefont {C.}~\bibnamefont {Kim}},\ }\bibfield
  {title} {\bibinfo {title} {{Broken Kramers Degeneracy in Altermagnetic
  MnTe}},\ }\href {https://doi.org/10.1103/PhysRevLett.132.036702} {\bibfield
  {journal} {\bibinfo  {journal} {Phys. Rev. Lett.}\ }\textbf {\bibinfo
  {volume} {132}},\ \bibinfo {pages} {036702} (\bibinfo {year}
  {2024}{\natexlab{a}})}\BibitemShut {NoStop}%
\bibitem [{\citenamefont {Osumi}\ \emph {et~al.}(2024)\citenamefont {Osumi},
  \citenamefont {Souma}, \citenamefont {Aoyama}, \citenamefont {Yamauchi},
  \citenamefont {Honma}, \citenamefont {Nakayama}, \citenamefont {Takahashi},
  \citenamefont {Ohgushi},\ and\ \citenamefont {Sato}}]{Osumi2024MnTe}%
  \BibitemOpen
  \bibfield  {author} {\bibinfo {author} {\bibfnamefont {T.}~\bibnamefont
  {Osumi}}, \bibinfo {author} {\bibfnamefont {S.}~\bibnamefont {Souma}},
  \bibinfo {author} {\bibfnamefont {T.}~\bibnamefont {Aoyama}}, \bibinfo
  {author} {\bibfnamefont {K.}~\bibnamefont {Yamauchi}}, \bibinfo {author}
  {\bibfnamefont {A.}~\bibnamefont {Honma}}, \bibinfo {author} {\bibfnamefont
  {K.}~\bibnamefont {Nakayama}}, \bibinfo {author} {\bibfnamefont
  {T.}~\bibnamefont {Takahashi}}, \bibinfo {author} {\bibfnamefont
  {K.}~\bibnamefont {Ohgushi}},\ and\ \bibinfo {author} {\bibfnamefont
  {T.}~\bibnamefont {Sato}},\ }\bibfield  {title} {\bibinfo {title}
  {{Observation of a giant band splitting in altermagnetic MnTe}},\ }\href
  {https://doi.org/10.1103/PhysRevB.109.115102} {\bibfield  {journal} {\bibinfo
   {journal} {Phys. Rev. B}\ }\textbf {\bibinfo {volume} {109}},\ \bibinfo
  {pages} {115102} (\bibinfo {year} {2024})}\BibitemShut {NoStop}%
\bibitem [{\citenamefont {Reimers}\ \emph {et~al.}(2024)\citenamefont
  {Reimers}, \citenamefont {Odenbreit}, \citenamefont {{\v{S}}mejkal},
  \citenamefont {Strocov}, \citenamefont {Constantinou}, \citenamefont
  {Hellenes}, \citenamefont {Jaeschke~Ubiergo}, \citenamefont {Campos},
  \citenamefont {Bharadwaj}, \citenamefont {Chakraborty}, \citenamefont
  {Denneulin}, \citenamefont {Shi}, \citenamefont {Dunin-Borkowski},
  \citenamefont {Das}, \citenamefont {Kl{\"a}ui}, \citenamefont {Sinova},\ and\
  \citenamefont {Jourdan}}]{Reimers2024}%
  \BibitemOpen
  \bibfield  {author} {\bibinfo {author} {\bibfnamefont {S.}~\bibnamefont
  {Reimers}}, \bibinfo {author} {\bibfnamefont {L.}~\bibnamefont {Odenbreit}},
  \bibinfo {author} {\bibfnamefont {L.}~\bibnamefont {{\v{S}}mejkal}}, \bibinfo
  {author} {\bibfnamefont {V.~N.}\ \bibnamefont {Strocov}}, \bibinfo {author}
  {\bibfnamefont {P.}~\bibnamefont {Constantinou}}, \bibinfo {author}
  {\bibfnamefont {A.~B.}\ \bibnamefont {Hellenes}}, \bibinfo {author}
  {\bibfnamefont {R.}~\bibnamefont {Jaeschke~Ubiergo}}, \bibinfo {author}
  {\bibfnamefont {W.~H.}\ \bibnamefont {Campos}}, \bibinfo {author}
  {\bibfnamefont {V.~K.}\ \bibnamefont {Bharadwaj}}, \bibinfo {author}
  {\bibfnamefont {A.}~\bibnamefont {Chakraborty}}, \bibinfo {author}
  {\bibfnamefont {T.}~\bibnamefont {Denneulin}}, \bibinfo {author}
  {\bibfnamefont {W.}~\bibnamefont {Shi}}, \bibinfo {author} {\bibfnamefont
  {R.~E.}\ \bibnamefont {Dunin-Borkowski}}, \bibinfo {author} {\bibfnamefont
  {S.}~\bibnamefont {Das}}, \bibinfo {author} {\bibfnamefont {M.}~\bibnamefont
  {Kl{\"a}ui}}, \bibinfo {author} {\bibfnamefont {J.}~\bibnamefont {Sinova}},\
  and\ \bibinfo {author} {\bibfnamefont {M.}~\bibnamefont {Jourdan}},\
  }\bibfield  {title} {\bibinfo {title} {{Direct observation of altermagnetic
  band splitting in CrSb thin films}},\ }\href
  {https://doi.org/10.1038/s41467-024-46476-5} {\bibfield  {journal} {\bibinfo
  {journal} {Nature Communications}\ }\textbf {\bibinfo {volume} {15}},\
  \bibinfo {pages} {2116} (\bibinfo {year} {2024})}\BibitemShut {NoStop}%
\bibitem [{\citenamefont {Krempask{\'y}}\ \emph {et~al.}(2024)\citenamefont
  {Krempask{\'y}}, \citenamefont {{\v{S}}mejkal}, \citenamefont {D'Souza},
  \citenamefont {Hajlaoui}, \citenamefont {Springholz}, \citenamefont
  {Uhl{\'i}{\v{r}}ov{\'a}}, \citenamefont {Alarab}, \citenamefont
  {Constantinou}, \citenamefont {Strocov}, \citenamefont {Usanov},
  \citenamefont {Pudelko}, \citenamefont {Gonz{\'a}lez-Hern{\'a}ndez},
  \citenamefont {Birk~Hellenes}, \citenamefont {Jansa}, \citenamefont
  {Reichlov{\'a}}, \citenamefont {{\v{S}}ob{\'a}{\v{n}}}, \citenamefont
  {Gonzalez~Betancourt}, \citenamefont {Wadley}, \citenamefont {Sinova},
  \citenamefont {Kriegner}, \citenamefont {Min{\'a}r}, \citenamefont {Dil},\
  and\ \citenamefont {Jungwirth}}]{Krempasky2024}%
  \BibitemOpen
  \bibfield  {author} {\bibinfo {author} {\bibfnamefont {J.}~\bibnamefont
  {Krempask{\'y}}}, \bibinfo {author} {\bibfnamefont {L.}~\bibnamefont
  {{\v{S}}mejkal}}, \bibinfo {author} {\bibfnamefont {S.~W.}\ \bibnamefont
  {D'Souza}}, \bibinfo {author} {\bibfnamefont {M.}~\bibnamefont {Hajlaoui}},
  \bibinfo {author} {\bibfnamefont {G.}~\bibnamefont {Springholz}}, \bibinfo
  {author} {\bibfnamefont {K.}~\bibnamefont {Uhl{\'i}{\v{r}}ov{\'a}}}, \bibinfo
  {author} {\bibfnamefont {F.}~\bibnamefont {Alarab}}, \bibinfo {author}
  {\bibfnamefont {P.~C.}\ \bibnamefont {Constantinou}}, \bibinfo {author}
  {\bibfnamefont {V.}~\bibnamefont {Strocov}}, \bibinfo {author} {\bibfnamefont
  {D.}~\bibnamefont {Usanov}}, \bibinfo {author} {\bibfnamefont {W.~R.}\
  \bibnamefont {Pudelko}}, \bibinfo {author} {\bibfnamefont {R.}~\bibnamefont
  {Gonz{\'a}lez-Hern{\'a}ndez}}, \bibinfo {author} {\bibfnamefont
  {A.}~\bibnamefont {Birk~Hellenes}}, \bibinfo {author} {\bibfnamefont
  {Z.}~\bibnamefont {Jansa}}, \bibinfo {author} {\bibfnamefont
  {H.}~\bibnamefont {Reichlov{\'a}}}, \bibinfo {author} {\bibfnamefont
  {Z.}~\bibnamefont {{\v{S}}ob{\'a}{\v{n}}}}, \bibinfo {author} {\bibfnamefont
  {R.~D.}\ \bibnamefont {Gonzalez~Betancourt}}, \bibinfo {author}
  {\bibfnamefont {P.}~\bibnamefont {Wadley}}, \bibinfo {author} {\bibfnamefont
  {J.}~\bibnamefont {Sinova}}, \bibinfo {author} {\bibfnamefont
  {D.}~\bibnamefont {Kriegner}}, \bibinfo {author} {\bibfnamefont
  {J.}~\bibnamefont {Min{\'a}r}}, \bibinfo {author} {\bibfnamefont {J.~H.}\
  \bibnamefont {Dil}},\ and\ \bibinfo {author} {\bibfnamefont {T.}~\bibnamefont
  {Jungwirth}},\ }\bibfield  {title} {\bibinfo {title} {{Altermagnetic lifting
  of Kramers spin degeneracy}},\ }\href
  {https://doi.org/10.1038/s41586-023-06907-7} {\bibfield  {journal} {\bibinfo
  {journal} {Nature}\ }\textbf {\bibinfo {volume} {626}},\ \bibinfo {pages}
  {517} (\bibinfo {year} {2024})}\BibitemShut {NoStop}%
\bibitem [{\citenamefont {Yang}\ \emph {et~al.}(2025)\citenamefont {Yang},
  \citenamefont {Li}, \citenamefont {Yang}, \citenamefont {Li}, \citenamefont
  {Zheng}, \citenamefont {Zhu}, \citenamefont {Pan}, \citenamefont {Xu},
  \citenamefont {Cao}, \citenamefont {Zhao}, \citenamefont {Jana},
  \citenamefont {Zhang}, \citenamefont {Ye}, \citenamefont {Song},
  \citenamefont {Hu}, \citenamefont {Yang}, \citenamefont {Fujii},
  \citenamefont {Vobornik}, \citenamefont {Shi}, \citenamefont {Yuan},
  \citenamefont {Zhang}, \citenamefont {Xu},\ and\ \citenamefont
  {Liu}}]{Yang2024CrSb}%
  \BibitemOpen
  \bibfield  {author} {\bibinfo {author} {\bibfnamefont {G.}~\bibnamefont
  {Yang}}, \bibinfo {author} {\bibfnamefont {Z.}~\bibnamefont {Li}}, \bibinfo
  {author} {\bibfnamefont {S.}~\bibnamefont {Yang}}, \bibinfo {author}
  {\bibfnamefont {J.}~\bibnamefont {Li}}, \bibinfo {author} {\bibfnamefont
  {H.}~\bibnamefont {Zheng}}, \bibinfo {author} {\bibfnamefont
  {W.}~\bibnamefont {Zhu}}, \bibinfo {author} {\bibfnamefont {Z.}~\bibnamefont
  {Pan}}, \bibinfo {author} {\bibfnamefont {Y.}~\bibnamefont {Xu}}, \bibinfo
  {author} {\bibfnamefont {S.}~\bibnamefont {Cao}}, \bibinfo {author}
  {\bibfnamefont {W.}~\bibnamefont {Zhao}}, \bibinfo {author} {\bibfnamefont
  {A.}~\bibnamefont {Jana}}, \bibinfo {author} {\bibfnamefont {J.}~\bibnamefont
  {Zhang}}, \bibinfo {author} {\bibfnamefont {M.}~\bibnamefont {Ye}}, \bibinfo
  {author} {\bibfnamefont {Y.}~\bibnamefont {Song}}, \bibinfo {author}
  {\bibfnamefont {L.-H.}\ \bibnamefont {Hu}}, \bibinfo {author} {\bibfnamefont
  {L.}~\bibnamefont {Yang}}, \bibinfo {author} {\bibfnamefont {J.}~\bibnamefont
  {Fujii}}, \bibinfo {author} {\bibfnamefont {I.}~\bibnamefont {Vobornik}},
  \bibinfo {author} {\bibfnamefont {M.}~\bibnamefont {Shi}}, \bibinfo {author}
  {\bibfnamefont {H.}~\bibnamefont {Yuan}}, \bibinfo {author} {\bibfnamefont
  {Y.}~\bibnamefont {Zhang}}, \bibinfo {author} {\bibfnamefont
  {Y.}~\bibnamefont {Xu}},\ and\ \bibinfo {author} {\bibfnamefont
  {Y.}~\bibnamefont {Liu}},\ }\bibfield  {title} {\bibinfo {title}
  {{Three-dimensional mapping of the altermagnetic spin splitting in CrSb}},\
  }\href {https://doi.org/10.1038/s41467-025-56647-7} {\bibfield  {journal}
  {\bibinfo  {journal} {Nature Communications}\ }\textbf {\bibinfo {volume}
  {16}},\ \bibinfo {pages} {1442} (\bibinfo {year} {2025})}\BibitemShut
  {NoStop}%
\bibitem [{\citenamefont {Zeng}\ \emph {et~al.}(2024)\citenamefont {Zeng},
  \citenamefont {Zhu}, \citenamefont {Zhu}, \citenamefont {Liu}, \citenamefont
  {Ma}, \citenamefont {Hao}, \citenamefont {Liu}, \citenamefont {Qu},
  \citenamefont {Yang}, \citenamefont {Jiang}, \citenamefont {Yamagami},
  \citenamefont {Arita}, \citenamefont {Zhang}, \citenamefont {Shao},
  \citenamefont {Dai}, \citenamefont {Shimada}, \citenamefont {Liu},
  \citenamefont {Ye}, \citenamefont {Huang}, \citenamefont {Liu},\ and\
  \citenamefont {Liu}}]{Zeng2024CrSb}%
  \BibitemOpen
  \bibfield  {author} {\bibinfo {author} {\bibfnamefont {M.}~\bibnamefont
  {Zeng}}, \bibinfo {author} {\bibfnamefont {M.-Y.}\ \bibnamefont {Zhu}},
  \bibinfo {author} {\bibfnamefont {Y.-P.}\ \bibnamefont {Zhu}}, \bibinfo
  {author} {\bibfnamefont {X.-R.}\ \bibnamefont {Liu}}, \bibinfo {author}
  {\bibfnamefont {X.-M.}\ \bibnamefont {Ma}}, \bibinfo {author} {\bibfnamefont
  {Y.-J.}\ \bibnamefont {Hao}}, \bibinfo {author} {\bibfnamefont
  {P.}~\bibnamefont {Liu}}, \bibinfo {author} {\bibfnamefont {G.}~\bibnamefont
  {Qu}}, \bibinfo {author} {\bibfnamefont {Y.}~\bibnamefont {Yang}}, \bibinfo
  {author} {\bibfnamefont {Z.}~\bibnamefont {Jiang}}, \bibinfo {author}
  {\bibfnamefont {K.}~\bibnamefont {Yamagami}}, \bibinfo {author}
  {\bibfnamefont {M.}~\bibnamefont {Arita}}, \bibinfo {author} {\bibfnamefont
  {X.}~\bibnamefont {Zhang}}, \bibinfo {author} {\bibfnamefont {T.-H.}\
  \bibnamefont {Shao}}, \bibinfo {author} {\bibfnamefont {Y.}~\bibnamefont
  {Dai}}, \bibinfo {author} {\bibfnamefont {K.}~\bibnamefont {Shimada}},
  \bibinfo {author} {\bibfnamefont {Z.}~\bibnamefont {Liu}}, \bibinfo {author}
  {\bibfnamefont {M.}~\bibnamefont {Ye}}, \bibinfo {author} {\bibfnamefont
  {Y.}~\bibnamefont {Huang}}, \bibinfo {author} {\bibfnamefont
  {Q.}~\bibnamefont {Liu}},\ and\ \bibinfo {author} {\bibfnamefont
  {C.}~\bibnamefont {Liu}},\ }\bibfield  {title} {\bibinfo {title}
  {{Observation of Spin Splitting in Room-Temperature Metallic Antiferromagnet
  CrSb}},\ }\href {https://doi.org/https://doi.org/10.1002/advs.202406529}
  {\bibfield  {journal} {\bibinfo  {journal} {Advanced Science}\ }\textbf
  {\bibinfo {volume} {11}},\ \bibinfo {pages} {2406529} (\bibinfo {year}
  {2024})}\BibitemShut {NoStop}%
\bibitem [{\citenamefont {Jiang}\ \emph {et~al.}(2025)\citenamefont {Jiang},
  \citenamefont {Hu}, \citenamefont {Bai}, \citenamefont {Song}, \citenamefont
  {Mu}, \citenamefont {Qu}, \citenamefont {Li}, \citenamefont {Zhu},
  \citenamefont {Pi}, \citenamefont {Wei}, \citenamefont {Sun}, \citenamefont
  {Huang}, \citenamefont {Zheng}, \citenamefont {Peng}, \citenamefont {He},
  \citenamefont {Li}, \citenamefont {Luo}, \citenamefont {Li}, \citenamefont
  {Chen}, \citenamefont {Li}, \citenamefont {Weng},\ and\ \citenamefont
  {Qian}}]{Jiang2024KV2Se2O}%
  \BibitemOpen
  \bibfield  {author} {\bibinfo {author} {\bibfnamefont {B.}~\bibnamefont
  {Jiang}}, \bibinfo {author} {\bibfnamefont {M.}~\bibnamefont {Hu}}, \bibinfo
  {author} {\bibfnamefont {J.}~\bibnamefont {Bai}}, \bibinfo {author}
  {\bibfnamefont {Z.}~\bibnamefont {Song}}, \bibinfo {author} {\bibfnamefont
  {C.}~\bibnamefont {Mu}}, \bibinfo {author} {\bibfnamefont {G.}~\bibnamefont
  {Qu}}, \bibinfo {author} {\bibfnamefont {W.}~\bibnamefont {Li}}, \bibinfo
  {author} {\bibfnamefont {W.}~\bibnamefont {Zhu}}, \bibinfo {author}
  {\bibfnamefont {H.}~\bibnamefont {Pi}}, \bibinfo {author} {\bibfnamefont
  {Z.}~\bibnamefont {Wei}}, \bibinfo {author} {\bibfnamefont {Y.-J.}\
  \bibnamefont {Sun}}, \bibinfo {author} {\bibfnamefont {Y.}~\bibnamefont
  {Huang}}, \bibinfo {author} {\bibfnamefont {X.}~\bibnamefont {Zheng}},
  \bibinfo {author} {\bibfnamefont {Y.}~\bibnamefont {Peng}}, \bibinfo {author}
  {\bibfnamefont {L.}~\bibnamefont {He}}, \bibinfo {author} {\bibfnamefont
  {S.}~\bibnamefont {Li}}, \bibinfo {author} {\bibfnamefont {J.}~\bibnamefont
  {Luo}}, \bibinfo {author} {\bibfnamefont {Z.}~\bibnamefont {Li}}, \bibinfo
  {author} {\bibfnamefont {G.}~\bibnamefont {Chen}}, \bibinfo {author}
  {\bibfnamefont {H.}~\bibnamefont {Li}}, \bibinfo {author} {\bibfnamefont
  {H.}~\bibnamefont {Weng}},\ and\ \bibinfo {author} {\bibfnamefont
  {T.}~\bibnamefont {Qian}},\ }\bibfield  {title} {\bibinfo {title} {A metallic
  room-temperature d-wave altermagnet},\ }\href
  {https://doi.org/10.1038/s41567-025-02822-y} {\bibfield  {journal} {\bibinfo
  {journal} {Nature Physics}\ }\textbf {\bibinfo {volume} {21}},\ \bibinfo
  {pages} {754} (\bibinfo {year} {2025})}\BibitemShut {NoStop}%
\bibitem [{\citenamefont {Zhang}\ \emph {et~al.}(2025)\citenamefont {Zhang},
  \citenamefont {Cheng}, \citenamefont {Yin}, \citenamefont {Liu},
  \citenamefont {Deng}, \citenamefont {Qiao}, \citenamefont {Shi},
  \citenamefont {Zhang}, \citenamefont {Lin}, \citenamefont {Liu},
  \citenamefont {Ye}, \citenamefont {Huang}, \citenamefont {Meng},
  \citenamefont {Zhang}, \citenamefont {Okuda}, \citenamefont {Shimada},
  \citenamefont {Cui}, \citenamefont {Zhao}, \citenamefont {Cao}, \citenamefont
  {Qiao}, \citenamefont {Liu},\ and\ \citenamefont {Chen}}]{zhangAM2025}%
  \BibitemOpen
  \bibfield  {author} {\bibinfo {author} {\bibfnamefont {F.}~\bibnamefont
  {Zhang}}, \bibinfo {author} {\bibfnamefont {X.}~\bibnamefont {Cheng}},
  \bibinfo {author} {\bibfnamefont {Z.}~\bibnamefont {Yin}}, \bibinfo {author}
  {\bibfnamefont {C.}~\bibnamefont {Liu}}, \bibinfo {author} {\bibfnamefont
  {L.}~\bibnamefont {Deng}}, \bibinfo {author} {\bibfnamefont {Y.}~\bibnamefont
  {Qiao}}, \bibinfo {author} {\bibfnamefont {Z.}~\bibnamefont {Shi}}, \bibinfo
  {author} {\bibfnamefont {S.}~\bibnamefont {Zhang}}, \bibinfo {author}
  {\bibfnamefont {J.}~\bibnamefont {Lin}}, \bibinfo {author} {\bibfnamefont
  {Z.}~\bibnamefont {Liu}}, \bibinfo {author} {\bibfnamefont {M.}~\bibnamefont
  {Ye}}, \bibinfo {author} {\bibfnamefont {Y.}~\bibnamefont {Huang}}, \bibinfo
  {author} {\bibfnamefont {X.}~\bibnamefont {Meng}}, \bibinfo {author}
  {\bibfnamefont {C.}~\bibnamefont {Zhang}}, \bibinfo {author} {\bibfnamefont
  {T.}~\bibnamefont {Okuda}}, \bibinfo {author} {\bibfnamefont
  {K.}~\bibnamefont {Shimada}}, \bibinfo {author} {\bibfnamefont
  {S.}~\bibnamefont {Cui}}, \bibinfo {author} {\bibfnamefont {Y.}~\bibnamefont
  {Zhao}}, \bibinfo {author} {\bibfnamefont {G.-H.}\ \bibnamefont {Cao}},
  \bibinfo {author} {\bibfnamefont {S.}~\bibnamefont {Qiao}}, \bibinfo {author}
  {\bibfnamefont {J.}~\bibnamefont {Liu}},\ and\ \bibinfo {author}
  {\bibfnamefont {C.}~\bibnamefont {Chen}},\ }\bibfield  {title} {\bibinfo
  {title} {Crystal-symmetry-paired spin–valley locking in a layered
  room-temperature metallic altermagnet candidate},\ }\href
  {https://doi.org/10.1038/s41567-025-02864-2} {\bibfield  {journal} {\bibinfo
  {journal} {Nature Physics}\ }\textbf {\bibinfo {volume} {21}},\ \bibinfo
  {pages} {760} (\bibinfo {year} {2025})}\BibitemShut {NoStop}%
\bibitem [{\citenamefont {Ding}\ \emph {et~al.}(2024)\citenamefont {Ding},
  \citenamefont {Jiang}, \citenamefont {Chen}, \citenamefont {Tao},
  \citenamefont {Liu}, \citenamefont {Li}, \citenamefont {Liu}, \citenamefont
  {Sun}, \citenamefont {Cheng}, \citenamefont {Liu}, \citenamefont {Yang},
  \citenamefont {Zhang}, \citenamefont {Deng}, \citenamefont {Jing},
  \citenamefont {Huang}, \citenamefont {Shi}, \citenamefont {Ye}, \citenamefont
  {Qiao}, \citenamefont {Wang}, \citenamefont {Guo}, \citenamefont {Feng},\
  and\ \citenamefont {Shen}}]{Ding2024AM}%
  \BibitemOpen
  \bibfield  {author} {\bibinfo {author} {\bibfnamefont {J.}~\bibnamefont
  {Ding}}, \bibinfo {author} {\bibfnamefont {Z.}~\bibnamefont {Jiang}},
  \bibinfo {author} {\bibfnamefont {X.}~\bibnamefont {Chen}}, \bibinfo {author}
  {\bibfnamefont {Z.}~\bibnamefont {Tao}}, \bibinfo {author} {\bibfnamefont
  {Z.}~\bibnamefont {Liu}}, \bibinfo {author} {\bibfnamefont {T.}~\bibnamefont
  {Li}}, \bibinfo {author} {\bibfnamefont {J.}~\bibnamefont {Liu}}, \bibinfo
  {author} {\bibfnamefont {J.}~\bibnamefont {Sun}}, \bibinfo {author}
  {\bibfnamefont {J.}~\bibnamefont {Cheng}}, \bibinfo {author} {\bibfnamefont
  {J.}~\bibnamefont {Liu}}, \bibinfo {author} {\bibfnamefont {Y.}~\bibnamefont
  {Yang}}, \bibinfo {author} {\bibfnamefont {R.}~\bibnamefont {Zhang}},
  \bibinfo {author} {\bibfnamefont {L.}~\bibnamefont {Deng}}, \bibinfo {author}
  {\bibfnamefont {W.}~\bibnamefont {Jing}}, \bibinfo {author} {\bibfnamefont
  {Y.}~\bibnamefont {Huang}}, \bibinfo {author} {\bibfnamefont
  {Y.}~\bibnamefont {Shi}}, \bibinfo {author} {\bibfnamefont {M.}~\bibnamefont
  {Ye}}, \bibinfo {author} {\bibfnamefont {S.}~\bibnamefont {Qiao}}, \bibinfo
  {author} {\bibfnamefont {Y.}~\bibnamefont {Wang}}, \bibinfo {author}
  {\bibfnamefont {Y.}~\bibnamefont {Guo}}, \bibinfo {author} {\bibfnamefont
  {D.}~\bibnamefont {Feng}},\ and\ \bibinfo {author} {\bibfnamefont
  {D.}~\bibnamefont {Shen}},\ }\bibfield  {title} {\bibinfo {title} {Large band
  splitting in $g$-wave altermagnet crsb},\ }\href
  {https://doi.org/10.1103/PhysRevLett.133.206401} {\bibfield  {journal}
  {\bibinfo  {journal} {Phys. Rev. Lett.}\ }\textbf {\bibinfo {volume} {133}},\
  \bibinfo {pages} {206401} (\bibinfo {year} {2024})}\BibitemShut {NoStop}%
\bibitem [{\citenamefont {Bai}\ \emph {et~al.}(2024)\citenamefont {Bai},
  \citenamefont {Feng}, \citenamefont {Liu}, \citenamefont {Šmejkal},
  \citenamefont {Mokrousov},\ and\ \citenamefont {Yao}}]{Bai2024AM}%
  \BibitemOpen
  \bibfield  {author} {\bibinfo {author} {\bibfnamefont {L.}~\bibnamefont
  {Bai}}, \bibinfo {author} {\bibfnamefont {W.}~\bibnamefont {Feng}}, \bibinfo
  {author} {\bibfnamefont {S.}~\bibnamefont {Liu}}, \bibinfo {author}
  {\bibfnamefont {L.}~\bibnamefont {Šmejkal}}, \bibinfo {author}
  {\bibfnamefont {Y.}~\bibnamefont {Mokrousov}},\ and\ \bibinfo {author}
  {\bibfnamefont {Y.}~\bibnamefont {Yao}},\ }\bibfield  {title} {\bibinfo
  {title} {Altermagnetism: Exploring new frontiers in magnetism and
  spintronics},\ }\href
  {https://doi.org/https://doi.org/10.1002/adfm.202409327} {\bibfield
  {journal} {\bibinfo  {journal} {Advanced Functional Materials}\ }\textbf
  {\bibinfo {volume} {34}},\ \bibinfo {pages} {2409327} (\bibinfo {year}
  {2024})}\BibitemShut {NoStop}%
\bibitem [{\citenamefont {Song}\ \emph {et~al.}(2025)\citenamefont {Song},
  \citenamefont {Bai}, \citenamefont {Zhou}, \citenamefont {Han}, \citenamefont
  {Reichlova}, \citenamefont {Dil}, \citenamefont {Liu}, \citenamefont {Chen},\
  and\ \citenamefont {Pan}}]{Song2025AMreview}%
  \BibitemOpen
  \bibfield  {author} {\bibinfo {author} {\bibfnamefont {C.}~\bibnamefont
  {Song}}, \bibinfo {author} {\bibfnamefont {H.}~\bibnamefont {Bai}}, \bibinfo
  {author} {\bibfnamefont {Z.}~\bibnamefont {Zhou}}, \bibinfo {author}
  {\bibfnamefont {L.}~\bibnamefont {Han}}, \bibinfo {author} {\bibfnamefont
  {H.}~\bibnamefont {Reichlova}}, \bibinfo {author} {\bibfnamefont {J.~H.}\
  \bibnamefont {Dil}}, \bibinfo {author} {\bibfnamefont {J.}~\bibnamefont
  {Liu}}, \bibinfo {author} {\bibfnamefont {X.}~\bibnamefont {Chen}},\ and\
  \bibinfo {author} {\bibfnamefont {F.}~\bibnamefont {Pan}},\ }\bibfield
  {title} {\bibinfo {title} {{Altermagnets as a new class of functional
  materials}},\ }\href {https://doi.org/10.1038/s41578-025-00779-1} {\bibfield
  {journal} {\bibinfo  {journal} {Nature Reviews Materials}\ }\textbf {\bibinfo
  {volume} {10}},\ \bibinfo {pages} {473} (\bibinfo {year} {2025})}\BibitemShut
  {NoStop}%
\bibitem [{\citenamefont {{Liu}}\ \emph {et~al.}(2025)\citenamefont {{Liu}},
  \citenamefont {{Hu}},\ and\ \citenamefont {{Liu}}}]{Liu2025AMreview}%
  \BibitemOpen
  \bibfield  {author} {\bibinfo {author} {\bibfnamefont {Z.}~\bibnamefont
  {{Liu}}}, \bibinfo {author} {\bibfnamefont {H.}~\bibnamefont {{Hu}}},\ and\
  \bibinfo {author} {\bibfnamefont {X.-J.}\ \bibnamefont {{Liu}}},\ }\bibfield
  {title} {\bibinfo {title} {{Altermagnetism and Superconductivity: A Short
  Historical Review}},\ }\href {https://doi.org/10.48550/arXiv.2510.09170}
  {\bibfield  {journal} {\bibinfo  {journal} {arXiv e-prints}\ ,\ \bibinfo
  {pages} {arXiv:2510.09170}} (\bibinfo {year} {2025})}\BibitemShut {NoStop}%
\bibitem [{\citenamefont {Fukaya}\ \emph {et~al.}(2025)\citenamefont {Fukaya},
  \citenamefont {Lu}, \citenamefont {Yada}, \citenamefont {Tanaka},\ and\
  \citenamefont {Cayao}}]{Fukaya2025}%
  \BibitemOpen
  \bibfield  {author} {\bibinfo {author} {\bibfnamefont {Y.}~\bibnamefont
  {Fukaya}}, \bibinfo {author} {\bibfnamefont {B.}~\bibnamefont {Lu}}, \bibinfo
  {author} {\bibfnamefont {K.}~\bibnamefont {Yada}}, \bibinfo {author}
  {\bibfnamefont {Y.}~\bibnamefont {Tanaka}},\ and\ \bibinfo {author}
  {\bibfnamefont {J.}~\bibnamefont {Cayao}},\ }\bibfield  {title} {\bibinfo
  {title} {{Superconducting phenomena in systems with unconventional
  magnets}},\ }\href {https://doi.org/10.1088/1361-648X/adf1cf} {\bibfield
  {journal} {\bibinfo  {journal} {Journal of Physics: Condensed Matter}\
  }\textbf {\bibinfo {volume} {37}},\ \bibinfo {pages} {313003} (\bibinfo
  {year} {2025})}\BibitemShut {NoStop}%
\bibitem [{\citenamefont {Jungwirth}\ \emph {et~al.}(2025)\citenamefont
  {Jungwirth}, \citenamefont {Fernandes}, \citenamefont {Fradkin},
  \citenamefont {MacDonald}, \citenamefont {Sinova},\ and\ \citenamefont
  {Šmejkal}}]{Jungwirth2025AMreview}%
  \BibitemOpen
  \bibfield  {author} {\bibinfo {author} {\bibfnamefont {T.}~\bibnamefont
  {Jungwirth}}, \bibinfo {author} {\bibfnamefont {R.~M.}\ \bibnamefont
  {Fernandes}}, \bibinfo {author} {\bibfnamefont {E.}~\bibnamefont {Fradkin}},
  \bibinfo {author} {\bibfnamefont {A.~H.}\ \bibnamefont {MacDonald}}, \bibinfo
  {author} {\bibfnamefont {J.}~\bibnamefont {Sinova}},\ and\ \bibinfo {author}
  {\bibfnamefont {L.}~\bibnamefont {Šmejkal}},\ }\bibfield  {title} {\bibinfo
  {title} {{Altermagnetism: An unconventional spin-ordered phase of matter}},\
  }\href {https://doi.org/https://doi.org/10.1016/j.newton.2025.100162}
  {\bibfield  {journal} {\bibinfo  {journal} {Newton}\ }\textbf {\bibinfo
  {volume} {1}},\ \bibinfo {pages} {100162} (\bibinfo {year}
  {2025})}\BibitemShut {NoStop}%
\bibitem [{\citenamefont {{Luo}}\ \emph {et~al.}(2025)\citenamefont {{Luo}},
  \citenamefont {{Hu}},\ and\ \citenamefont {{Law}}}]{Luo2025pwave}%
  \BibitemOpen
  \bibfield  {author} {\bibinfo {author} {\bibfnamefont {X.-J.}\ \bibnamefont
  {{Luo}}}, \bibinfo {author} {\bibfnamefont {J.-X.}\ \bibnamefont {{Hu}}},\
  and\ \bibinfo {author} {\bibfnamefont {K.~T.}\ \bibnamefont {{Law}}},\
  }\bibfield  {title} {\bibinfo {title} {{Spin Symmetry Criteria for Odd-parity
  Magnets}},\ }\href {https://doi.org/10.48550/arXiv.2510.05512} {\bibfield
  {journal} {\bibinfo  {journal} {arXiv e-prints}\ ,\ \bibinfo {eid}
  {arXiv:2510.05512}} (\bibinfo {year} {2025})},\ \Eprint
  {https://arxiv.org/abs/2510.05512} {arXiv:2510.05512 [cond-mat.other]}
  \BibitemShut {NoStop}%
\bibitem [{\citenamefont {Brekke}\ \emph {et~al.}(2024)\citenamefont {Brekke},
  \citenamefont {Sukhachov}, \citenamefont {Giil}, \citenamefont {Brataas},\
  and\ \citenamefont {Linder}}]{Brekke2024pwave}%
  \BibitemOpen
  \bibfield  {author} {\bibinfo {author} {\bibfnamefont {B.}~\bibnamefont
  {Brekke}}, \bibinfo {author} {\bibfnamefont {P.}~\bibnamefont {Sukhachov}},
  \bibinfo {author} {\bibfnamefont {H.~G.}\ \bibnamefont {Giil}}, \bibinfo
  {author} {\bibfnamefont {A.}~\bibnamefont {Brataas}},\ and\ \bibinfo {author}
  {\bibfnamefont {J.}~\bibnamefont {Linder}},\ }\bibfield  {title} {\bibinfo
  {title} {Minimal models and transport properties of unconventional $p$-wave
  magnets},\ }\href {https://doi.org/10.1103/PhysRevLett.133.236703} {\bibfield
   {journal} {\bibinfo  {journal} {Phys. Rev. Lett.}\ }\textbf {\bibinfo
  {volume} {133}},\ \bibinfo {pages} {236703} (\bibinfo {year}
  {2024})}\BibitemShut {NoStop}%
\bibitem [{\citenamefont {Ezawa}(2024)}]{Ezawa2024pwave}%
  \BibitemOpen
  \bibfield  {author} {\bibinfo {author} {\bibfnamefont {M.}~\bibnamefont
  {Ezawa}},\ }\bibfield  {title} {\bibinfo {title} {{Topological insulators and
  superconductors based on $p$-wave magnets: Electrical control and detection
  of a domain wall}},\ }\href {https://doi.org/10.1103/PhysRevB.110.165429}
  {\bibfield  {journal} {\bibinfo  {journal} {Phys. Rev. B}\ }\textbf {\bibinfo
  {volume} {110}},\ \bibinfo {pages} {165429} (\bibinfo {year}
  {2024})}\BibitemShut {NoStop}%
\bibitem [{\citenamefont {Hedayati}\ and\ \citenamefont
  {Salehi}(2025)}]{Hedayati2025}%
  \BibitemOpen
  \bibfield  {author} {\bibinfo {author} {\bibfnamefont {A.~A.}\ \bibnamefont
  {Hedayati}}\ and\ \bibinfo {author} {\bibfnamefont {M.}~\bibnamefont
  {Salehi}},\ }\bibfield  {title} {\bibinfo {title} {{Transverse spin current
  at normal-metal /$p$-wave magnet junctions}},\ }\href
  {https://doi.org/10.1103/PhysRevB.111.035404} {\bibfield  {journal} {\bibinfo
   {journal} {Phys. Rev. B}\ }\textbf {\bibinfo {volume} {111}},\ \bibinfo
  {pages} {035404} (\bibinfo {year} {2025})}\BibitemShut {NoStop}%
\bibitem [{\citenamefont {Soori}(2025)}]{Soori2025pwave}%
  \BibitemOpen
  \bibfield  {author} {\bibinfo {author} {\bibfnamefont {A.}~\bibnamefont
  {Soori}},\ }\bibfield  {title} {\bibinfo {title} {{Crossed Andreev reflection
  in collinear $p$-wave magnet/triplet superconductor junctions}},\ }\href
  {https://doi.org/10.1103/PhysRevB.111.165413} {\bibfield  {journal} {\bibinfo
   {journal} {Phys. Rev. B}\ }\textbf {\bibinfo {volume} {111}},\ \bibinfo
  {pages} {165413} (\bibinfo {year} {2025})}\BibitemShut {NoStop}%
\bibitem [{\citenamefont {Nagae}\ \emph {et~al.}(2025)\citenamefont {Nagae},
  \citenamefont {Katayama},\ and\ \citenamefont {Ikegaya}}]{Nagae2025}%
  \BibitemOpen
  \bibfield  {author} {\bibinfo {author} {\bibfnamefont {Y.}~\bibnamefont
  {Nagae}}, \bibinfo {author} {\bibfnamefont {L.}~\bibnamefont {Katayama}},\
  and\ \bibinfo {author} {\bibfnamefont {S.}~\bibnamefont {Ikegaya}},\
  }\bibfield  {title} {\bibinfo {title} {{Flat-band zero-energy states and
  anomalous proximity effects in $p$-wave magnet--superconductor hybrid
  systems}},\ }\href {https://doi.org/10.1103/PhysRevB.111.174519} {\bibfield
  {journal} {\bibinfo  {journal} {Phys. Rev. B}\ }\textbf {\bibinfo {volume}
  {111}},\ \bibinfo {pages} {174519} (\bibinfo {year} {2025})}\BibitemShut
  {NoStop}%
\bibitem [{\citenamefont {{Fukaya}}\ \emph {et~al.}(2025)\citenamefont
  {{Fukaya}}, \citenamefont {{Lu}}, \citenamefont {{Yada}}, \citenamefont
  {{Tanaka}},\ and\ \citenamefont {{Cayao}}}]{Fukaya2025AM}%
  \BibitemOpen
  \bibfield  {author} {\bibinfo {author} {\bibfnamefont {Y.}~\bibnamefont
  {{Fukaya}}}, \bibinfo {author} {\bibfnamefont {B.}~\bibnamefont {{Lu}}},
  \bibinfo {author} {\bibfnamefont {K.}~\bibnamefont {{Yada}}}, \bibinfo
  {author} {\bibfnamefont {Y.}~\bibnamefont {{Tanaka}}},\ and\ \bibinfo
  {author} {\bibfnamefont {J.}~\bibnamefont {{Cayao}}},\ }\bibfield  {title}
  {\bibinfo {title} {{Crossed surface flat bands in three-dimensional
  superconducting altermagnets}},\ }\href
  {https://doi.org/10.48550/arXiv.2510.14724} {\bibfield  {journal} {\bibinfo
  {journal} {arXiv e-prints}\ ,\ \bibinfo {pages} {arXiv:2510.14724}} (\bibinfo
  {year} {2025})}\BibitemShut {NoStop}%
\bibitem [{\citenamefont {Ouassou}\ \emph {et~al.}(2023)\citenamefont
  {Ouassou}, \citenamefont {Brataas},\ and\ \citenamefont
  {Linder}}]{Ouassou2023AM}%
  \BibitemOpen
  \bibfield  {author} {\bibinfo {author} {\bibfnamefont {J.~A.}\ \bibnamefont
  {Ouassou}}, \bibinfo {author} {\bibfnamefont {A.}~\bibnamefont {Brataas}},\
  and\ \bibinfo {author} {\bibfnamefont {J.}~\bibnamefont {Linder}},\
  }\bibfield  {title} {\bibinfo {title} {{dc Josephson Effect in
  Altermagnets}},\ }\href {https://doi.org/10.1103/PhysRevLett.131.076003}
  {\bibfield  {journal} {\bibinfo  {journal} {Phys. Rev. Lett.}\ }\textbf
  {\bibinfo {volume} {131}},\ \bibinfo {pages} {076003} (\bibinfo {year}
  {2023})}\BibitemShut {NoStop}%
\bibitem [{\citenamefont {Cheng}\ and\ \citenamefont
  {Sun}(2024)}]{Cheng2024AM}%
  \BibitemOpen
  \bibfield  {author} {\bibinfo {author} {\bibfnamefont {Q.}~\bibnamefont
  {Cheng}}\ and\ \bibinfo {author} {\bibfnamefont {Q.-F.}\ \bibnamefont
  {Sun}},\ }\bibfield  {title} {\bibinfo {title} {{Orientation-dependent
  Josephson effect in spin-singlet superconductor/altermagnet/spin-triplet
  superconductor junctions}},\ }\href
  {https://doi.org/10.1103/PhysRevB.109.024517} {\bibfield  {journal} {\bibinfo
   {journal} {Phys. Rev. B}\ }\textbf {\bibinfo {volume} {109}},\ \bibinfo
  {pages} {024517} (\bibinfo {year} {2024})}\BibitemShut {NoStop}%
\bibitem [{\citenamefont {Lu}\ \emph {et~al.}(2024)\citenamefont {Lu},
  \citenamefont {Maeda}, \citenamefont {Ito}, \citenamefont {Yada},\ and\
  \citenamefont {Tanaka}}]{Lu2024AM}%
  \BibitemOpen
  \bibfield  {author} {\bibinfo {author} {\bibfnamefont {B.}~\bibnamefont
  {Lu}}, \bibinfo {author} {\bibfnamefont {K.}~\bibnamefont {Maeda}}, \bibinfo
  {author} {\bibfnamefont {H.}~\bibnamefont {Ito}}, \bibinfo {author}
  {\bibfnamefont {K.}~\bibnamefont {Yada}},\ and\ \bibinfo {author}
  {\bibfnamefont {Y.}~\bibnamefont {Tanaka}},\ }\bibfield  {title} {\bibinfo
  {title} {{$\ensuremath{\varphi}$ Josephson Junction Induced by
  Altermagnetism}},\ }\href {https://doi.org/10.1103/PhysRevLett.133.226002}
  {\bibfield  {journal} {\bibinfo  {journal} {Phys. Rev. Lett.}\ }\textbf
  {\bibinfo {volume} {133}},\ \bibinfo {pages} {226002} (\bibinfo {year}
  {2024})}\BibitemShut {NoStop}%
\bibitem [{\citenamefont {Lin}\ \emph {et~al.}(2025)\citenamefont {Lin},
  \citenamefont {Zhang}, \citenamefont {Lu},\ and\ \citenamefont
  {Xie}}]{Lin2024AM12}%
  \BibitemOpen
  \bibfield  {author} {\bibinfo {author} {\bibfnamefont {H.-J.}\ \bibnamefont
  {Lin}}, \bibinfo {author} {\bibfnamefont {S.-B.}\ \bibnamefont {Zhang}},
  \bibinfo {author} {\bibfnamefont {H.-Z.}\ \bibnamefont {Lu}},\ and\ \bibinfo
  {author} {\bibfnamefont {X.~C.}\ \bibnamefont {Xie}},\ }\bibfield  {title}
  {\bibinfo {title} {Coulomb drag in altermagnets},\ }\href
  {https://doi.org/10.1103/PhysRevLett.134.136301} {\bibfield  {journal}
  {\bibinfo  {journal} {Phys. Rev. Lett.}\ }\textbf {\bibinfo {volume} {134}},\
  \bibinfo {pages} {136301} (\bibinfo {year} {2025})}\BibitemShut {NoStop}%
\bibitem [{\citenamefont {Zhang}\ \emph {et~al.}(2024)\citenamefont {Zhang},
  \citenamefont {Hu},\ and\ \citenamefont {Neupert}}]{Zhang2024AM}%
  \BibitemOpen
  \bibfield  {author} {\bibinfo {author} {\bibfnamefont {S.-B.}\ \bibnamefont
  {Zhang}}, \bibinfo {author} {\bibfnamefont {L.-H.}\ \bibnamefont {Hu}},\ and\
  \bibinfo {author} {\bibfnamefont {T.}~\bibnamefont {Neupert}},\ }\bibfield
  {title} {\bibinfo {title} {{Finite-momentum Cooper pairing in proximitized
  altermagnets}},\ }\href {https://doi.org/10.1038/s41467-024-45951-3}
  {\bibfield  {journal} {\bibinfo  {journal} {Nature Communications}\ }\textbf
  {\bibinfo {volume} {15}},\ \bibinfo {pages} {1801} (\bibinfo {year}
  {2024})}\BibitemShut {NoStop}%
\bibitem [{\citenamefont {Zhu}\ \emph {et~al.}(2023)\citenamefont {Zhu},
  \citenamefont {Zhuang}, \citenamefont {Wu},\ and\ \citenamefont
  {Yan}}]{Zhu2023TSC}%
  \BibitemOpen
  \bibfield  {author} {\bibinfo {author} {\bibfnamefont {D.}~\bibnamefont
  {Zhu}}, \bibinfo {author} {\bibfnamefont {Z.-Y.}\ \bibnamefont {Zhuang}},
  \bibinfo {author} {\bibfnamefont {Z.}~\bibnamefont {Wu}},\ and\ \bibinfo
  {author} {\bibfnamefont {Z.}~\bibnamefont {Yan}},\ }\bibfield  {title}
  {\bibinfo {title} {Topological superconductivity in two-dimensional
  altermagnetic metals},\ }\href {https://doi.org/10.1103/PhysRevB.108.184505}
  {\bibfield  {journal} {\bibinfo  {journal} {Phys. Rev. B}\ }\textbf {\bibinfo
  {volume} {108}},\ \bibinfo {pages} {184505} (\bibinfo {year}
  {2023})}\BibitemShut {NoStop}%
\bibitem [{\citenamefont {Zhu}\ \emph {et~al.}(2024{\natexlab{b}})\citenamefont
  {Zhu}, \citenamefont {Liu}, \citenamefont {Zhuang}, \citenamefont {Wu},\ and\
  \citenamefont {Yan}}]{Zhu2024dislocation}%
  \BibitemOpen
  \bibfield  {author} {\bibinfo {author} {\bibfnamefont {D.}~\bibnamefont
  {Zhu}}, \bibinfo {author} {\bibfnamefont {D.}~\bibnamefont {Liu}}, \bibinfo
  {author} {\bibfnamefont {Z.-Y.}\ \bibnamefont {Zhuang}}, \bibinfo {author}
  {\bibfnamefont {Z.}~\bibnamefont {Wu}},\ and\ \bibinfo {author}
  {\bibfnamefont {Z.}~\bibnamefont {Yan}},\ }\bibfield  {title} {\bibinfo
  {title} {{Field-sensitive dislocation bound states in two-dimensional
  $d$-wave altermagnets}},\ }\href
  {https://doi.org/10.1103/PhysRevB.110.165141} {\bibfield  {journal} {\bibinfo
   {journal} {Phys. Rev. B}\ }\textbf {\bibinfo {volume} {110}},\ \bibinfo
  {pages} {165141} (\bibinfo {year} {2024}{\natexlab{b}})}\BibitemShut
  {NoStop}%
\bibitem [{\citenamefont {Li}\ and\ \citenamefont {Liu}(2023)}]{Li2023AMHOTSC}%
  \BibitemOpen
  \bibfield  {author} {\bibinfo {author} {\bibfnamefont {Y.-X.}\ \bibnamefont
  {Li}}\ and\ \bibinfo {author} {\bibfnamefont {C.-C.}\ \bibnamefont {Liu}},\
  }\bibfield  {title} {\bibinfo {title} {{Majorana corner modes and tunable
  patterns in an altermagnet heterostructure}},\ }\href
  {https://doi.org/10.1103/PhysRevB.108.205410} {\bibfield  {journal} {\bibinfo
   {journal} {Phys. Rev. B}\ }\textbf {\bibinfo {volume} {108}},\ \bibinfo
  {pages} {205410} (\bibinfo {year} {2023})}\BibitemShut {NoStop}%
\bibitem [{\citenamefont {Li}\ \emph {et~al.}(2024)\citenamefont {Li},
  \citenamefont {Liu},\ and\ \citenamefont {Liu}}]{Li2024AMHOTI}%
  \BibitemOpen
  \bibfield  {author} {\bibinfo {author} {\bibfnamefont {Y.-X.}\ \bibnamefont
  {Li}}, \bibinfo {author} {\bibfnamefont {Y.}~\bibnamefont {Liu}},\ and\
  \bibinfo {author} {\bibfnamefont {C.-C.}\ \bibnamefont {Liu}},\ }\bibfield
  {title} {\bibinfo {title} {{Creation and manipulation of higher-order
  topological states by altermagnets}},\ }\href
  {https://doi.org/10.1103/PhysRevB.109.L201109} {\bibfield  {journal}
  {\bibinfo  {journal} {Phys. Rev. B}\ }\textbf {\bibinfo {volume} {109}},\
  \bibinfo {pages} {L201109} (\bibinfo {year} {2024})}\BibitemShut {NoStop}%
\bibitem [{\citenamefont {Ghorashi}\ \emph {et~al.}(2024)\citenamefont
  {Ghorashi}, \citenamefont {Hughes},\ and\ \citenamefont
  {Cano}}]{Ghorashi2024AM}%
  \BibitemOpen
  \bibfield  {author} {\bibinfo {author} {\bibfnamefont {S.~A.~A.}\
  \bibnamefont {Ghorashi}}, \bibinfo {author} {\bibfnamefont {T.~L.}\
  \bibnamefont {Hughes}},\ and\ \bibinfo {author} {\bibfnamefont
  {J.}~\bibnamefont {Cano}},\ }\bibfield  {title} {\bibinfo {title}
  {{Altermagnetic Routes to Majorana Modes in Zero Net Magnetization}},\ }\href
  {https://doi.org/10.1103/PhysRevLett.133.106601} {\bibfield  {journal}
  {\bibinfo  {journal} {Phys. Rev. Lett.}\ }\textbf {\bibinfo {volume} {133}},\
  \bibinfo {pages} {106601} (\bibinfo {year} {2024})}\BibitemShut {NoStop}%
\bibitem [{\citenamefont {Antonenko}\ \emph {et~al.}(2025)\citenamefont
  {Antonenko}, \citenamefont {Fernandes},\ and\ \citenamefont
  {Venderbos}}]{Antonenko2025AM}%
  \BibitemOpen
  \bibfield  {author} {\bibinfo {author} {\bibfnamefont {D.~S.}\ \bibnamefont
  {Antonenko}}, \bibinfo {author} {\bibfnamefont {R.~M.}\ \bibnamefont
  {Fernandes}},\ and\ \bibinfo {author} {\bibfnamefont {J.~W.~F.}\ \bibnamefont
  {Venderbos}},\ }\bibfield  {title} {\bibinfo {title} {{Mirror Chern Bands and
  Weyl Nodal Loops in Altermagnets}},\ }\href
  {https://doi.org/10.1103/PhysRevLett.134.096703} {\bibfield  {journal}
  {\bibinfo  {journal} {Phys. Rev. Lett.}\ }\textbf {\bibinfo {volume} {134}},\
  \bibinfo {pages} {096703} (\bibinfo {year} {2025})}\BibitemShut {NoStop}%
\bibitem [{\citenamefont {Parshukov}\ \emph {et~al.}(2025)\citenamefont
  {Parshukov}, \citenamefont {Wiedmann},\ and\ \citenamefont
  {Schnyder}}]{Parshukov2025AM7}%
  \BibitemOpen
  \bibfield  {author} {\bibinfo {author} {\bibfnamefont {K.}~\bibnamefont
  {Parshukov}}, \bibinfo {author} {\bibfnamefont {R.}~\bibnamefont
  {Wiedmann}},\ and\ \bibinfo {author} {\bibfnamefont {A.~P.}\ \bibnamefont
  {Schnyder}},\ }\bibfield  {title} {\bibinfo {title} {Topological crossings in
  two-dimensional altermagnets: Symmetry classification and topological
  responses},\ }\href {https://doi.org/10.1103/PhysRevB.111.224406} {\bibfield
  {journal} {\bibinfo  {journal} {Phys. Rev. B}\ }\textbf {\bibinfo {volume}
  {111}},\ \bibinfo {pages} {224406} (\bibinfo {year} {2025})}\BibitemShut
  {NoStop}%
\bibitem [{\citenamefont {{Yang}}\ \emph {et~al.}(2025)\citenamefont {{Yang}},
  \citenamefont {{Sun}}, \citenamefont {{Xie}},\ and\ \citenamefont
  {{Law}}}]{Yang2025AM2}%
  \BibitemOpen
  \bibfield  {author} {\bibinfo {author} {\bibfnamefont {G.~Z.~X.}\
  \bibnamefont {{Yang}}}, \bibinfo {author} {\bibfnamefont {Z.-T.}\
  \bibnamefont {{Sun}}}, \bibinfo {author} {\bibfnamefont {Y.-M.}\ \bibnamefont
  {{Xie}}},\ and\ \bibinfo {author} {\bibfnamefont {K.~T.}\ \bibnamefont
  {{Law}}},\ }\bibfield  {title} {\bibinfo {title} {{Topological altermagnetic
  Josephson junctions}},\ }\href {https://doi.org/10.48550/arXiv.2502.20283}
  {\bibfield  {journal} {\bibinfo  {journal} {arXiv e-prints}\ ,\ \bibinfo
  {pages} {arXiv:2502.20283}} (\bibinfo {year} {2025})}\BibitemShut {NoStop}%
\bibitem [{\citenamefont {Qu}\ \emph {et~al.}(2025)\citenamefont {Qu},
  \citenamefont {Hou}, \citenamefont {Liu}, \citenamefont {Guo},\ and\
  \citenamefont {Lu}}]{Qu2025AM}%
  \BibitemOpen
  \bibfield  {author} {\bibinfo {author} {\bibfnamefont {S.}~\bibnamefont
  {Qu}}, \bibinfo {author} {\bibfnamefont {X.-Y.}\ \bibnamefont {Hou}},
  \bibinfo {author} {\bibfnamefont {Z.-X.}\ \bibnamefont {Liu}}, \bibinfo
  {author} {\bibfnamefont {P.-J.}\ \bibnamefont {Guo}},\ and\ \bibinfo {author}
  {\bibfnamefont {Z.-Y.}\ \bibnamefont {Lu}},\ }\bibfield  {title} {\bibinfo
  {title} {{Altermagnetic Weyl node-network metals protected by spin
  symmetry}},\ }\href {https://doi.org/10.1103/PhysRevB.111.195138} {\bibfield
  {journal} {\bibinfo  {journal} {Phys. Rev. B}\ }\textbf {\bibinfo {volume}
  {111}},\ \bibinfo {pages} {195138} (\bibinfo {year} {2025})}\BibitemShut
  {NoStop}%
\bibitem [{\citenamefont {Li}\ and\ \citenamefont {Zhang}(2025)}]{Li2025AM1}%
  \BibitemOpen
  \bibfield  {author} {\bibinfo {author} {\bibfnamefont {Y.-Y.}\ \bibnamefont
  {Li}}\ and\ \bibinfo {author} {\bibfnamefont {S.-B.}\ \bibnamefont {Zhang}},\
  }\bibfield  {title} {\bibinfo {title} {{Floating edge bands in the
  Bernevig-Hughes-Zhang model with altermagnetism}},\ }\href
  {https://doi.org/10.1103/PhysRevB.111.045106} {\bibfield  {journal} {\bibinfo
   {journal} {Phys. Rev. B}\ }\textbf {\bibinfo {volume} {111}},\ \bibinfo
  {pages} {045106} (\bibinfo {year} {2025})}\BibitemShut {NoStop}%
\bibitem [{\citenamefont {Rao}\ \emph {et~al.}(2024)\citenamefont {Rao},
  \citenamefont {Mook},\ and\ \citenamefont {Knolle}}]{Rao2024AM7}%
  \BibitemOpen
  \bibfield  {author} {\bibinfo {author} {\bibfnamefont {P.}~\bibnamefont
  {Rao}}, \bibinfo {author} {\bibfnamefont {A.}~\bibnamefont {Mook}},\ and\
  \bibinfo {author} {\bibfnamefont {J.}~\bibnamefont {Knolle}},\ }\bibfield
  {title} {\bibinfo {title} {{Tunable band topology and optical conductivity in
  altermagnets}},\ }\href {https://doi.org/10.1103/PhysRevB.110.024425}
  {\bibfield  {journal} {\bibinfo  {journal} {Phys. Rev. B}\ }\textbf {\bibinfo
  {volume} {110}},\ \bibinfo {pages} {024425} (\bibinfo {year}
  {2024})}\BibitemShut {NoStop}%
\bibitem [{\citenamefont {Ma}\ and\ \citenamefont {Jia}(2024)}]{Ma2024AM8}%
  \BibitemOpen
  \bibfield  {author} {\bibinfo {author} {\bibfnamefont {H.-Y.}\ \bibnamefont
  {Ma}}\ and\ \bibinfo {author} {\bibfnamefont {J.-F.}\ \bibnamefont {Jia}},\
  }\bibfield  {title} {\bibinfo {title} {{Altermagnetic topological insulator
  and the selection rules}},\ }\href
  {https://doi.org/10.1103/PhysRevB.110.064426} {\bibfield  {journal} {\bibinfo
   {journal} {Phys. Rev. B}\ }\textbf {\bibinfo {volume} {110}},\ \bibinfo
  {pages} {064426} (\bibinfo {year} {2024})}\BibitemShut {NoStop}%
\bibitem [{\citenamefont {Fernandes}\ \emph {et~al.}(2024)\citenamefont
  {Fernandes}, \citenamefont {de~Carvalho}, \citenamefont {Birol},\ and\
  \citenamefont {Pereira}}]{Fernandes2024AM}%
  \BibitemOpen
  \bibfield  {author} {\bibinfo {author} {\bibfnamefont {R.~M.}\ \bibnamefont
  {Fernandes}}, \bibinfo {author} {\bibfnamefont {V.~S.}\ \bibnamefont
  {de~Carvalho}}, \bibinfo {author} {\bibfnamefont {T.}~\bibnamefont {Birol}},\
  and\ \bibinfo {author} {\bibfnamefont {R.~G.}\ \bibnamefont {Pereira}},\
  }\bibfield  {title} {\bibinfo {title} {{Topological transition from nodal to
  nodeless Zeeman splitting in altermagnets}},\ }\href
  {https://doi.org/10.1103/PhysRevB.109.024404} {\bibfield  {journal} {\bibinfo
   {journal} {Phys. Rev. B}\ }\textbf {\bibinfo {volume} {109}},\ \bibinfo
  {pages} {024404} (\bibinfo {year} {2024})}\BibitemShut {NoStop}%
\bibitem [{\citenamefont {{Zhuang}}\ \emph
  {et~al.}(2025{\natexlab{a}})\citenamefont {{Zhuang}}, \citenamefont {{Zhu}},
  \citenamefont {{Wu}},\ and\ \citenamefont {{Yan}}}]{Zhuang2025SNL}%
  \BibitemOpen
  \bibfield  {author} {\bibinfo {author} {\bibfnamefont {Z.-Y.}\ \bibnamefont
  {{Zhuang}}}, \bibinfo {author} {\bibfnamefont {D.}~\bibnamefont {{Zhu}}},
  \bibinfo {author} {\bibfnamefont {Z.}~\bibnamefont {{Wu}}},\ and\ \bibinfo
  {author} {\bibfnamefont {Z.}~\bibnamefont {{Yan}}},\ }\bibfield  {title}
  {\bibinfo {title} {{Cartesian Nodal Lines and Magnetic Kramers Weyl Nodes in
  Spin-Split Antiferromagnets}},\ }\href
  {https://doi.org/10.48550/arXiv.2502.13212} {\bibfield  {journal} {\bibinfo
  {journal} {arXiv e-prints}\ ,\ \bibinfo {pages} {arXiv:2502.13212}} (\bibinfo
  {year} {2025}{\natexlab{a}})}\BibitemShut {NoStop}%
\bibitem [{\citenamefont {Liu}\ \emph {et~al.}(2025)\citenamefont {Liu},
  \citenamefont {Sun},\ and\ \citenamefont {Zhang}}]{Liu2025AM}%
  \BibitemOpen
  \bibfield  {author} {\bibinfo {author} {\bibfnamefont {L.}~\bibnamefont
  {Liu}}, \bibinfo {author} {\bibfnamefont {Q.-F.}\ \bibnamefont {Sun}},\ and\
  \bibinfo {author} {\bibfnamefont {Y.-T.}\ \bibnamefont {Zhang}},\ }\bibfield
  {title} {\bibinfo {title} {Tunable two-dimensional dirac-weyl semimetal phase
  induced by altermagnetism},\ }\href {https://doi.org/10.1103/3hxt-fynf}
  {\bibfield  {journal} {\bibinfo  {journal} {Phys. Rev. B}\ }\textbf {\bibinfo
  {volume} {112}},\ \bibinfo {pages} {L161411} (\bibinfo {year}
  {2025})}\BibitemShut {NoStop}%
\bibitem [{\citenamefont {Chakraborty}\ and\ \citenamefont
  {Black-Schaffer}(2024)}]{Chakraborty2024AM8}%
  \BibitemOpen
  \bibfield  {author} {\bibinfo {author} {\bibfnamefont {D.}~\bibnamefont
  {Chakraborty}}\ and\ \bibinfo {author} {\bibfnamefont {A.~M.}\ \bibnamefont
  {Black-Schaffer}},\ }\bibfield  {title} {\bibinfo {title} {Zero-field
  finite-momentum and field-induced superconductivity in altermagnets},\ }\href
  {https://doi.org/10.1103/PhysRevB.110.L060508} {\bibfield  {journal}
  {\bibinfo  {journal} {Phys. Rev. B}\ }\textbf {\bibinfo {volume} {110}},\
  \bibinfo {pages} {L060508} (\bibinfo {year} {2024})}\BibitemShut {NoStop}%
\bibitem [{\citenamefont {Brekke}\ \emph {et~al.}(2023)\citenamefont {Brekke},
  \citenamefont {Brataas},\ and\ \citenamefont {Sudb\o{}}}]{Brekke2023AM}%
  \BibitemOpen
  \bibfield  {author} {\bibinfo {author} {\bibfnamefont {B.}~\bibnamefont
  {Brekke}}, \bibinfo {author} {\bibfnamefont {A.}~\bibnamefont {Brataas}},\
  and\ \bibinfo {author} {\bibfnamefont {A.}~\bibnamefont {Sudb\o{}}},\
  }\bibfield  {title} {\bibinfo {title} {{Two-dimensional altermagnets:
  Superconductivity in a minimal microscopic model}},\ }\href
  {https://doi.org/10.1103/PhysRevB.108.224421} {\bibfield  {journal} {\bibinfo
   {journal} {Phys. Rev. B}\ }\textbf {\bibinfo {volume} {108}},\ \bibinfo
  {pages} {224421} (\bibinfo {year} {2023})}\BibitemShut {NoStop}%
\bibitem [{\citenamefont {de~Carvalho}\ and\ \citenamefont
  {Freire}(2024)}]{Carvalho2024AM12}%
  \BibitemOpen
  \bibfield  {author} {\bibinfo {author} {\bibfnamefont {V.~S.}\ \bibnamefont
  {de~Carvalho}}\ and\ \bibinfo {author} {\bibfnamefont {H.}~\bibnamefont
  {Freire}},\ }\bibfield  {title} {\bibinfo {title} {Unconventional
  superconductivity in altermagnets with spin-orbit coupling},\ }\href
  {https://doi.org/10.1103/PhysRevB.110.L220503} {\bibfield  {journal}
  {\bibinfo  {journal} {Phys. Rev. B}\ }\textbf {\bibinfo {volume} {110}},\
  \bibinfo {pages} {L220503} (\bibinfo {year} {2024})}\BibitemShut {NoStop}%
\bibitem [{\citenamefont {Maeda}\ \emph {et~al.}(2025)\citenamefont {Maeda},
  \citenamefont {Fukaya}, \citenamefont {Yada}, \citenamefont {Lu},
  \citenamefont {Tanaka},\ and\ \citenamefont {Cayao}}]{Maeda2025AM4}%
  \BibitemOpen
  \bibfield  {author} {\bibinfo {author} {\bibfnamefont {K.}~\bibnamefont
  {Maeda}}, \bibinfo {author} {\bibfnamefont {Y.}~\bibnamefont {Fukaya}},
  \bibinfo {author} {\bibfnamefont {K.}~\bibnamefont {Yada}}, \bibinfo {author}
  {\bibfnamefont {B.}~\bibnamefont {Lu}}, \bibinfo {author} {\bibfnamefont
  {Y.}~\bibnamefont {Tanaka}},\ and\ \bibinfo {author} {\bibfnamefont
  {J.}~\bibnamefont {Cayao}},\ }\bibfield  {title} {\bibinfo {title}
  {{Classification of pair symmetries in superconductors with unconventional
  magnetism}},\ }\href {https://doi.org/10.1103/PhysRevB.111.144508} {\bibfield
   {journal} {\bibinfo  {journal} {Phys. Rev. B}\ }\textbf {\bibinfo {volume}
  {111}},\ \bibinfo {pages} {144508} (\bibinfo {year} {2025})}\BibitemShut
  {NoStop}%
\bibitem [{\citenamefont {{Liu}}\ \emph {et~al.}(2025)\citenamefont {{Liu}},
  \citenamefont {{Hu}},\ and\ \citenamefont {{Liu}}}]{Liu2025AM8}%
  \BibitemOpen
  \bibfield  {author} {\bibinfo {author} {\bibfnamefont {Z.}~\bibnamefont
  {{Liu}}}, \bibinfo {author} {\bibfnamefont {H.}~\bibnamefont {{Hu}}},\ and\
  \bibinfo {author} {\bibfnamefont {X.-j.}\ \bibnamefont {{Liu}}},\ }\bibfield
  {title} {\bibinfo {title} {{Fulde-Ferrell-Larkin-Ovchinnikov States and
  Topological Bogoliubov Fermi Surfaces in Altermagnets: an Analytical
  Study}},\ }\href {https://doi.org/10.48550/arXiv.2508.07813} {\bibfield
  {journal} {\bibinfo  {journal} {arXiv e-prints}\ ,\ \bibinfo {pages}
  {arXiv:2508.07813}} (\bibinfo {year} {2025})}\BibitemShut {NoStop}%
\bibitem [{\citenamefont {Bose}\ \emph {et~al.}(2024)\citenamefont {Bose},
  \citenamefont {Vadnais},\ and\ \citenamefont {Paramekanti}}]{Bose2024AM11}%
  \BibitemOpen
  \bibfield  {author} {\bibinfo {author} {\bibfnamefont {A.}~\bibnamefont
  {Bose}}, \bibinfo {author} {\bibfnamefont {S.}~\bibnamefont {Vadnais}},\ and\
  \bibinfo {author} {\bibfnamefont {A.}~\bibnamefont {Paramekanti}},\
  }\bibfield  {title} {\bibinfo {title} {{Altermagnetism and superconductivity
  in a multiorbital $t\ensuremath{-}J$ model}},\ }\href
  {https://doi.org/10.1103/PhysRevB.110.205120} {\bibfield  {journal} {\bibinfo
   {journal} {Phys. Rev. B}\ }\textbf {\bibinfo {volume} {110}},\ \bibinfo
  {pages} {205120} (\bibinfo {year} {2024})}\BibitemShut {NoStop}%
\bibitem [{\citenamefont {Hong}\ \emph {et~al.}(2025)\citenamefont {Hong},
  \citenamefont {Park},\ and\ \citenamefont {Kim}}]{Hong2025AM4}%
  \BibitemOpen
  \bibfield  {author} {\bibinfo {author} {\bibfnamefont {S.}~\bibnamefont
  {Hong}}, \bibinfo {author} {\bibfnamefont {M.~J.}\ \bibnamefont {Park}},\
  and\ \bibinfo {author} {\bibfnamefont {K.-M.}\ \bibnamefont {Kim}},\
  }\bibfield  {title} {\bibinfo {title} {Unconventional $p$-wave and
  finite-momentum superconductivity induced by altermagnetism through the
  formation of bogoliubov fermi surface},\ }\href
  {https://doi.org/10.1103/PhysRevB.111.054501} {\bibfield  {journal} {\bibinfo
   {journal} {Phys. Rev. B}\ }\textbf {\bibinfo {volume} {111}},\ \bibinfo
  {pages} {054501} (\bibinfo {year} {2025})}\BibitemShut {NoStop}%
\bibitem [{\citenamefont {{Parshukov}}\ and\ \citenamefont
  {{Schnyder}}(2025)}]{Parshukov2025}%
  \BibitemOpen
  \bibfield  {author} {\bibinfo {author} {\bibfnamefont {K.}~\bibnamefont
  {{Parshukov}}}\ and\ \bibinfo {author} {\bibfnamefont {A.~P.}\ \bibnamefont
  {{Schnyder}}},\ }\bibfield  {title} {\bibinfo {title} {{Exotic
  superconducting states in altermagnets}},\ }\href
  {https://doi.org/10.48550/arXiv.2507.10700} {\bibfield  {journal} {\bibinfo
  {journal} {arXiv e-prints}\ ,\ \bibinfo {pages} {arXiv:2507.10700}} (\bibinfo
  {year} {2025})}\BibitemShut {NoStop}%
\bibitem [{\citenamefont {Fang}\ \emph {et~al.}(2024)\citenamefont {Fang},
  \citenamefont {Cano},\ and\ \citenamefont {Ghorashi}}]{Fang2023NHE}%
  \BibitemOpen
  \bibfield  {author} {\bibinfo {author} {\bibfnamefont {Y.}~\bibnamefont
  {Fang}}, \bibinfo {author} {\bibfnamefont {J.}~\bibnamefont {Cano}},\ and\
  \bibinfo {author} {\bibfnamefont {S.~A.~A.}\ \bibnamefont {Ghorashi}},\
  }\bibfield  {title} {\bibinfo {title} {Quantum geometry induced nonlinear
  transport in altermagnets},\ }\href
  {https://doi.org/10.1103/PhysRevLett.133.106701} {\bibfield  {journal}
  {\bibinfo  {journal} {Phys. Rev. Lett.}\ }\textbf {\bibinfo {volume} {133}},\
  \bibinfo {pages} {106701} (\bibinfo {year} {2024})}\BibitemShut {NoStop}%
\bibitem [{\citenamefont {Hu}\ \emph {et~al.}(2025{\natexlab{b}})\citenamefont
  {Hu}, \citenamefont {Matsyshyn},\ and\ \citenamefont {Song}}]{Hu2025NLME}%
  \BibitemOpen
  \bibfield  {author} {\bibinfo {author} {\bibfnamefont {J.-X.}\ \bibnamefont
  {Hu}}, \bibinfo {author} {\bibfnamefont {O.}~\bibnamefont {Matsyshyn}},\ and\
  \bibinfo {author} {\bibfnamefont {J.~C.~W.}\ \bibnamefont {Song}},\
  }\bibfield  {title} {\bibinfo {title} {Nonlinear superconducting
  magnetoelectric effect},\ }\href
  {https://doi.org/10.1103/PhysRevLett.134.026001} {\bibfield  {journal}
  {\bibinfo  {journal} {Phys. Rev. Lett.}\ }\textbf {\bibinfo {volume} {134}},\
  \bibinfo {pages} {026001} (\bibinfo {year} {2025}{\natexlab{b}})}\BibitemShut
  {NoStop}%
\bibitem [{\citenamefont {Ezawa}(2025{\natexlab{a}})}]{Ezawa2025magnet}%
  \BibitemOpen
  \bibfield  {author} {\bibinfo {author} {\bibfnamefont {M.}~\bibnamefont
  {Ezawa}},\ }\bibfield  {title} {\bibinfo {title} {{Almost half-quantized
  planar Hall effects in $X$-wave magnets with $X=p,d,f,g,i$}},\ }\href
  {https://doi.org/10.1103/zt4l-y18j} {\bibfield  {journal} {\bibinfo
  {journal} {Phys. Rev. B}\ }\textbf {\bibinfo {volume} {112}},\ \bibinfo
  {pages} {235307} (\bibinfo {year} {2025}{\natexlab{a}})}\BibitemShut
  {NoStop}%
\bibitem [{\citenamefont {Ezawa}(2025{\natexlab{b}})}]{Ezawa2025nonlinear}%
  \BibitemOpen
  \bibfield  {author} {\bibinfo {author} {\bibfnamefont {M.}~\bibnamefont
  {Ezawa}},\ }\bibfield  {title} {\bibinfo {title} {Third-order and fifth-order
  nonlinear spin-current generation in $g$-wave and $i$-wave altermagnets and
  perfectly nonreciprocal spin current in $f$-wave magnets},\ }\href
  {https://doi.org/10.1103/PhysRevB.111.125420} {\bibfield  {journal} {\bibinfo
   {journal} {Phys. Rev. B}\ }\textbf {\bibinfo {volume} {111}},\ \bibinfo
  {pages} {125420} (\bibinfo {year} {2025}{\natexlab{b}})}\BibitemShut
  {NoStop}%
\bibitem [{\citenamefont {Zarzuela}\ \emph {et~al.}(2025)\citenamefont
  {Zarzuela}, \citenamefont {Jaeschke-Ubiergo}, \citenamefont {Gomonay},
  \citenamefont {\ifmmode~\check{S}\else \v{S}\fi{}mejkal},\ and\ \citenamefont
  {Sinova}}]{Zarzuela2025transport}%
  \BibitemOpen
  \bibfield  {author} {\bibinfo {author} {\bibfnamefont {R.}~\bibnamefont
  {Zarzuela}}, \bibinfo {author} {\bibfnamefont {R.}~\bibnamefont
  {Jaeschke-Ubiergo}}, \bibinfo {author} {\bibfnamefont {O.}~\bibnamefont
  {Gomonay}}, \bibinfo {author} {\bibfnamefont {L.}~\bibnamefont
  {\ifmmode~\check{S}\else \v{S}\fi{}mejkal}},\ and\ \bibinfo {author}
  {\bibfnamefont {J.}~\bibnamefont {Sinova}},\ }\bibfield  {title} {\bibinfo
  {title} {Transport theory and spin-transfer physics in $d$-wave
  altermagnets},\ }\href {https://doi.org/10.1103/PhysRevB.111.064422}
  {\bibfield  {journal} {\bibinfo  {journal} {Phys. Rev. B}\ }\textbf {\bibinfo
  {volume} {111}},\ \bibinfo {pages} {064422} (\bibinfo {year}
  {2025})}\BibitemShut {NoStop}%
\bibitem [{\citenamefont {{Zhu}}\ \emph
  {et~al.}(2025{\natexlab{a}})\citenamefont {{Zhu}}, \citenamefont {{Huang}},
  \citenamefont {{Chen}}, \citenamefont {{Duan}}, \citenamefont {{Zhang}},
  \citenamefont {{Zutic}},\ and\ \citenamefont {{Zhou}}}]{Zhu2025AMTopo}%
  \BibitemOpen
  \bibfield  {author} {\bibinfo {author} {\bibfnamefont {Z.}~\bibnamefont
  {{Zhu}}}, \bibinfo {author} {\bibfnamefont {R.}~\bibnamefont {{Huang}}},
  \bibinfo {author} {\bibfnamefont {X.}~\bibnamefont {{Chen}}}, \bibinfo
  {author} {\bibfnamefont {X.}~\bibnamefont {{Duan}}}, \bibinfo {author}
  {\bibfnamefont {J.}~\bibnamefont {{Zhang}}}, \bibinfo {author} {\bibfnamefont
  {I.}~\bibnamefont {{Zutic}}},\ and\ \bibinfo {author} {\bibfnamefont
  {T.}~\bibnamefont {{Zhou}}},\ }\bibfield  {title} {\bibinfo {title}
  {{Altermagnetic Proximity Effect}},\ }\href
  {https://doi.org/10.48550/arXiv.2509.06790} {\bibfield  {journal} {\bibinfo
  {journal} {arXiv e-prints}\ ,\ \bibinfo {pages} {arXiv:2509.06790}} (\bibinfo
  {year} {2025}{\natexlab{a}})}\BibitemShut {NoStop}%
\bibitem [{\citenamefont {{Zhu}}\ \emph
  {et~al.}(2025{\natexlab{b}})\citenamefont {{Zhu}}, \citenamefont {{Chen}},
  \citenamefont {{Duan}}, \citenamefont {{Cui}}, \citenamefont {{Zhang}},
  \citenamefont {{Zutic}},\ and\ \citenamefont {{Zhou}}}]{Zhu2025AMFET}%
  \BibitemOpen
  \bibfield  {author} {\bibinfo {author} {\bibfnamefont {Z.}~\bibnamefont
  {{Zhu}}}, \bibinfo {author} {\bibfnamefont {X.}~\bibnamefont {{Chen}}},
  \bibinfo {author} {\bibfnamefont {X.}~\bibnamefont {{Duan}}}, \bibinfo
  {author} {\bibfnamefont {Z.}~\bibnamefont {{Cui}}}, \bibinfo {author}
  {\bibfnamefont {J.}~\bibnamefont {{Zhang}}}, \bibinfo {author} {\bibfnamefont
  {I.}~\bibnamefont {{Zutic}}},\ and\ \bibinfo {author} {\bibfnamefont
  {T.}~\bibnamefont {{Zhou}}},\ }\bibfield  {title} {\bibinfo {title}
  {{Altermagnetoelectric Spin Field Effect Transistor}},\ }\href
  {https://doi.org/10.48550/arXiv.2512.02974} {\bibfield  {journal} {\bibinfo
  {journal} {arXiv e-prints}\ ,\ \bibinfo {pages} {arXiv:2512.02974}} (\bibinfo
  {year} {2025}{\natexlab{b}})}\BibitemShut {NoStop}%
\bibitem [{\citenamefont {Fu}\ \emph {et~al.}(2025)\citenamefont {Fu},
  \citenamefont {Lv}, \citenamefont {Xu}, \citenamefont {Cayao}, \citenamefont
  {Liu},\ and\ \citenamefont {Yu}}]{Fu2025AM}%
  \BibitemOpen
  \bibfield  {author} {\bibinfo {author} {\bibfnamefont {P.-H.}\ \bibnamefont
  {Fu}}, \bibinfo {author} {\bibfnamefont {Q.}~\bibnamefont {Lv}}, \bibinfo
  {author} {\bibfnamefont {Y.}~\bibnamefont {Xu}}, \bibinfo {author}
  {\bibfnamefont {J.}~\bibnamefont {Cayao}}, \bibinfo {author} {\bibfnamefont
  {J.-F.}\ \bibnamefont {Liu}},\ and\ \bibinfo {author} {\bibfnamefont {X.-L.}\
  \bibnamefont {Yu}},\ }\bibfield  {title} {\bibinfo {title} {All-electrically
  controlled spintronics in altermagnetic heterostructures},\ }\href
  {https://doi.org/10.1038/s41535-025-00827-7} {\bibfield  {journal} {\bibinfo
  {journal} {npj Quantum Materials}\ }\textbf {\bibinfo {volume} {10}},\
  \bibinfo {pages} {111} (\bibinfo {year} {2025})}\BibitemShut {NoStop}%
\bibitem [{\citenamefont {Litvin}\ and\ \citenamefont
  {Opechowski}(1974)}]{Livtin1974}%
  \BibitemOpen
  \bibfield  {author} {\bibinfo {author} {\bibfnamefont {D.}~\bibnamefont
  {Litvin}}\ and\ \bibinfo {author} {\bibfnamefont {W.}~\bibnamefont
  {Opechowski}},\ }\bibfield  {title} {\bibinfo {title} {{Spin groups}},\
  }\href {https://doi.org/https://doi.org/10.1016/0031-8914(74)90157-8}
  {\bibfield  {journal} {\bibinfo  {journal} {Physica}\ }\textbf {\bibinfo
  {volume} {76}},\ \bibinfo {pages} {538} (\bibinfo {year} {1974})}\BibitemShut
  {NoStop}%
\bibitem [{\citenamefont {Litvin}(1977)}]{Litvin1977}%
  \BibitemOpen
  \bibfield  {author} {\bibinfo {author} {\bibfnamefont {D.~B.}\ \bibnamefont
  {Litvin}},\ }\bibfield  {title} {\bibinfo {title} {{Spin point groups}},\
  }\href {https://doi.org/10.1107/S0567739477000709} {\bibfield  {journal}
  {\bibinfo  {journal} {Acta Crystallographica Section A}\ }\textbf {\bibinfo
  {volume} {33}},\ \bibinfo {pages} {279} (\bibinfo {year} {1977})}\BibitemShut
  {NoStop}%
\bibitem [{\citenamefont {Hayami}\ \emph
  {et~al.}(2020{\natexlab{a}})\citenamefont {Hayami}, \citenamefont {Yanagi},\
  and\ \citenamefont {Kusunose}}]{Hayami2020AM1}%
  \BibitemOpen
  \bibfield  {author} {\bibinfo {author} {\bibfnamefont {S.}~\bibnamefont
  {Hayami}}, \bibinfo {author} {\bibfnamefont {Y.}~\bibnamefont {Yanagi}},\
  and\ \bibinfo {author} {\bibfnamefont {H.}~\bibnamefont {Kusunose}},\
  }\bibfield  {title} {\bibinfo {title} {Bottom-up design of spin-split and
  reshaped electronic band structures in antiferromagnets without spin-orbit
  coupling: Procedure on the basis of augmented multipoles},\ }\href
  {https://doi.org/10.1103/PhysRevB.102.144441} {\bibfield  {journal} {\bibinfo
   {journal} {Phys. Rev. B}\ }\textbf {\bibinfo {volume} {102}},\ \bibinfo
  {pages} {144441} (\bibinfo {year} {2020}{\natexlab{a}})}\BibitemShut
  {NoStop}%
\bibitem [{\citenamefont {Hayami}\ \emph
  {et~al.}(2020{\natexlab{b}})\citenamefont {Hayami}, \citenamefont {Yanagi},\
  and\ \citenamefont {Kusunose}}]{Hayami2020AM2}%
  \BibitemOpen
  \bibfield  {author} {\bibinfo {author} {\bibfnamefont {S.}~\bibnamefont
  {Hayami}}, \bibinfo {author} {\bibfnamefont {Y.}~\bibnamefont {Yanagi}},\
  and\ \bibinfo {author} {\bibfnamefont {H.}~\bibnamefont {Kusunose}},\
  }\bibfield  {title} {\bibinfo {title} {Spontaneous antisymmetric spin
  splitting in noncollinear antiferromagnets without spin-orbit coupling},\
  }\href {https://doi.org/10.1103/PhysRevB.101.220403} {\bibfield  {journal}
  {\bibinfo  {journal} {Phys. Rev. B}\ }\textbf {\bibinfo {volume} {101}},\
  \bibinfo {pages} {220403} (\bibinfo {year} {2020}{\natexlab{b}})}\BibitemShut
  {NoStop}%
\bibitem [{\citenamefont {Lee}\ \emph {et~al.}(2024{\natexlab{b}})\citenamefont
  {Lee}, \citenamefont {Qian},\ and\ \citenamefont {Yang}}]{Lee2024dwaveNAFM}%
  \BibitemOpen
  \bibfield  {author} {\bibinfo {author} {\bibfnamefont {S.~H.}\ \bibnamefont
  {Lee}}, \bibinfo {author} {\bibfnamefont {Y.}~\bibnamefont {Qian}},\ and\
  \bibinfo {author} {\bibfnamefont {B.-J.}\ \bibnamefont {Yang}},\ }\bibfield
  {title} {\bibinfo {title} {Fermi surface spin texture and topological
  superconductivity in spin-orbit free noncollinear antiferromagnets},\ }\href
  {https://doi.org/10.1103/PhysRevLett.132.196602} {\bibfield  {journal}
  {\bibinfo  {journal} {Phys. Rev. Lett.}\ }\textbf {\bibinfo {volume} {132}},\
  \bibinfo {pages} {196602} (\bibinfo {year} {2024}{\natexlab{b}})}\BibitemShut
  {NoStop}%
\bibitem [{\citenamefont {{Birk Hellenes}}\ \emph {et~al.}(2023)\citenamefont
  {{Birk Hellenes}}, \citenamefont {{Jungwirth}}, \citenamefont
  {{Jaeschke-Ubiergo}}, \citenamefont {{Chakraborty}}, \citenamefont
  {{Sinova}},\ and\ \citenamefont {{{\v{S}}mejkal}}}]{Birk2023}%
  \BibitemOpen
  \bibfield  {author} {\bibinfo {author} {\bibfnamefont {A.}~\bibnamefont
  {{Birk Hellenes}}}, \bibinfo {author} {\bibfnamefont {T.}~\bibnamefont
  {{Jungwirth}}}, \bibinfo {author} {\bibfnamefont {R.}~\bibnamefont
  {{Jaeschke-Ubiergo}}}, \bibinfo {author} {\bibfnamefont {A.}~\bibnamefont
  {{Chakraborty}}}, \bibinfo {author} {\bibfnamefont {J.}~\bibnamefont
  {{Sinova}}},\ and\ \bibinfo {author} {\bibfnamefont {L.}~\bibnamefont
  {{{\v{S}}mejkal}}},\ }\bibfield  {title} {\bibinfo {title} {{P-wave
  magnets}},\ }\href {https://doi.org/10.48550/arXiv.2309.01607} {\bibfield
  {journal} {\bibinfo  {journal} {arXiv e-prints}\ ,\ \bibinfo {pages}
  {arXiv:2309.01607}} (\bibinfo {year} {2023})}\BibitemShut {NoStop}%
\bibitem [{\citenamefont {Yamada}\ \emph {et~al.}(2025)\citenamefont {Yamada},
  \citenamefont {Birch}, \citenamefont {Baral}, \citenamefont {Okumura},
  \citenamefont {Nakano}, \citenamefont {Gao}, \citenamefont {Ezawa},
  \citenamefont {Nomoto}, \citenamefont {Masell}, \citenamefont {Ishihara},
  \citenamefont {Kolincio}, \citenamefont {Belopolski}, \citenamefont
  {Sagayama}, \citenamefont {Nakao}, \citenamefont {Ohishi}, \citenamefont
  {Ohhara}, \citenamefont {Kiyanagi}, \citenamefont {Nakajima}, \citenamefont
  {Tokura}, \citenamefont {Arima}, \citenamefont {Motome}, \citenamefont
  {Hirschmann},\ and\ \citenamefont {Hirschberger}}]{Yamada2025}%
  \BibitemOpen
  \bibfield  {author} {\bibinfo {author} {\bibfnamefont {R.}~\bibnamefont
  {Yamada}}, \bibinfo {author} {\bibfnamefont {M.~T.}\ \bibnamefont {Birch}},
  \bibinfo {author} {\bibfnamefont {P.~R.}\ \bibnamefont {Baral}}, \bibinfo
  {author} {\bibfnamefont {S.}~\bibnamefont {Okumura}}, \bibinfo {author}
  {\bibfnamefont {R.}~\bibnamefont {Nakano}}, \bibinfo {author} {\bibfnamefont
  {S.}~\bibnamefont {Gao}}, \bibinfo {author} {\bibfnamefont {M.}~\bibnamefont
  {Ezawa}}, \bibinfo {author} {\bibfnamefont {T.}~\bibnamefont {Nomoto}},
  \bibinfo {author} {\bibfnamefont {J.}~\bibnamefont {Masell}}, \bibinfo
  {author} {\bibfnamefont {Y.}~\bibnamefont {Ishihara}}, \bibinfo {author}
  {\bibfnamefont {K.~K.}\ \bibnamefont {Kolincio}}, \bibinfo {author}
  {\bibfnamefont {I.}~\bibnamefont {Belopolski}}, \bibinfo {author}
  {\bibfnamefont {H.}~\bibnamefont {Sagayama}}, \bibinfo {author}
  {\bibfnamefont {H.}~\bibnamefont {Nakao}}, \bibinfo {author} {\bibfnamefont
  {K.}~\bibnamefont {Ohishi}}, \bibinfo {author} {\bibfnamefont
  {T.}~\bibnamefont {Ohhara}}, \bibinfo {author} {\bibfnamefont
  {R.}~\bibnamefont {Kiyanagi}}, \bibinfo {author} {\bibfnamefont
  {T.}~\bibnamefont {Nakajima}}, \bibinfo {author} {\bibfnamefont
  {Y.}~\bibnamefont {Tokura}}, \bibinfo {author} {\bibfnamefont {T.-h.}\
  \bibnamefont {Arima}}, \bibinfo {author} {\bibfnamefont {Y.}~\bibnamefont
  {Motome}}, \bibinfo {author} {\bibfnamefont {M.~M.}\ \bibnamefont
  {Hirschmann}},\ and\ \bibinfo {author} {\bibfnamefont {M.}~\bibnamefont
  {Hirschberger}},\ }\bibfield  {title} {\bibinfo {title} {{A metallic p-wave
  magnet with commensurate spin helix}},\ }\href
  {https://doi.org/10.1038/s41586-025-09633-4} {\bibfield  {journal} {\bibinfo
  {journal} {Nature}\ }\textbf {\bibinfo {volume} {646}},\ \bibinfo {pages}
  {837} (\bibinfo {year} {2025})}\BibitemShut {NoStop}%
\bibitem [{\citenamefont {Yu}\ \emph {et~al.}(2025)\citenamefont {Yu},
  \citenamefont {Lyngby}, \citenamefont {Shishidou}, \citenamefont {Roig},
  \citenamefont {Kreisel}, \citenamefont {Weinert}, \citenamefont {Andersen},\
  and\ \citenamefont {Agterberg}}]{Yu2025pwave}%
  \BibitemOpen
  \bibfield  {author} {\bibinfo {author} {\bibfnamefont {Y.}~\bibnamefont
  {Yu}}, \bibinfo {author} {\bibfnamefont {M.~B.}\ \bibnamefont {Lyngby}},
  \bibinfo {author} {\bibfnamefont {T.}~\bibnamefont {Shishidou}}, \bibinfo
  {author} {\bibfnamefont {M.}~\bibnamefont {Roig}}, \bibinfo {author}
  {\bibfnamefont {A.}~\bibnamefont {Kreisel}}, \bibinfo {author} {\bibfnamefont
  {M.}~\bibnamefont {Weinert}}, \bibinfo {author} {\bibfnamefont {B.~M.}\
  \bibnamefont {Andersen}},\ and\ \bibinfo {author} {\bibfnamefont {D.~F.}\
  \bibnamefont {Agterberg}},\ }\bibfield  {title} {\bibinfo {title} {Odd-parity
  magnetism driven by antiferromagnetic exchange},\ }\href
  {https://doi.org/10.1103/zk69-k6b2} {\bibfield  {journal} {\bibinfo
  {journal} {Phys. Rev. Lett.}\ }\textbf {\bibinfo {volume} {135}},\ \bibinfo
  {pages} {046701} (\bibinfo {year} {2025})}\BibitemShut {NoStop}%
\bibitem [{\citenamefont {{Lin}}(2025)}]{Lin2025OAM}%
  \BibitemOpen
  \bibfield  {author} {\bibinfo {author} {\bibfnamefont {Y.-P.}\ \bibnamefont
  {{Lin}}},\ }\bibfield  {title} {\bibinfo {title} {{Odd-parity altermagnetism
  through sublattice currents: From Haldane-Hubbard model to general bipartite
  lattices}},\ }\href {https://doi.org/10.48550/arXiv.2503.09602} {\bibfield
  {journal} {\bibinfo  {journal} {arXiv e-prints}\ ,\ \bibinfo {pages}
  {arXiv:2503.09602}} (\bibinfo {year} {2025})}\BibitemShut {NoStop}%
\bibitem [{\citenamefont {{Zeng}}\ \emph
  {et~al.}(2025{\natexlab{a}})\citenamefont {{Zeng}}, \citenamefont {{Qin}},
  \citenamefont {{Qin}}, \citenamefont {{Feng}}, \citenamefont {{Xu}},\ and\
  \citenamefont {{Wang}}}]{Zeng2025OPAMa}%
  \BibitemOpen
  \bibfield  {author} {\bibinfo {author} {\bibfnamefont {M.}~\bibnamefont
  {{Zeng}}}, \bibinfo {author} {\bibfnamefont {Z.}~\bibnamefont {{Qin}}},
  \bibinfo {author} {\bibfnamefont {L.}~\bibnamefont {{Qin}}}, \bibinfo
  {author} {\bibfnamefont {S.}~\bibnamefont {{Feng}}}, \bibinfo {author}
  {\bibfnamefont {D.-H.}\ \bibnamefont {{Xu}}},\ and\ \bibinfo {author}
  {\bibfnamefont {R.}~\bibnamefont {{Wang}}},\ }\bibfield  {title} {\bibinfo
  {title} {{The odd-parity altermagnetism: A spin group study}},\ }\href
  {https://doi.org/10.48550/arXiv.2507.09906} {\bibfield  {journal} {\bibinfo
  {journal} {arXiv e-prints}\ ,\ \bibinfo {pages} {arXiv:2507.09906}} (\bibinfo
  {year} {2025}{\natexlab{a}})}\BibitemShut {NoStop}%
\bibitem [{\citenamefont {{Zeng}}\ \emph
  {et~al.}(2025{\natexlab{b}})\citenamefont {{Zeng}}, \citenamefont {{Qin}},
  \citenamefont {{Feng}}, \citenamefont {{Xu}},\ and\ \citenamefont
  {{Wang}}}]{Zeng2025OPAMb}%
  \BibitemOpen
  \bibfield  {author} {\bibinfo {author} {\bibfnamefont {M.}~\bibnamefont
  {{Zeng}}}, \bibinfo {author} {\bibfnamefont {L.}~\bibnamefont {{Qin}}},
  \bibinfo {author} {\bibfnamefont {S.}~\bibnamefont {{Feng}}}, \bibinfo
  {author} {\bibfnamefont {D.-H.}\ \bibnamefont {{Xu}}},\ and\ \bibinfo
  {author} {\bibfnamefont {R.}~\bibnamefont {{Wang}}},\ }\bibfield  {title}
  {\bibinfo {title} {{The spin Hall conductivity in the hole-doped bilayer
  Haldane-Hubbard model with odd-parity ALM}},\ }\href
  {https://doi.org/10.48550/arXiv.2510.12602} {\bibfield  {journal} {\bibinfo
  {journal} {arXiv e-prints}\ ,\ \bibinfo {pages} {arXiv:2510.12602}} (\bibinfo
  {year} {2025}{\natexlab{b}})}\BibitemShut {NoStop}%
\bibitem [{\citenamefont {{Zhuang}}\ \emph
  {et~al.}(2025{\natexlab{b}})\citenamefont {{Zhuang}}, \citenamefont {{Zhu}},
  \citenamefont {{Liu}}, \citenamefont {{Wu}},\ and\ \citenamefont
  {{Yan}}}]{zhuang2025AM9}%
  \BibitemOpen
  \bibfield  {author} {\bibinfo {author} {\bibfnamefont {Z.-Y.}\ \bibnamefont
  {{Zhuang}}}, \bibinfo {author} {\bibfnamefont {D.}~\bibnamefont {{Zhu}}},
  \bibinfo {author} {\bibfnamefont {D.}~\bibnamefont {{Liu}}}, \bibinfo
  {author} {\bibfnamefont {Z.}~\bibnamefont {{Wu}}},\ and\ \bibinfo {author}
  {\bibfnamefont {Z.}~\bibnamefont {{Yan}}},\ }\bibfield  {title} {\bibinfo
  {title} {{Odd-Parity Altermagnetism Originated from Orbital Orders}},\ }\href
  {https://doi.org/10.48550/arXiv.2508.18361} {\bibfield  {journal} {\bibinfo
  {journal} {arXiv e-prints}\ ,\ \bibinfo {pages} {arXiv:2508.18361}} (\bibinfo
  {year} {2025}{\natexlab{b}})}\BibitemShut {NoStop}%
\bibitem [{\citenamefont {Huang}\ \emph {et~al.}(2026)\citenamefont {Huang},
  \citenamefont {Qin}, \citenamefont {Zhan}, \citenamefont {Xu}, \citenamefont
  {Ma},\ and\ \citenamefont {Wang}}]{Huang2025AM7}%
  \BibitemOpen
  \bibfield  {author} {\bibinfo {author} {\bibfnamefont {S.}~\bibnamefont
  {Huang}}, \bibinfo {author} {\bibfnamefont {Z.}~\bibnamefont {Qin}}, \bibinfo
  {author} {\bibfnamefont {F.}~\bibnamefont {Zhan}}, \bibinfo {author}
  {\bibfnamefont {D.-H.}\ \bibnamefont {Xu}}, \bibinfo {author} {\bibfnamefont
  {D.-S.}\ \bibnamefont {Ma}},\ and\ \bibinfo {author} {\bibfnamefont
  {R.}~\bibnamefont {Wang}},\ }\bibfield  {title} {\bibinfo {title}
  {Light-induced odd-parity magnetism in conventional antiferromagnetism},\
  }\href {https://doi.org/10.1103/9346-9jpf} {\bibfield  {journal} {\bibinfo
  {journal} {Phys. Rev. Lett.}\ }\textbf {\bibinfo {volume} {136}},\ \bibinfo
  {pages} {126703} (\bibinfo {year} {2026})}\BibitemShut {NoStop}%
\bibitem [{\citenamefont {Li}\ \emph {et~al.}(2026)\citenamefont {Li},
  \citenamefont {Shao},\ and\ \citenamefont {Kovalev}}]{Li2025AM7}%
  \BibitemOpen
  \bibfield  {author} {\bibinfo {author} {\bibfnamefont {B.}~\bibnamefont
  {Li}}, \bibinfo {author} {\bibfnamefont {D.-F.}\ \bibnamefont {Shao}},\ and\
  \bibinfo {author} {\bibfnamefont {A.~A.}\ \bibnamefont {Kovalev}},\
  }\bibfield  {title} {\bibinfo {title} {Floquet spin splitting and spin
  generation in antiferromagnets},\ }\href {https://doi.org/10.1103/xzm1-l6yf}
  {\bibfield  {journal} {\bibinfo  {journal} {Phys. Rev. Lett.}\ }\textbf
  {\bibinfo {volume} {136}},\ \bibinfo {pages} {166701} (\bibinfo {year}
  {2026})}\BibitemShut {NoStop}%
\bibitem [{\citenamefont {Zhu}\ \emph {et~al.}(2026{\natexlab{a}})\citenamefont
  {Zhu}, \citenamefont {Zhou}, \citenamefont {Wang}, \citenamefont {Wei},\ and\
  \citenamefont {Ruan}}]{Zhu2025AM8}%
  \BibitemOpen
  \bibfield  {author} {\bibinfo {author} {\bibfnamefont {T.}~\bibnamefont
  {Zhu}}, \bibinfo {author} {\bibfnamefont {D.}~\bibnamefont {Zhou}}, \bibinfo
  {author} {\bibfnamefont {H.}~\bibnamefont {Wang}}, \bibinfo {author}
  {\bibfnamefont {S.-H.}\ \bibnamefont {Wei}},\ and\ \bibinfo {author}
  {\bibfnamefont {J.}~\bibnamefont {Ruan}},\ }\bibfield  {title} {\bibinfo
  {title} {Floquet odd-parity collinear magnets},\ }\href
  {https://doi.org/10.1103/7ywb-ml2q} {\bibfield  {journal} {\bibinfo
  {journal} {Phys. Rev. Lett.}\ }\textbf {\bibinfo {volume} {136}},\ \bibinfo
  {pages} {126704} (\bibinfo {year} {2026}{\natexlab{a}})}\BibitemShut
  {NoStop}%
\bibitem [{\citenamefont {Liu}\ \emph {et~al.}(2026)\citenamefont {Liu},
  \citenamefont {Zhuang}, \citenamefont {Zhu}, \citenamefont {Wu},\ and\
  \citenamefont {Yan}}]{liu2025AM10}%
  \BibitemOpen
  \bibfield  {author} {\bibinfo {author} {\bibfnamefont {D.}~\bibnamefont
  {Liu}}, \bibinfo {author} {\bibfnamefont {Z.-Y.}\ \bibnamefont {Zhuang}},
  \bibinfo {author} {\bibfnamefont {D.}~\bibnamefont {Zhu}}, \bibinfo {author}
  {\bibfnamefont {Z.}~\bibnamefont {Wu}},\ and\ \bibinfo {author}
  {\bibfnamefont {Z.}~\bibnamefont {Yan}},\ }\bibfield  {title} {\bibinfo
  {title} {Light-induced odd-parity altermagnets on dimerized lattices},\
  }\href {https://doi.org/10.1103/wnqs-3djt} {\bibfield  {journal} {\bibinfo
  {journal} {Phys. Rev. B}\ }\textbf {\bibinfo {volume} {113}},\ \bibinfo
  {pages} {L060409} (\bibinfo {year} {2026})}\BibitemShut {NoStop}%
\bibitem [{\citenamefont {Pan}\ \emph {et~al.}(2025)\citenamefont {Pan},
  \citenamefont {Zhou}, \citenamefont {Hu}, \citenamefont {Liu}, \citenamefont
  {Zhou}, \citenamefont {Xiao}, \citenamefont {Yang},\ and\ \citenamefont
  {Sun}}]{Pan2025OPAM}%
  \BibitemOpen
  \bibfield  {author} {\bibinfo {author} {\bibfnamefont {B.}~\bibnamefont
  {Pan}}, \bibinfo {author} {\bibfnamefont {P.}~\bibnamefont {Zhou}}, \bibinfo
  {author} {\bibfnamefont {Y.}~\bibnamefont {Hu}}, \bibinfo {author}
  {\bibfnamefont {S.}~\bibnamefont {Liu}}, \bibinfo {author} {\bibfnamefont
  {B.}~\bibnamefont {Zhou}}, \bibinfo {author} {\bibfnamefont {H.}~\bibnamefont
  {Xiao}}, \bibinfo {author} {\bibfnamefont {X.}~\bibnamefont {Yang}},\ and\
  \bibinfo {author} {\bibfnamefont {L.}~\bibnamefont {Sun}},\ }\bibfield
  {title} {\bibinfo {title} {{Floquet-induced altermagnetic transition in
  $A$-type antiferromagnetic bilayers}},\ }\href
  {https://doi.org/10.1103/nn5t-kmln} {\bibfield  {journal} {\bibinfo
  {journal} {Phys. Rev. B}\ }\textbf {\bibinfo {volume} {112}},\ \bibinfo
  {pages} {224430} (\bibinfo {year} {2025})}\BibitemShut {NoStop}%
\bibitem [{\citenamefont {{Li}}\ \emph {et~al.}(2025)\citenamefont {{Li}},
  \citenamefont {{Li}}, \citenamefont {{Guan}},\ and\ \citenamefont
  {{Meng}}}]{Li2025FloAM}%
  \BibitemOpen
  \bibfield  {author} {\bibinfo {author} {\bibfnamefont {Z.}~\bibnamefont
  {{Li}}}, \bibinfo {author} {\bibfnamefont {L.}~\bibnamefont {{Li}}}, \bibinfo
  {author} {\bibfnamefont {M.}~\bibnamefont {{Guan}}},\ and\ \bibinfo {author}
  {\bibfnamefont {S.}~\bibnamefont {{Meng}}},\ }\bibfield  {title} {\bibinfo
  {title} {{Stacking-sliding and irradiation-direction invariant Floquet
  altermagnets in A-type antiferromagnetic bilayers}},\ }\href
  {https://doi.org/10.48550/arXiv.2512.06416} {\bibfield  {journal} {\bibinfo
  {journal} {arXiv e-prints}\ ,\ \bibinfo {pages} {arXiv:2512.06416}} (\bibinfo
  {year} {2025})}\BibitemShut {NoStop}%
\bibitem [{\citenamefont {Oka}\ and\ \citenamefont {Aoki}(2009)}]{Oka2009}%
  \BibitemOpen
  \bibfield  {author} {\bibinfo {author} {\bibfnamefont {T.}~\bibnamefont
  {Oka}}\ and\ \bibinfo {author} {\bibfnamefont {H.}~\bibnamefont {Aoki}},\
  }\bibfield  {title} {\bibinfo {title} {{Photovoltaic Hall effect in
  graphene}},\ }\href {https://doi.org/10.1103/PhysRevB.79.081406} {\bibfield
  {journal} {\bibinfo  {journal} {Phys. Rev. B}\ }\textbf {\bibinfo {volume}
  {79}},\ \bibinfo {pages} {081406} (\bibinfo {year} {2009})}\BibitemShut
  {NoStop}%
\bibitem [{\citenamefont {Lindner}\ \emph {et~al.}(2011)\citenamefont
  {Lindner}, \citenamefont {Refael},\ and\ \citenamefont
  {Galitski}}]{lindner2011floquet}%
  \BibitemOpen
  \bibfield  {author} {\bibinfo {author} {\bibfnamefont {N.~H.}\ \bibnamefont
  {Lindner}}, \bibinfo {author} {\bibfnamefont {G.}~\bibnamefont {Refael}},\
  and\ \bibinfo {author} {\bibfnamefont {V.}~\bibnamefont {Galitski}},\
  }\bibfield  {title} {\bibinfo {title} {{Floquet topological insulator in
  semiconductor quantum wells}},\ }\href
  {https://doi.org/https://doi.org/10.1038/nphys1926} {\bibfield  {journal}
  {\bibinfo  {journal} {Nature Physics}\ }\textbf {\bibinfo {volume} {7}},\
  \bibinfo {pages} {490} (\bibinfo {year} {2011})}\BibitemShut {NoStop}%
\bibitem [{\citenamefont {Yan}\ and\ \citenamefont
  {Wang}(2016)}]{yan2016tunable}%
  \BibitemOpen
  \bibfield  {author} {\bibinfo {author} {\bibfnamefont {Z.}~\bibnamefont
  {Yan}}\ and\ \bibinfo {author} {\bibfnamefont {Z.}~\bibnamefont {Wang}},\
  }\bibfield  {title} {\bibinfo {title} {{Tunable Weyl Points in Periodically
  Driven Nodal Line Semimetals}},\ }\href
  {https://doi.org/10.1103/PhysRevLett.117.087402} {\bibfield  {journal}
  {\bibinfo  {journal} {Phys. Rev. Lett.}\ }\textbf {\bibinfo {volume} {117}},\
  \bibinfo {pages} {087402} (\bibinfo {year} {2016})}\BibitemShut {NoStop}%
\bibitem [{\citenamefont {Chan}\ \emph {et~al.}(2016)\citenamefont {Chan},
  \citenamefont {Oh}, \citenamefont {Han},\ and\ \citenamefont
  {Lee}}]{Chan2016type}%
  \BibitemOpen
  \bibfield  {author} {\bibinfo {author} {\bibfnamefont {C.-K.}\ \bibnamefont
  {Chan}}, \bibinfo {author} {\bibfnamefont {Y.-T.}\ \bibnamefont {Oh}},
  \bibinfo {author} {\bibfnamefont {J.~H.}\ \bibnamefont {Han}},\ and\ \bibinfo
  {author} {\bibfnamefont {P.~A.}\ \bibnamefont {Lee}},\ }\bibfield  {title}
  {\bibinfo {title} {{Type-II Weyl cone transitions in driven semimetals}},\
  }\href {https://doi.org/10.1103/PhysRevB.94.121106} {\bibfield  {journal}
  {\bibinfo  {journal} {Phys. Rev. B}\ }\textbf {\bibinfo {volume} {94}},\
  \bibinfo {pages} {121106} (\bibinfo {year} {2016})}\BibitemShut {NoStop}%
\bibitem [{\citenamefont {Zhou}\ \emph
  {et~al.}(2023{\natexlab{a}})\citenamefont {Zhou}, \citenamefont {Bao},
  \citenamefont {Fan}, \citenamefont {Zhou}, \citenamefont {Gao}, \citenamefont
  {Zhong}, \citenamefont {Lin}, \citenamefont {Liu}, \citenamefont {Yu},
  \citenamefont {Tang}, \citenamefont {Meng}, \citenamefont {Duan},\ and\
  \citenamefont {Zhou}}]{Zhou2023Floquet}%
  \BibitemOpen
  \bibfield  {author} {\bibinfo {author} {\bibfnamefont {S.}~\bibnamefont
  {Zhou}}, \bibinfo {author} {\bibfnamefont {C.}~\bibnamefont {Bao}}, \bibinfo
  {author} {\bibfnamefont {B.}~\bibnamefont {Fan}}, \bibinfo {author}
  {\bibfnamefont {H.}~\bibnamefont {Zhou}}, \bibinfo {author} {\bibfnamefont
  {Q.}~\bibnamefont {Gao}}, \bibinfo {author} {\bibfnamefont {H.}~\bibnamefont
  {Zhong}}, \bibinfo {author} {\bibfnamefont {T.}~\bibnamefont {Lin}}, \bibinfo
  {author} {\bibfnamefont {H.}~\bibnamefont {Liu}}, \bibinfo {author}
  {\bibfnamefont {P.}~\bibnamefont {Yu}}, \bibinfo {author} {\bibfnamefont
  {P.}~\bibnamefont {Tang}}, \bibinfo {author} {\bibfnamefont {S.}~\bibnamefont
  {Meng}}, \bibinfo {author} {\bibfnamefont {W.}~\bibnamefont {Duan}},\ and\
  \bibinfo {author} {\bibfnamefont {S.}~\bibnamefont {Zhou}},\ }\bibfield
  {title} {\bibinfo {title} {{Pseudospin-selective Floquet band engineering in
  black phosphorus}},\ }\href {https://doi.org/10.1038/s41586-022-05610-3}
  {\bibfield  {journal} {\bibinfo  {journal} {Nature}\ }\textbf {\bibinfo
  {volume} {614}},\ \bibinfo {pages} {75} (\bibinfo {year}
  {2023}{\natexlab{a}})}\BibitemShut {NoStop}%
\bibitem [{\citenamefont {Pal}\ \emph {et~al.}(2025)\citenamefont {Pal},
  \citenamefont {Mondal}, \citenamefont {Nag},\ and\ \citenamefont
  {Saha}}]{Pal2025FloAM}%
  \BibitemOpen
  \bibfield  {author} {\bibinfo {author} {\bibfnamefont {A.}~\bibnamefont
  {Pal}}, \bibinfo {author} {\bibfnamefont {D.}~\bibnamefont {Mondal}},
  \bibinfo {author} {\bibfnamefont {T.}~\bibnamefont {Nag}},\ and\ \bibinfo
  {author} {\bibfnamefont {A.}~\bibnamefont {Saha}},\ }\bibfield  {title}
  {\bibinfo {title} {{Josephson current signature of Floquet Majorana and
  topological accidental zero modes in altermagnet heterostructures}},\ }\href
  {https://doi.org/10.1103/prnx-47mk} {\bibfield  {journal} {\bibinfo
  {journal} {Phys. Rev. B}\ }\textbf {\bibinfo {volume} {112}},\ \bibinfo
  {pages} {L201408} (\bibinfo {year} {2025})}\BibitemShut {NoStop}%
\bibitem [{\citenamefont {{Fu}}\ \emph
  {et~al.}(2025{\natexlab{a}})\citenamefont {{Fu}}, \citenamefont {{Mondal}},
  \citenamefont {{Liu}},\ and\ \citenamefont {{Cayao}}}]{Fu2025FloAM}%
  \BibitemOpen
  \bibfield  {author} {\bibinfo {author} {\bibfnamefont {P.-H.}\ \bibnamefont
  {{Fu}}}, \bibinfo {author} {\bibfnamefont {S.}~\bibnamefont {{Mondal}}},
  \bibinfo {author} {\bibfnamefont {J.-F.}\ \bibnamefont {{Liu}}},\ and\
  \bibinfo {author} {\bibfnamefont {J.}~\bibnamefont {{Cayao}}},\ }\bibfield
  {title} {\bibinfo {title} {{Light-induced Floquet spin-triplet Cooper pairs
  in unconventional magnets}},\ }\href
  {https://doi.org/10.48550/arXiv.2506.10590} {\bibfield  {journal} {\bibinfo
  {journal} {arXiv e-prints}\ ,\ \bibinfo {pages} {arXiv:2506.10590}} (\bibinfo
  {year} {2025}{\natexlab{a}})}\BibitemShut {NoStop}%
\bibitem [{\citenamefont {{Fu}}\ \emph
  {et~al.}(2025{\natexlab{b}})\citenamefont {{Fu}}, \citenamefont {{Mondal}},
  \citenamefont {{Liu}}, \citenamefont {{Tanaka}},\ and\ \citenamefont
  {{Cayao}}}]{Fu2025FloAM2}%
  \BibitemOpen
  \bibfield  {author} {\bibinfo {author} {\bibfnamefont {P.-H.}\ \bibnamefont
  {{Fu}}}, \bibinfo {author} {\bibfnamefont {S.}~\bibnamefont {{Mondal}}},
  \bibinfo {author} {\bibfnamefont {J.-F.}\ \bibnamefont {{Liu}}}, \bibinfo
  {author} {\bibfnamefont {Y.}~\bibnamefont {{Tanaka}}},\ and\ \bibinfo
  {author} {\bibfnamefont {J.}~\bibnamefont {{Cayao}}},\ }\bibfield  {title}
  {\bibinfo {title} {{Floquet engineering spin triplet states in unconventional
  magnets}},\ }\href {https://doi.org/10.48550/arXiv.2505.20205} {\bibfield
  {journal} {\bibinfo  {journal} {arXiv e-prints}\ ,\ \bibinfo {pages}
  {arXiv:2505.20205}} (\bibinfo {year} {2025}{\natexlab{b}})}\BibitemShut
  {NoStop}%
\bibitem [{\citenamefont {Ghorashi}\ and\ \citenamefont
  {Li}(2025)}]{Ghorashi2025FloAM}%
  \BibitemOpen
  \bibfield  {author} {\bibinfo {author} {\bibfnamefont {S.~A.~A.}\
  \bibnamefont {Ghorashi}}\ and\ \bibinfo {author} {\bibfnamefont
  {Q.}~\bibnamefont {Li}},\ }\bibfield  {title} {\bibinfo {title} {{Dynamical
  Generation of Higher-Order Spin-Orbit Coupling, Topology, and Persistent Spin
  Texture in Light-Irradiated Altermagnets}},\ }\href
  {https://doi.org/10.1103/tm58-lbdl} {\bibfield  {journal} {\bibinfo
  {journal} {Phys. Rev. Lett.}\ }\textbf {\bibinfo {volume} {135}},\ \bibinfo
  {pages} {236702} (\bibinfo {year} {2025})}\BibitemShut {NoStop}%
\bibitem [{\citenamefont {Yokoyama}(2025)}]{Yokoyama2025FloAM}%
  \BibitemOpen
  \bibfield  {author} {\bibinfo {author} {\bibfnamefont {T.}~\bibnamefont
  {Yokoyama}},\ }\bibfield  {title} {\bibinfo {title} {{Floquet engineering
  triplet superconductivity in superconductors with spin-orbit coupling or
  altermagnetism}},\ }\href {https://doi.org/10.1103/4tng-rhc4} {\bibfield
  {journal} {\bibinfo  {journal} {Phys. Rev. B}\ }\textbf {\bibinfo {volume}
  {112}},\ \bibinfo {pages} {024512} (\bibinfo {year} {2025})}\BibitemShut
  {NoStop}%
\bibitem [{\citenamefont {{Ganguli}}\ \emph {et~al.}(2025)\citenamefont
  {{Ganguli}}, \citenamefont {{Jana}},\ and\ \citenamefont
  {{Narayan}}}]{Ganguli2025FloAM}%
  \BibitemOpen
  \bibfield  {author} {\bibinfo {author} {\bibfnamefont {M.}~\bibnamefont
  {{Ganguli}}}, \bibinfo {author} {\bibfnamefont {A.}~\bibnamefont {{Jana}}},\
  and\ \bibinfo {author} {\bibfnamefont {A.}~\bibnamefont {{Narayan}}},\
  }\bibfield  {title} {\bibinfo {title} {{Tunable topology, Hall response, and
  spin-textures in bicircularly polarized light illuminated altermagnets}},\
  }\href {https://doi.org/10.48550/arXiv.2509.06349} {\bibfield  {journal}
  {\bibinfo  {journal} {arXiv e-prints}\ ,\ \bibinfo {pages}
  {arXiv:2509.06349}} (\bibinfo {year} {2025})}\BibitemShut {NoStop}%
\bibitem [{\citenamefont {Liu}\ \emph {et~al.}(2025)\citenamefont {Liu},
  \citenamefont {Zhuang},\ and\ \citenamefont {Yan}}]{Liu2025FWSM}%
  \BibitemOpen
  \bibfield  {author} {\bibinfo {author} {\bibfnamefont {D.}~\bibnamefont
  {Liu}}, \bibinfo {author} {\bibfnamefont {Z.-Y.}\ \bibnamefont {Zhuang}},\
  and\ \bibinfo {author} {\bibfnamefont {Z.}~\bibnamefont {Yan}},\ }\bibfield
  {title} {\bibinfo {title} {{Floquet Weyl Semimetals with Linked Fermi
  Arcs}},\ }\href
  {http://iopscience.iop.org/article/10.1088/0256-307X/43/1/010709} {\bibfield
  {journal} {\bibinfo  {journal} {Chinese Physics Letters}\ } (\bibinfo {year}
  {2025})}\BibitemShut {NoStop}%
\bibitem [{\citenamefont {Yarmohammadi}\ \emph {et~al.}(2025)\citenamefont
  {Yarmohammadi}, \citenamefont {Z\"ulicke}, \citenamefont {Berakdar},
  \citenamefont {Linder},\ and\ \citenamefont
  {Freericks}}]{Yarmohammadi2025FloAM}%
  \BibitemOpen
  \bibfield  {author} {\bibinfo {author} {\bibfnamefont {M.}~\bibnamefont
  {Yarmohammadi}}, \bibinfo {author} {\bibfnamefont {U.}~\bibnamefont
  {Z\"ulicke}}, \bibinfo {author} {\bibfnamefont {J.}~\bibnamefont {Berakdar}},
  \bibinfo {author} {\bibfnamefont {J.}~\bibnamefont {Linder}},\ and\ \bibinfo
  {author} {\bibfnamefont {J.~K.}\ \bibnamefont {Freericks}},\ }\bibfield
  {title} {\bibinfo {title} {Anisotropic light-tailored rkky interaction in
  two-dimensional $d$-wave altermagnets},\ }\href
  {https://doi.org/10.1103/k3xb-8pts} {\bibfield  {journal} {\bibinfo
  {journal} {Phys. Rev. B}\ }\textbf {\bibinfo {volume} {111}},\ \bibinfo
  {pages} {224412} (\bibinfo {year} {2025})}\BibitemShut {NoStop}%
\bibitem [{\citenamefont {Yarmohammadi}\ \emph {et~al.}(2026)\citenamefont
  {Yarmohammadi}, \citenamefont {Berritta}, \citenamefont {Bukov},
  \citenamefont {\ifmmode~\check{S}\else \v{S}\fi{}mejkal}, \citenamefont
  {Linder},\ and\ \citenamefont {Oppeneer}}]{Yarmohammadi2026FloAM}%
  \BibitemOpen
  \bibfield  {author} {\bibinfo {author} {\bibfnamefont {M.}~\bibnamefont
  {Yarmohammadi}}, \bibinfo {author} {\bibfnamefont {M.}~\bibnamefont
  {Berritta}}, \bibinfo {author} {\bibfnamefont {M.}~\bibnamefont {Bukov}},
  \bibinfo {author} {\bibfnamefont {L.}~\bibnamefont {\ifmmode~\check{S}\else
  \v{S}\fi{}mejkal}}, \bibinfo {author} {\bibfnamefont {J.}~\bibnamefont
  {Linder}},\ and\ \bibinfo {author} {\bibfnamefont {P.~M.}\ \bibnamefont
  {Oppeneer}},\ }\bibfield  {title} {\bibinfo {title} {Spin polarization
  engineering in $d$-wave altermagnets},\ }\href
  {https://doi.org/10.1103/xt23-9pnv} {\bibfield  {journal} {\bibinfo
  {journal} {Phys. Rev. B}\ }\textbf {\bibinfo {volume} {113}},\ \bibinfo
  {pages} {L060403} (\bibinfo {year} {2026})}\BibitemShut {NoStop}%
\bibitem [{\citenamefont {Kitagawa}\ \emph {et~al.}(2011)\citenamefont
  {Kitagawa}, \citenamefont {Oka}, \citenamefont {Brataas}, \citenamefont
  {Fu},\ and\ \citenamefont {Demler}}]{Kitagawa2011Floquet}%
  \BibitemOpen
  \bibfield  {author} {\bibinfo {author} {\bibfnamefont {T.}~\bibnamefont
  {Kitagawa}}, \bibinfo {author} {\bibfnamefont {T.}~\bibnamefont {Oka}},
  \bibinfo {author} {\bibfnamefont {A.}~\bibnamefont {Brataas}}, \bibinfo
  {author} {\bibfnamefont {L.}~\bibnamefont {Fu}},\ and\ \bibinfo {author}
  {\bibfnamefont {E.}~\bibnamefont {Demler}},\ }\bibfield  {title} {\bibinfo
  {title} {{Transport properties of nonequilibrium systems under the
  application of light: Photoinduced quantum Hall insulators without Landau
  levels}},\ }\href {https://doi.org/10.1103/PhysRevB.84.235108} {\bibfield
  {journal} {\bibinfo  {journal} {Phys. Rev. B}\ }\textbf {\bibinfo {volume}
  {84}},\ \bibinfo {pages} {235108} (\bibinfo {year} {2011})}\BibitemShut
  {NoStop}%
\bibitem [{\citenamefont {Goldman}\ and\ \citenamefont
  {Dalibard}(2014)}]{Goldman2014}%
  \BibitemOpen
  \bibfield  {author} {\bibinfo {author} {\bibfnamefont {N.}~\bibnamefont
  {Goldman}}\ and\ \bibinfo {author} {\bibfnamefont {J.}~\bibnamefont
  {Dalibard}},\ }\bibfield  {title} {\bibinfo {title} {{Periodically Driven
  Quantum Systems: Effective Hamiltonians and Engineered Gauge Fields}},\
  }\href {https://doi.org/10.1103/PhysRevX.4.031027} {\bibfield  {journal}
  {\bibinfo  {journal} {Phys. Rev. X}\ }\textbf {\bibinfo {volume} {4}},\
  \bibinfo {pages} {031027} (\bibinfo {year} {2014})}\BibitemShut {NoStop}%
\bibitem [{sup()}]{supplemental}%
  \BibitemOpen
  \href@noop {} {}\bibinfo {howpublished} {This supplemental material contains
  three sections, including: (I) Derivation of the Floquet lattice Hamiltonian;
  (II) Detailed symmetry analysis of the Hamiltonian before and after driving;
  (III) Analytical method for determining the spin texture.}\BibitemShut
  {Stop}%
\bibitem [{\citenamefont {Hu}\ and\ \citenamefont {Zhang}(2025)}]{Hu2025NAFM2}%
  \BibitemOpen
  \bibfield  {author} {\bibinfo {author} {\bibfnamefont {L.-H.}\ \bibnamefont
  {Hu}}\ and\ \bibinfo {author} {\bibfnamefont {S.-B.}\ \bibnamefont {Zhang}},\
  }\bibfield  {title} {\bibinfo {title} {Spin magnetization in unconventional
  antiferromagnets with collinear and non-collinear spins},\ }\href
  {https://doi.org/10.1007/s11433-024-2567-6} {\bibfield  {journal} {\bibinfo
  {journal} {Science China Physics, Mechanics \& Astronomy}\ }\textbf {\bibinfo
  {volume} {68}},\ \bibinfo {pages} {247211} (\bibinfo {year}
  {2025})}\BibitemShut {NoStop}%
\bibitem [{\citenamefont {Sentef}\ \emph {et~al.}(2015)\citenamefont {Sentef},
  \citenamefont {Claassen}, \citenamefont {Kemper}, \citenamefont {Moritz},
  \citenamefont {Oka}, \citenamefont {Freericks},\ and\ \citenamefont
  {Devereaux}}]{SentefTrARPES}%
  \BibitemOpen
  \bibfield  {author} {\bibinfo {author} {\bibfnamefont {M.}~\bibnamefont
  {Sentef}}, \bibinfo {author} {\bibfnamefont {M.}~\bibnamefont {Claassen}},
  \bibinfo {author} {\bibfnamefont {A.}~\bibnamefont {Kemper}}, \bibinfo
  {author} {\bibfnamefont {B.}~\bibnamefont {Moritz}}, \bibinfo {author}
  {\bibfnamefont {T.}~\bibnamefont {Oka}}, \bibinfo {author} {\bibfnamefont
  {J.}~\bibnamefont {Freericks}},\ and\ \bibinfo {author} {\bibfnamefont
  {T.}~\bibnamefont {Devereaux}},\ }\bibfield  {title} {\bibinfo {title}
  {Theory of {Floquet} band formation and local pseudospin textures in
  pump-probe photoemission of graphene},\ }\href
  {https://doi.org/10.1038/ncomms8047} {\bibfield  {journal} {\bibinfo
  {journal} {Nature Communications}\ }\textbf {\bibinfo {volume} {6}},\
  \bibinfo {pages} {7047} (\bibinfo {year} {2015})}\BibitemShut {NoStop}%
\bibitem [{\citenamefont {Freericks}\ \emph {et~al.}(2009)\citenamefont
  {Freericks}, \citenamefont {Krishnamurthy},\ and\ \citenamefont
  {Pruschke}}]{FreeeicksTrARPES}%
  \BibitemOpen
  \bibfield  {author} {\bibinfo {author} {\bibfnamefont {J.~K.}\ \bibnamefont
  {Freericks}}, \bibinfo {author} {\bibfnamefont {H.~R.}\ \bibnamefont
  {Krishnamurthy}},\ and\ \bibinfo {author} {\bibfnamefont {T.}~\bibnamefont
  {Pruschke}},\ }\bibfield  {title} {\bibinfo {title} {Theoretical description
  of time-resolved photoemission spectroscopy: Application to pump-probe
  experiments},\ }\href {https://doi.org/10.1103/PhysRevLett.102.136401}
  {\bibfield  {journal} {\bibinfo  {journal} {Phys. Rev. Lett.}\ }\textbf
  {\bibinfo {volume} {102}},\ \bibinfo {pages} {136401} (\bibinfo {year}
  {2009})}\BibitemShut {NoStop}%
\bibitem [{\citenamefont {Farrell}\ \emph {et~al.}(2016)\citenamefont
  {Farrell}, \citenamefont {Arsenault},\ and\ \citenamefont
  {Pereg-Barnea}}]{FarrellTrARPES}%
  \BibitemOpen
  \bibfield  {author} {\bibinfo {author} {\bibfnamefont {A.}~\bibnamefont
  {Farrell}}, \bibinfo {author} {\bibfnamefont {A.}~\bibnamefont {Arsenault}},\
  and\ \bibinfo {author} {\bibfnamefont {T.}~\bibnamefont {Pereg-Barnea}},\
  }\bibfield  {title} {\bibinfo {title} {Dirac cones, floquet side bands, and
  theory of time-resolved angle-resolved photoemission},\ }\href
  {https://doi.org/10.1103/PhysRevB.94.155304} {\bibfield  {journal} {\bibinfo
  {journal} {Phys. Rev. B}\ }\textbf {\bibinfo {volume} {94}},\ \bibinfo
  {pages} {155304} (\bibinfo {year} {2016})}\BibitemShut {NoStop}%
\bibitem [{\citenamefont {{Song}}\ \emph {et~al.}(2025)\citenamefont {{Song}},
  \citenamefont {{Qi}}, \citenamefont {{Fang}}, \citenamefont {{Fang}},\ and\
  \citenamefont {{Weng}}}]{Song2025coplanar}%
  \BibitemOpen
  \bibfield  {author} {\bibinfo {author} {\bibfnamefont {Z.}~\bibnamefont
  {{Song}}}, \bibinfo {author} {\bibfnamefont {Z.}~\bibnamefont {{Qi}}},
  \bibinfo {author} {\bibfnamefont {C.}~\bibnamefont {{Fang}}}, \bibinfo
  {author} {\bibfnamefont {Z.}~\bibnamefont {{Fang}}},\ and\ \bibinfo {author}
  {\bibfnamefont {H.}~\bibnamefont {{Weng}}},\ }\bibfield  {title} {\bibinfo
  {title} {{A Unified Symmetry Classification of Magnetic Orders via Spin Space
  Groups: Prediction of Coplanar Even-Wave Phases}},\ }\href
  {https://doi.org/10.48550/arXiv.2512.08901} {\bibfield  {journal} {\bibinfo
  {journal} {arXiv e-prints}\ ,\ \bibinfo {pages} {arXiv:2512.08901}} (\bibinfo
  {year} {2025})}\BibitemShut {NoStop}%
\bibitem [{\citenamefont {Eckardt}\ and\ \citenamefont
  {Anisimovas}(2015)}]{Eckardt2015Highf}%
  \BibitemOpen
  \bibfield  {author} {\bibinfo {author} {\bibfnamefont {A.}~\bibnamefont
  {Eckardt}}\ and\ \bibinfo {author} {\bibfnamefont {E.}~\bibnamefont
  {Anisimovas}},\ }\bibfield  {title} {\bibinfo {title} {High-frequency
  approximation for periodically driven quantum systems from a floquet-space
  perspective},\ }\href@noop {} {\bibfield  {journal} {\bibinfo  {journal} {New
  journal of physics}\ }\textbf {\bibinfo {volume} {17}},\ \bibinfo {pages}
  {093039} (\bibinfo {year} {2015})}\BibitemShut {NoStop}%
\bibitem [{\citenamefont {Chen}\ \emph {et~al.}(2022)\citenamefont {Chen},
  \citenamefont {Yang}, \citenamefont {Zhou}, \citenamefont {Wu}, \citenamefont
  {Duan}, \citenamefont {Deng},\ and\ \citenamefont {Wang}}]{ChenFDrude2022}%
  \BibitemOpen
  \bibfield  {author} {\bibinfo {author} {\bibfnamefont {J.-N.}\ \bibnamefont
  {Chen}}, \bibinfo {author} {\bibfnamefont {Y.-Y.}\ \bibnamefont {Yang}},
  \bibinfo {author} {\bibfnamefont {Y.-L.}\ \bibnamefont {Zhou}}, \bibinfo
  {author} {\bibfnamefont {Y.-J.}\ \bibnamefont {Wu}}, \bibinfo {author}
  {\bibfnamefont {H.-J.}\ \bibnamefont {Duan}}, \bibinfo {author}
  {\bibfnamefont {M.-X.}\ \bibnamefont {Deng}},\ and\ \bibinfo {author}
  {\bibfnamefont {R.-Q.}\ \bibnamefont {Wang}},\ }\bibfield  {title} {\bibinfo
  {title} {{Photon-modulated linear and nonlinear anomalous Hall effects in
  type-II semi-Dirac semimetals}},\ }\href
  {https://doi.org/10.1103/PhysRevB.105.085124} {\bibfield  {journal} {\bibinfo
   {journal} {Phys. Rev. B}\ }\textbf {\bibinfo {volume} {105}},\ \bibinfo
  {pages} {085124} (\bibinfo {year} {2022})}\BibitemShut {NoStop}%
\bibitem [{\citenamefont {Zhu}\ \emph {et~al.}(2026{\natexlab{b}})\citenamefont
  {Zhu}, \citenamefont {Chen},\ and\ \citenamefont {Zhou}}]{ZhuDrudesc2026}%
  \BibitemOpen
  \bibfield  {author} {\bibinfo {author} {\bibfnamefont {J.-Y.}\ \bibnamefont
  {Zhu}}, \bibinfo {author} {\bibfnamefont {R.}~\bibnamefont {Chen}},\ and\
  \bibinfo {author} {\bibfnamefont {B.}~\bibnamefont {Zhou}},\ }\bibfield
  {title} {\bibinfo {title} {Nonlinear hall effect induced by two-frequency
  drives},\ }\href {https://doi.org/10.1103/17rp-cx9f} {\bibfield  {journal}
  {\bibinfo  {journal} {Phys. Rev. B}\ }\textbf {\bibinfo {volume} {113}},\
  \bibinfo {pages} {195422} (\bibinfo {year} {2026}{\natexlab{b}})}\BibitemShut
  {NoStop}%
\bibitem [{\citenamefont {Xiao}\ \emph {et~al.}(2010)\citenamefont {Xiao},
  \citenamefont {Chang},\ and\ \citenamefont {Niu}}]{Xiao2010review}%
  \BibitemOpen
  \bibfield  {author} {\bibinfo {author} {\bibfnamefont {D.}~\bibnamefont
  {Xiao}}, \bibinfo {author} {\bibfnamefont {M.-C.}\ \bibnamefont {Chang}},\
  and\ \bibinfo {author} {\bibfnamefont {Q.}~\bibnamefont {Niu}},\ }\bibfield
  {title} {\bibinfo {title} {Berry phase effects on electronic properties},\
  }\href {https://doi.org/10.1103/RevModPhys.82.1959} {\bibfield  {journal}
  {\bibinfo  {journal} {Rev. Mod. Phys.}\ }\textbf {\bibinfo {volume} {82}},\
  \bibinfo {pages} {1959} (\bibinfo {year} {2010})}\BibitemShut {NoStop}%
\bibitem [{\citenamefont {\ifmmode~\check{Z}\else \v{Z}\fi{}elezn\'y}\ \emph
  {et~al.}(2017)\citenamefont {\ifmmode~\check{Z}\else \v{Z}\fi{}elezn\'y},
  \citenamefont {Zhang}, \citenamefont {Felser},\ and\ \citenamefont
  {Yan}}]{Yan2017PRLMn3Ir}%
  \BibitemOpen
  \bibfield  {author} {\bibinfo {author} {\bibfnamefont {J.}~\bibnamefont
  {\ifmmode~\check{Z}\else \v{Z}\fi{}elezn\'y}}, \bibinfo {author}
  {\bibfnamefont {Y.}~\bibnamefont {Zhang}}, \bibinfo {author} {\bibfnamefont
  {C.}~\bibnamefont {Felser}},\ and\ \bibinfo {author} {\bibfnamefont
  {B.}~\bibnamefont {Yan}},\ }\bibfield  {title} {\bibinfo {title}
  {Spin-polarized current in noncollinear antiferromagnets},\ }\href
  {https://doi.org/10.1103/PhysRevLett.119.187204} {\bibfield  {journal}
  {\bibinfo  {journal} {Phys. Rev. Lett.}\ }\textbf {\bibinfo {volume} {119}},\
  \bibinfo {pages} {187204} (\bibinfo {year} {2017})}\BibitemShut {NoStop}%
\bibitem [{\citenamefont {Liu}\ \emph {et~al.}(2024)\citenamefont {Liu},
  \citenamefont {Yu},\ and\ \citenamefont {Liu}}]{LiuTwistAM2024}%
  \BibitemOpen
  \bibfield  {author} {\bibinfo {author} {\bibfnamefont {Y.}~\bibnamefont
  {Liu}}, \bibinfo {author} {\bibfnamefont {J.}~\bibnamefont {Yu}},\ and\
  \bibinfo {author} {\bibfnamefont {C.-C.}\ \bibnamefont {Liu}},\ }\bibfield
  {title} {\bibinfo {title} {Twisted magnetic van der waals bilayers: An ideal
  platform for altermagnetism},\ }\href
  {https://doi.org/10.1103/PhysRevLett.133.206702} {\bibfield  {journal}
  {\bibinfo  {journal} {Phys. Rev. Lett.}\ }\textbf {\bibinfo {volume} {133}},\
  \bibinfo {pages} {206702} (\bibinfo {year} {2024})}\BibitemShut {NoStop}%
\bibitem [{\citenamefont {Bai}\ \emph {et~al.}(2025)\citenamefont {Bai},
  \citenamefont {Zhang}, \citenamefont {Feng},\ and\ \citenamefont
  {Yao}}]{BaiTypeIVAFM2025}%
  \BibitemOpen
  \bibfield  {author} {\bibinfo {author} {\bibfnamefont {L.}~\bibnamefont
  {Bai}}, \bibinfo {author} {\bibfnamefont {R.-W.}\ \bibnamefont {Zhang}},
  \bibinfo {author} {\bibfnamefont {W.}~\bibnamefont {Feng}},\ and\ \bibinfo
  {author} {\bibfnamefont {Y.}~\bibnamefont {Yao}},\ }\bibfield  {title}
  {\bibinfo {title} {Anomalous hall effect in type iv 2d collinear magnets},\
  }\href {https://doi.org/10.1103/rn1l-d6cq} {\bibfield  {journal} {\bibinfo
  {journal} {Phys. Rev. Lett.}\ }\textbf {\bibinfo {volume} {135}},\ \bibinfo
  {pages} {036702} (\bibinfo {year} {2025})}\BibitemShut {NoStop}%
\bibitem [{\citenamefont {de~la Torre}\ \emph {et~al.}(2021)\citenamefont
  {de~la Torre}, \citenamefont {Kennes}, \citenamefont {Claassen},
  \citenamefont {Gerber}, \citenamefont {McIver},\ and\ \citenamefont
  {Sentef}}]{Torre2021RMPheating}%
  \BibitemOpen
  \bibfield  {author} {\bibinfo {author} {\bibfnamefont {A.}~\bibnamefont
  {de~la Torre}}, \bibinfo {author} {\bibfnamefont {D.~M.}\ \bibnamefont
  {Kennes}}, \bibinfo {author} {\bibfnamefont {M.}~\bibnamefont {Claassen}},
  \bibinfo {author} {\bibfnamefont {S.}~\bibnamefont {Gerber}}, \bibinfo
  {author} {\bibfnamefont {J.~W.}\ \bibnamefont {McIver}},\ and\ \bibinfo
  {author} {\bibfnamefont {M.~A.}\ \bibnamefont {Sentef}},\ }\bibfield  {title}
  {\bibinfo {title} {Colloquium: Nonthermal pathways to ultrafast control in
  quantum materials},\ }\href {https://doi.org/10.1103/RevModPhys.93.041002}
  {\bibfield  {journal} {\bibinfo  {journal} {Rev. Mod. Phys.}\ }\textbf
  {\bibinfo {volume} {93}},\ \bibinfo {pages} {041002} (\bibinfo {year}
  {2021})}\BibitemShut {NoStop}%
\bibitem [{\citenamefont {Mentink}\ \emph {et~al.}(2015)\citenamefont
  {Mentink}, \citenamefont {Balzer},\ and\ \citenamefont
  {Eckstein}}]{Mentink2015}%
  \BibitemOpen
  \bibfield  {author} {\bibinfo {author} {\bibfnamefont {J.~H.}\ \bibnamefont
  {Mentink}}, \bibinfo {author} {\bibfnamefont {K.}~\bibnamefont {Balzer}},\
  and\ \bibinfo {author} {\bibfnamefont {M.}~\bibnamefont {Eckstein}},\
  }\bibfield  {title} {\bibinfo {title} {Ultrafast and reversible control of
  the exchange interaction in {Mott} insulators},\ }\href
  {https://doi.org/10.1038/ncomms7708} {\bibfield  {journal} {\bibinfo
  {journal} {Nature Communications}\ }\textbf {\bibinfo {volume} {6}},\
  \bibinfo {pages} {6708} (\bibinfo {year} {2015})}\BibitemShut {NoStop}%
\bibitem [{\citenamefont {Boschini}\ \emph {et~al.}(2024)\citenamefont
  {Boschini}, \citenamefont {Zonno},\ and\ \citenamefont
  {Damascelli}}]{Boschini2024RMPTrARPES}%
  \BibitemOpen
  \bibfield  {author} {\bibinfo {author} {\bibfnamefont {F.}~\bibnamefont
  {Boschini}}, \bibinfo {author} {\bibfnamefont {M.}~\bibnamefont {Zonno}},\
  and\ \bibinfo {author} {\bibfnamefont {A.}~\bibnamefont {Damascelli}},\
  }\bibfield  {title} {\bibinfo {title} {Time-resolved arpes studies of quantum
  materials},\ }\href {https://doi.org/10.1103/RevModPhys.96.015003} {\bibfield
   {journal} {\bibinfo  {journal} {Rev. Mod. Phys.}\ }\textbf {\bibinfo
  {volume} {96}},\ \bibinfo {pages} {015003} (\bibinfo {year}
  {2024})}\BibitemShut {NoStop}%
\bibitem [{\citenamefont {Zhou}\ \emph
  {et~al.}(2023{\natexlab{b}})\citenamefont {Zhou}, \citenamefont {Bao},
  \citenamefont {Fan}, \citenamefont {Zhou}, \citenamefont {Gao}, \citenamefont
  {Zhong}, \citenamefont {Lin}, \citenamefont {Liu}, \citenamefont {Yu},
  \citenamefont {Tang}, \citenamefont {Meng}, \citenamefont {Duan},\ and\
  \citenamefont {Zhou}}]{Zhou2023TrARPES}%
  \BibitemOpen
  \bibfield  {author} {\bibinfo {author} {\bibfnamefont {S.}~\bibnamefont
  {Zhou}}, \bibinfo {author} {\bibfnamefont {C.}~\bibnamefont {Bao}}, \bibinfo
  {author} {\bibfnamefont {B.}~\bibnamefont {Fan}}, \bibinfo {author}
  {\bibfnamefont {H.}~\bibnamefont {Zhou}}, \bibinfo {author} {\bibfnamefont
  {Q.}~\bibnamefont {Gao}}, \bibinfo {author} {\bibfnamefont {H.}~\bibnamefont
  {Zhong}}, \bibinfo {author} {\bibfnamefont {T.}~\bibnamefont {Lin}}, \bibinfo
  {author} {\bibfnamefont {H.}~\bibnamefont {Liu}}, \bibinfo {author}
  {\bibfnamefont {P.}~\bibnamefont {Yu}}, \bibinfo {author} {\bibfnamefont
  {P.}~\bibnamefont {Tang}}, \bibinfo {author} {\bibfnamefont {S.}~\bibnamefont
  {Meng}}, \bibinfo {author} {\bibfnamefont {W.}~\bibnamefont {Duan}},\ and\
  \bibinfo {author} {\bibfnamefont {S.}~\bibnamefont {Zhou}},\ }\bibfield
  {title} {\bibinfo {title} {Pseudospin-selective {Floquet} band engineering in
  black phosphorus},\ }\href {https://doi.org/10.1038/s41586-022-05610-3}
  {\bibfield  {journal} {\bibinfo  {journal} {Nature}\ }\textbf {\bibinfo
  {volume} {614}},\ \bibinfo {pages} {75} (\bibinfo {year}
  {2023}{\natexlab{b}})}\BibitemShut {NoStop}%
\bibitem [{\citenamefont {Wang}\ \emph {et~al.}(2026)\citenamefont {Wang},
  \citenamefont {Cai}, \citenamefont {Tang}, \citenamefont {Lu}, \citenamefont
  {Chen}, \citenamefont {Sheng}, \citenamefont {Feng}, \citenamefont {Zhong},
  \citenamefont {Zhang}, \citenamefont {Yu},\ and\ \citenamefont
  {Zhou}}]{Wang2026Floquet}%
  \BibitemOpen
  \bibfield  {author} {\bibinfo {author} {\bibfnamefont {F.}~\bibnamefont
  {Wang}}, \bibinfo {author} {\bibfnamefont {X.}~\bibnamefont {Cai}}, \bibinfo
  {author} {\bibfnamefont {X.}~\bibnamefont {Tang}}, \bibinfo {author}
  {\bibfnamefont {J.}~\bibnamefont {Lu}}, \bibinfo {author} {\bibfnamefont
  {W.}~\bibnamefont {Chen}}, \bibinfo {author} {\bibfnamefont {T.}~\bibnamefont
  {Sheng}}, \bibinfo {author} {\bibfnamefont {R.}~\bibnamefont {Feng}},
  \bibinfo {author} {\bibfnamefont {H.}~\bibnamefont {Zhong}}, \bibinfo
  {author} {\bibfnamefont {H.}~\bibnamefont {Zhang}}, \bibinfo {author}
  {\bibfnamefont {P.}~\bibnamefont {Yu}},\ and\ \bibinfo {author}
  {\bibfnamefont {S.}~\bibnamefont {Zhou}},\ }\bibfield  {title} {\bibinfo
  {title} {Observation of floquet-induced gap in graphene},\ }\bibfield
  {journal} {\bibinfo  {journal} {Nature Materials}\ }\href
  {https://doi.org/10.1038/s41563-026-02549-y} {10.1038/s41563-026-02549-y}
  (\bibinfo {year} {2026})\BibitemShut {NoStop}%
\bibitem [{\citenamefont {Mori}\ \emph {et~al.}(2016)\citenamefont {Mori},
  \citenamefont {Kuwahara},\ and\ \citenamefont {Saito}}]{Mori2016heating}%
  \BibitemOpen
  \bibfield  {author} {\bibinfo {author} {\bibfnamefont {T.}~\bibnamefont
  {Mori}}, \bibinfo {author} {\bibfnamefont {T.}~\bibnamefont {Kuwahara}},\
  and\ \bibinfo {author} {\bibfnamefont {K.}~\bibnamefont {Saito}},\ }\bibfield
   {title} {\bibinfo {title} {Rigorous bound on energy absorption and generic
  relaxation in periodically driven quantum systems},\ }\href
  {https://doi.org/10.1103/PhysRevLett.116.120401} {\bibfield  {journal}
  {\bibinfo  {journal} {Phys. Rev. Lett.}\ }\textbf {\bibinfo {volume} {116}},\
  \bibinfo {pages} {120401} (\bibinfo {year} {2016})}\BibitemShut {NoStop}%
\bibitem [{\citenamefont {Dehghani}\ \emph {et~al.}(2014)\citenamefont
  {Dehghani}, \citenamefont {Oka},\ and\ \citenamefont
  {Mitra}}]{DehghaniDisspation}%
  \BibitemOpen
  \bibfield  {author} {\bibinfo {author} {\bibfnamefont {H.}~\bibnamefont
  {Dehghani}}, \bibinfo {author} {\bibfnamefont {T.}~\bibnamefont {Oka}},\ and\
  \bibinfo {author} {\bibfnamefont {A.}~\bibnamefont {Mitra}},\ }\bibfield
  {title} {\bibinfo {title} {Dissipative floquet topological systems},\ }\href
  {https://doi.org/10.1103/PhysRevB.90.195429} {\bibfield  {journal} {\bibinfo
  {journal} {Phys. Rev. B}\ }\textbf {\bibinfo {volume} {90}},\ \bibinfo
  {pages} {195429} (\bibinfo {year} {2014})}\BibitemShut {NoStop}%
\end{thebibliography}%

{\allowdisplaybreaks
        \onecolumngrid
        \clearpage
        \begin{center}
        \textbf{\large Supplemental Materials for `` Light-Induced  Even-Parity Unidirectionally Polarized Spin Splitting in Coplanar Antiferromagnets''}\\
        \vspace{4mm}
	     {Di Zhu,$^{1,\ast}$ Dongling Liu,$^1$ Zheng-Yang Zhuang,$^1$ Zhigang Wu,$^2$ and Zhongbo Yan$^{1,\dagger}$}\\
         \vspace{2mm}
	{\em $^1${Guangdong Provincial Key Laboratory of Magnetoelectric Physics and Devices, State Key Laboratory of Optoelectronic Materials and Technologies, School of Physics, Sun Yat-sen University, Guangzhou 510275, China}}\\
    {\em $^2$ Quantum Science Center of Guangdong-Hong Kong-Macao Greater Bay Area (Guangdong), Shenzhen 508045, China}
        \end{center}
    \setcounter{equation}{0}
		\renewcommand{\theequation}{S\arabic{equation}}
		\setcounter{figure}{0}
		\renewcommand{\thefigure}{S\arabic{figure}}
		\setcounter{section}{0}
\renewcommand{\thesection}{S\arabic{section}}
\setcounter{subsection}{0}
\renewcommand{\thesubsection}{\Alph{subsection}}
\setcounter{secnumdepth}{2}
		\setcounter{table}{0}
		\renewcommand{\thetable}{S\arabic{table}}

        his supplemental material contains three sections, including:
(I) Derivation of the Floquet lattice Hamiltonian; 
(II) Detailed symmetry analysis of the Hamiltonian before and after driving; 
(III) Analytical method for determining the spin texture.
(IV) Low-frequency Floquet Hamiltonian and band structure. (V) Generalization to other coplanar AFM structures. (VI)  Time-, spin-, and angle-resolved photoemission photocurrent. (VII)  Drude spin conductivity. (VIII)  Feasibility of the experimental realization and potential challenges.


\section{I. Derivation of the Floquet lattice Hamiltonian}

The lattice Hamiltonian for the coplanar AFM concerned in the main article is given by
\begin{align}
\mathcal{H}(\bm{k})=&2t\cos{k_x}\tau_0\sigma_xs_0+2t\cos{k_y}\tau_x\sigma_0s_0+4t_{s}\cos{k_x}\cos{k_y}\tau_x\sigma_xs_0\nonumber\\
&+4t_{a}\cos{k_x}\cos{k_y}\tau_y\sigma_ys_0-M\tau_0\sigma_zs_x+M\tau_z\sigma_0s_y,
\end{align}
where $\tau_{0}$, $\sigma_{0}$ and $s_{0}$ are all two-by-two identity matrices. Here, we restore these identity matrices for a clear presentation 
of their orders in the matrix product. 
Under the irradiation of CPL, which is described by a time-dependent vector potential $\bm{A}=A_0(\cos{\omega t},\sin{\omega t)}$, the effect of CPL is incorporated into the Hamiltonian through minimal coupling, i.e., replacing $\bm{k}$ by $\bm{k}+e \bm{A}(t)/\hbar$. 
For notational simplicity, in the following we set $e=\hbar=1$. 
The resulting time-periodic Hamiltonian, with period $T=\frac{2\pi}{\omega}$, admits a Fourier expansion: $\mathcal{H}(\bm{k},t)=\sum_{n}\mathcal{H}_n(\bm{k})e^{in\omega t}$ with $n\in \mathbb{Z}$. Here, we show 
the explicit expressions of $\mathcal{H}_0$, $\mathcal{H}_{\pm1}$, which make the leading-order 
contributions to the Floquet Hamiltonian. Specifically, their forms are:
\begin{align}
\mathcal{H}_0(\bm{k})=&\frac{1}{T}\int_0^T\mathcal{H}(\bm{k}+\bm{A}(t))dt\nonumber\\
=&J_0(A_0)(2t\cos{k_x}\tau_0\sigma_xs_0+2t\cos{k_y\tau_x\sigma_0s_0})+J_0(\sqrt{2}A_0)(4t_{s}\cos{k_x}\cos{k_y}\tau_x\sigma_xs_0\nonumber\\
&+4t_{a}\cos{k_x}\cos{k_y}\tau_y\sigma_ys_0)-M\tau_0\sigma_zs_x+M\tau_z\sigma_0s_y,\\
\mathcal{H}_1(\bm{k})=&\frac{1}{T}\int^T_0\mathcal{H}(\bm{k}+\bm{A}(t))e^{-i\omega t}dt\nonumber\\
=&-2tJ_1(A_0)\sin{k_x}\tau_0\sigma_xs_0+i2tJ_1(A_0)\sin{k_y}\tau_x\sigma_0s_0\nonumber\\
&+\frac{4t_{s}}{\sqrt{2}}J_1(\sqrt{2}A_0)(i\sin{k_y}\cos{k_x}-\sin{k_x}\cos{k_y})\tau_x\sigma_xs_0\nonumber\\
&+\frac{4t_{a}}{\sqrt{2}}J_1(\sqrt{2}A_0)(i\sin{k_y}\cos{k_x}-\sin{k_x}\cos{k_y})\tau_y\sigma_ys_0,\\
\mathcal{H}_{-1}(\bm{k})=&\frac{1}{T}\int^T_0\mathcal{H}(\bm{k}+\bm{A}(t))e^{i\omega t}dt\nonumber\\
=&-2tJ_{1}(A_0)\sin{k_x}\tau_0\sigma_xs_0-2itJ_1(A_0)\sin{k_y}\tau_x\sigma_0s_0\nonumber\\
&+\frac{4t_{s}}{\sqrt{2}}t_{s}J_1(\sqrt{2}A_0)(-i\sin{k_y}\cos{k_x}-\sin{k_x}\cos{k_y})\tau_x\sigma_xs_0\nonumber\\
&+\frac{4t_{s}}{\sqrt{2}}t_{a}J_1(\sqrt{2}A_0)(-i\sin{k_y}\cos{k_x}-\sin{k_x}\cos{k_y})\tau_y\sigma_ys_0,
\end{align}
where $J_n(x)$ represents the Bessel functions of the first kind, arising from the following equalities:
\begin{align}
J_n(x)&=\frac{1}{2\pi i^n}\int^{2\pi}_0e^{ix\cos{\theta}}e^{-in\theta}d\theta,\nonumber\\
J_n(x)&=\frac{1}{2\pi }\int^{2\pi}_0e^{ix\sin{\theta}}e^{-in\theta}d\theta.
\end{align}
As an illustrative example, we show the derivation details for $\mathcal{H}_1$. Following the definition, we have~\cite{Liu2025FWSM}
\begin{align}
\mathcal{H}_1(\bm{k})=&\frac{1}{T}\int^T_0\mathcal{H}(\bm{k}+\bm{A}(t'))e^{-i\omega t'}dt'\nonumber\\
=&\frac{1}{T}\int^{T}_0e^{-i\omega t'}t(e^{ik_x}e^{iA_0\cos{\omega t'}}+e^{-ik_x}e^{-iA_0\cos{\omega t'}})\tau_0\sigma_xs_0 dt'\nonumber\\
&+\frac{1}{T}\int^{T}_0e^{-i\omega t'}t(e^{ik_y}e^{iA_0\sin{\omega t'}}+e^{-ik_y}e^{-iA_0\sin{\omega t'}})\tau_x\sigma_0s_0dt'\nonumber\\
&+\frac{1}{T}\int^{T}_0e^{-i\omega t'}t_{s}(e^{ik_x}e^{iA_0\cos{\omega t'}}+e^{-ik_x}e^{-iA_0\cos{\omega t'}})(e^{ik_y}e^{iA_0\sin{\omega t'}}+e^{-ik_y}e^{-ieA_0\sin{\omega t'}})\tau_x\sigma_xs_0dt'\nonumber\\
&+\frac{1}{T}\int^{T}_0e^{-i\omega t'}t_{a}(e^{ik_x}e^{iA_0\cos{\omega t'}}+e^{-ik_x}e^{-iA_0\cos{\omega t'}})(e^{ik_y}e^{iA_0\sin{\omega t'}}+e^{-ik_y}e^{-iA_0\sin{\omega t'}})\tau_y\sigma_ys_0dt'\nonumber\\
=&-2tJ_1(A_0)\sin{k_x}\tau_0\sigma_xs_0+2itJ_1(A_0)\sin{k_y}\tau_x\sigma_0s_0\nonumber\\
&+t_{s}J_1(\sqrt{2}A_0)(e^{i\pi/4}(e^{i(k_x+k_y)}-e^{-i(k_x+k_y)})+e^{-i\pi/4}(e^{i(-k_x+k_y)}-e^{i(k_x-k_y)}))\tau_x\sigma_xs_0\nonumber\\
&+t_{s}J_1(\sqrt{2}A_0)(e^{i\pi/4}(e^{i(k_x+k_y)}-e^{-i(k_x+k_y)})+e^{-i\pi/4}(e^{i(-k_x+k_y)}-e^{i(k_x-k_y)}))\tau_y\sigma_ys_0\nonumber\\
=&-2tJ_1(A_0)\sin{k_x}\tau_0\sigma_xs_0+i2tJ_1(A_0)\sin{k_y}\tau_x\sigma_0s_0\nonumber\\
&+\frac{4t_{s}}{\sqrt{2}}J_1(\sqrt{2}A_0)(i\sin{k_y}\cos{k_x}-\sin{k_x}\cos{k_y})\tau_x\sigma_xs_0\nonumber\\
&+\frac{4t_{a}}{\sqrt{2}}J_1(\sqrt{2}A_0)(i\sin{k_y}\cos{k_x}-\sin{k_x}\cos{k_y})\tau_y\sigma_ys_0.
\end{align}
Other components can be similarly derived.

In the high-frequency off-resonant regime, the system is described by an effective static Hamiltonian given by
\begin{align}
\mathcal{H}_{\rm eff}(\bm{k})=&\mathcal{H}_0+\sum_{n\ge1}\frac{[\mathcal{H}_n,\mathcal{H}_{-n}]}{n\omega}+O(\omega^{-2}).
\end{align}
Typically,  the band structure modification is dominated by one-photon processes, implying 
the contributions from $n\ge 2$ can be neglected. Consequently, the Floquet Hamiltonian is given by 
\begin{align}
\mathcal{H}_{\rm eff}(\bm{k})=&\mathcal{H}_0+\frac{[\mathcal{H}_1,\mathcal{H}_{-1}]}{\omega}\nonumber\\
=&2J_0(A_0)(t\cos{k_x}\tau_0\sigma_xs_0+t\cos{k_y\tau_x\sigma_0s_0})\nonumber\\
&+4J_0(\sqrt{2}A_0)(t_{s}\cos{k_x}\cos{k_y}\tau_x\sigma_xs_0+t_{a}\cos{k_x}\cos{k_y}\tau_y\sigma_ys_0)\nonumber\\
&-\frac{16\sqrt{2}tt_{a}J_1(A_0)J_1(\sqrt{2}A_0)}{\omega}\sin{k_x}\sin{k_y}\cos{k_x}\tau_y\sigma_zs_0\nonumber\\
&+\frac{16\sqrt{2}tt_{a}J_1(A_0)J_1(\sqrt{2}A_0)}{\omega}\sin{k_x}\sin{k_y}\cos{k_y}\tau_z\sigma_ys_0\nonumber\\
&-M\tau_0\sigma_zs_x+M\tau_z\sigma_0s_y,
\end{align}
which is Eq.~(5) of the main text.

\begin{figure}[t]
\centering
\includegraphics[width=0.65\textwidth]{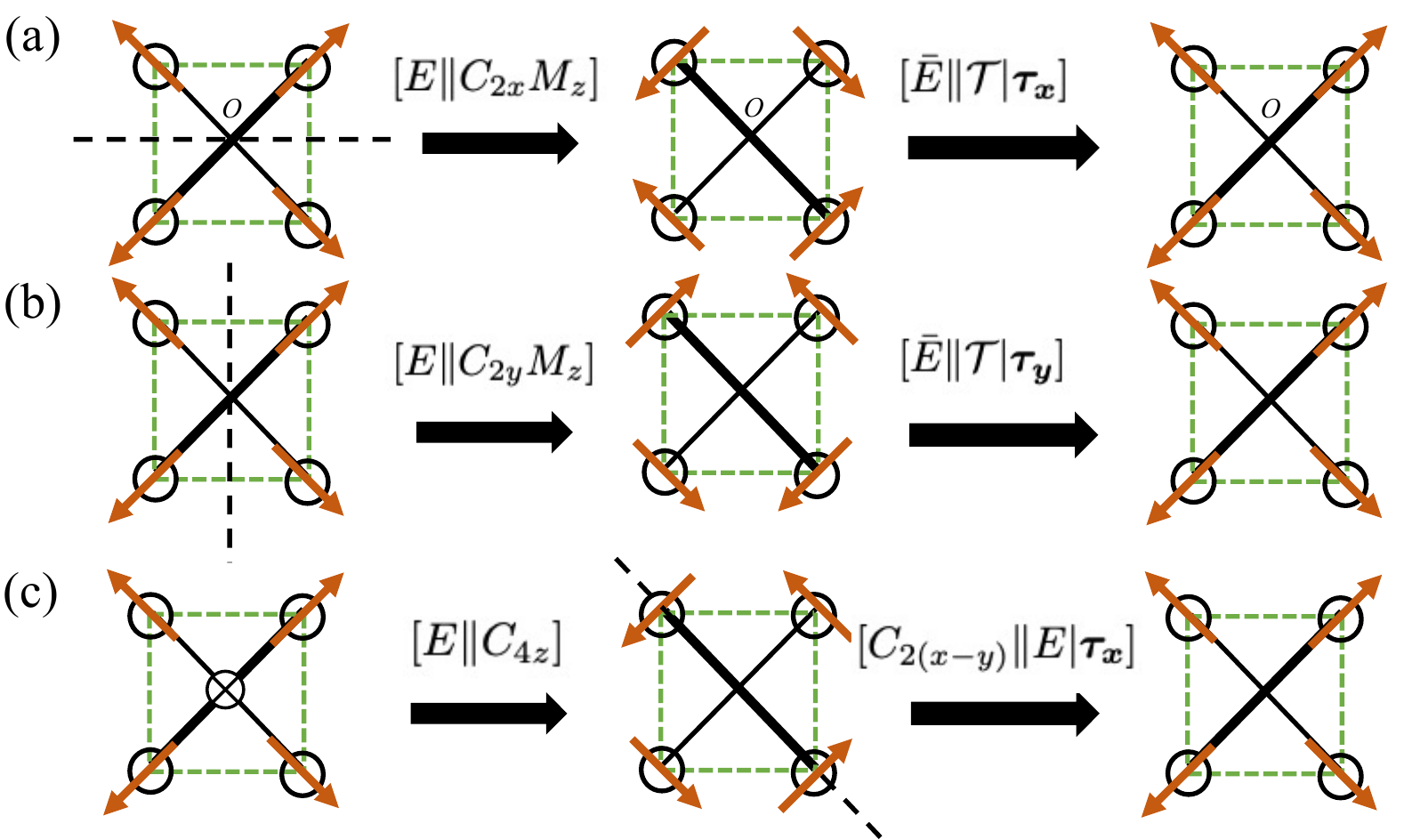}
\caption{Illustration of the symmetry operations  $[\bar{E}\|\mathcal{T}C_{2x}\mathcal{M}_{z}|\boldsymbol{\tau}_{x}]$, 
 $[\bar{E}\|\mathcal{T}C_{2y}\mathcal{M}_{z}|\boldsymbol{\tau}_{y}]$ and $[C_{2(x-y)}\|C_{4z}|\boldsymbol{\tau}_{x}]$. 
 The symbol $O$ in the top-row figures labels the unit cell at the origin (implicit in the 
 middle and bottom rows). In (a)-(c), the dashed lines indicate the $C_{2x}$-, $C_{2y}$- and $C_{2(x-y)}$-rotation axes, 
 respectively. In the left panel of (c), the symbol $\bigotimes$ indicates that the $C_{4z}$-rotation axis is perpendicular
 to the plane.  
}\label{figs1}
\end{figure}

\section{II. Detailed symmetry analysis of the Hamiltonian before and after driving}

We start from the pre-driven Hamiltonian:
\begin{eqnarray}
\mathcal{H}(\bm{k})=&2t\cos{k_x}\tau_0\sigma_xs_0+2t\cos{k_y}\tau_x\sigma_0s_0+4t_{s}\cos{k_x}\cos{k_y}\tau_x\sigma_xs_0\nonumber\\
&+4t_{a}\cos{k_x}\cos{k_y}\tau_y\sigma_ys_0-M\tau_0\sigma_zs_x+M\tau_z\sigma_0s_y.
\end{eqnarray}
As discussed in the main text, this 
Hamiltonian possesses a series of symmetries, as summarized in Table.~\ref{table of symmetries}, including 
$[\bar{C}_{2z}\|\mathcal{T}]$, $[\bar{E}\|\mathcal{T}C_{2z}]$, $[\bar{E}\|\mathcal{T}|\bm{\tau}_{d}]$, 
$[C_{2z}\|C_{2z}]$, $[C_{2z}\|E|\bm{\tau}_{d}]$, $[\bar{E}\|\mathcal{T}C_{2x}\mathcal{M}_{z}|\boldsymbol{\tau}_{x}]$,
$[\bar{E}\|\mathcal{T}C_{2y}\mathcal{M}_{z}|\boldsymbol{\tau}_{y}]$ and $[C_{2(x-y)}\|C_{4z}|\boldsymbol{\tau}_{x}]$, where $\bm{\tau}_{d}=a(1,-1)$,
$\bm{\tau}_{x}=a(1,0)$, and $\bm{\tau}_{y}=a(0,1)$ ($a$ is the nearest-neighbor lattice 
constant in the top-viewed unit cell and has been set to unity in the Bloch Hamiltonian).
The existence of these symmetries can be verified directly by examining the evolution of magnetic 
configurations and bond patterns in real space. As an illustration, Fig.~\ref{figs1} shows the 
transformations induced by the three symmetry operations $[\bar{E}\|\mathcal{T}C_{2x}\mathcal{M}_{z}|\boldsymbol{\tau}_{x}]$,
$[\bar{E}\|\mathcal{T}C_{2y}\mathcal{M}_{z}|\boldsymbol{\tau}_{y}]$, and $[C_{2(x-y)}\|C_{4z}|\boldsymbol{\tau}_{x}]$, 
which clearly show that the system is invariant after performing these symmetry operations.

\begin{table}[h]
  \centering

  \begin{tabular}{c|c|c}
    \hline
    \multirow{2}{*}{Symmetry} & Preserved or Broken  & \multirow{2}{*}{Constraint on spin polarization}\\
    &after applying CPL& \\
    \hline
    $[\bar{C}_{2z}\|\mathcal{T}]$& Broken & $\langle s_{z}(\bm{k})\rangle=-\langle s_{z}(\bm{-k})\rangle,\quad\langle s_{x,y}(\bm{k})\rangle=\langle s_{x,y}(\bm{-k})\rangle$\\
    $[\bar{E}\|\mathcal{T}C_{2z}]$&Broken&  $\langle s_{x,y,z}(\bm{k})\rangle=-\langle s_{x,y,z}(\bm{k})\rangle=0$\\
    $[\bar{E}\|\mathcal{T}|\bm{\tau}_d]$&Broken& $\langle s_{x,y,z}(\bm{k})\rangle=-\langle s_{x,y,z}(\bm{-k})\rangle$\\
    $[C_{2z}\|C_{2z}]$&Preserved&$\langle s_{z}(\bm{k})\rangle=\langle s_{z}(\bm{-k})\rangle,\quad \langle s_{x,y}(\bm{k})\rangle=-\langle s_{x,y}(\bm{-k})\rangle$\\
    $[C_{2z}\|E|\bm{\tau}_d]$&Preserved& $\quad\langle s_{x,y}(\bm{k})\rangle=-\langle s_{x,y}(\bm{k})\rangle=0$\\
    $[\bar{E}\|\mathcal{T}C_{2x}\mathcal{M}_z|\bm{\tau}_x]$&Preserved & $\langle s_{ x,y,z}(k_x,k_y)\rangle=-\langle s_{x,y,z}(-k_x,k_y)\rangle$(nodal line at $k_x=0$)\\
    $[\bar{E}\|\mathcal{T}C_{2y}\mathcal{M}_z|\bm{\tau}_y]$&Preserved & $\langle s_{x,y,z}(k_x,k_y)\rangle=-\langle s_{x,y,z}(k_x,-k_y)\rangle$(nodal line at $k_y=0$)\\
    $[C_{2(x-y)}\|C_{4z}|\bm{\tau}_x]$&Preserved& $\langle s_{z}(k_x,k_y)\rangle=-\langle s_{z}(k_y,-k_x)\rangle$(d-wave spin polarization)\\
    \hline
  \end{tabular}

  \caption{Table of key symmetries. Spin group notation $[g_{s}\|g_{r}]$ is used, where $g_{s}$ acts solely in spin space and $g_{r}$ solely in real space. The operators in the bracket have the following meanings: $C_{2a}$ denotes a $180^\circ$ rotation about the $a$ axis, $C_{4z}$ denotes a $90^\circ$ rotation about the $z$ axis, $\mathcal{T}$ denotes the time-reversal operator, $E$ is the identity operator, and $M_z$ is the mirror reflection about the midplane of the bilayer system. An overbar (e.g., $\bar{E}$) signifies the additional action of time reversal, which reverses spin. The fractional translation vectors are $\bm{\tau}_d=a(1,-1)$, $\bm{\tau}_x=a(1,0)$ and $\bm{\tau}_y=a(0,1)$.} \label{table of symmetries}
\end{table}

These symmetries can also be verified at the level of the Bloch Hamiltonian. 
This requires writing down their explicit representations in the Pauli matrix basis and examining their action on the Hamiltonian.
Specifically, we have $[\bar{C}_{2z}\|\mathcal{T}]=s_{x}\mathcal{K}$, $[\bar{E}\|\mathcal{T}C_{2z}]=\tau_{x}\sigma_{x}s_{y}\mathcal{K}$, 
$[\bar{E}\|\mathcal{T}|\bm{\tau}_{d}]=\tau_{x}\sigma_{x}s_{y}\mathcal{K}$, $[C_{2z}\|C_{2z}]=\tau_{x}\sigma_{x}s_{z}$, 
and $[\bar{E}\|\mathcal{T}|\bm{\tau}_{d}]=\tau_{x}\sigma_{x}s_{z}$, $[\bar{E}\|\mathcal{T}C_{2x}\mathcal{M}_{z}|\boldsymbol{\tau}_{x}]
=\tau_{x}\sigma_{x}s_{y}\mathcal{K}$, 
$[\bar{E}\|\mathcal{T}C_{2y}\mathcal{M}_{z}|\boldsymbol{\tau}_{y}]
=\tau_{x}\sigma_{x}s_{y}\mathcal{K}$, 
and $[C_{2(x-y)}\|C_{4z}|\boldsymbol{\tau}_{x}]=\frac{i}{2\sqrt{2}}(\tau_0\sigma_0+\tau_z\sigma_z+\tau_x\sigma_x+\tau_y\sigma_y)
(s_x-s_y)$, where $\mathcal{K}$ is the complex conjugate operator. It is noteworthy that some symmetry operators share identical matrix representations. 
This occurs because distinct symmetry operations can produce the same effect on the combined spin 
and sublattice degrees of freedom. For example, the operators $[\bar{E}\|\mathcal{T}C_{2z}]$ and $[\bar{E}\|\mathcal{T}|\bm{\tau}_{d}]$
have the same matrix form. The reason is that their core spatial operations---$C_{2z}$ and $[E|\bm{\tau}_{d}]$---both induce 
the identical sublattice exchange: $A\leftrightarrow D$ and  $B\leftrightarrow C$. Despite this equivalence in matrix 
representation, their actions on the Bloch Hamiltonian differ because they act differently on the crystal momentum. 

It is readily verified that these operators and the Hamiltonian satisfy the following relations: 
\begin{align}
[\bar{C}_{2z}\|\mathcal{T}]\mathcal{H}(\bm{k})[\bar{C}_{2z}\|\mathcal{T}]^{-1}&=\mathcal{H}(-\bm{k}),\nonumber\\
[\bar{E}\|\mathcal{T}C_{2z}]\mathcal{H}(\bm{k})[\bar{E}\|\mathcal{T}C_{2z}]^{-1}&=\mathcal{H}(\bm{k}),\nonumber\\
[\bar{E}\|\mathcal{T}|\bm{\tau}_{d}]\mathcal{H}(\bm{k})[\bar{E}\|\mathcal{T}|\bm{\tau}_{d}]^{-1}&=\mathcal{H}(-\bm{k}),\nonumber\\
[C_{2z}\|C_{2z}]\mathcal{H}(\bm{k})[C_{2z}\|C_{2z}]^{-1}&=\mathcal{H}(-\bm{k}),\nonumber\\
[C_{2z}\|E|\bm{\tau}_{d}]\mathcal{H}(\bm{k})[C_{2z}\|E|\bm{\tau}_{d}]^{-1}&=\mathcal{H}(\bm{k}),\nonumber\\
[\bar{E}\|\mathcal{T}C_{2x}\mathcal{M}_{z}|\bm{\tau}_{x}]\mathcal{H}(k_{x},k_{y})[\bar{E}\|\mathcal{T}C_{2x}\mathcal{M}_{z}|\bm{\tau}_{x}]^{-1}
&=\mathcal{H}(-k_{x},k_{y}),\nonumber\\
[\bar{E}\|\mathcal{T}C_{2y}\mathcal{M}_{z}|\bm{\tau}_{y}]\mathcal{H}(k_{x},k_{y})[\bar{E}\|\mathcal{T}C_{2y}\mathcal{M}_{z}|\bm{\tau}_{y}]^{-1}
&=\mathcal{H}(k_{x},-k_{y}),\nonumber\\
[C_{2(x-y)}\|C_{4z}|\boldsymbol{\tau}_{x}]\mathcal{H}(k_{x},k_{y})[C_{2(x-y)}\|C_{4z}|\boldsymbol{\tau}_{x}]^{-1}&=\mathcal{H}(k_{y},-k_{x}),
\end{align}
confirming the existence of these symmetries. 

The Floquet Hamiltonian is given by 
\begin{align}
\mathcal{H}_{\rm eff}(\bm{k})=&2J_0(A_0)(t\cos{k_x}\tau_0\sigma_xs_0+t\cos{k_y\tau_x\sigma_0s_0}),\nonumber\\
&+4J_0(\sqrt{2}A_0)(t_{s}\cos{k_x}\cos{k_y}\tau_x\sigma_xs_0+t_{a}\cos{k_x}\cos{k_y}\tau_y\sigma_ys_0)\nonumber\\
&-F(A_0,\omega)\sin{k_x}\sin{k_y}\cos{k_x}\tau_y\sigma_zs_0\nonumber\\
&+F(A_0,\omega)\sin{k_x}\sin{k_y}\cos{k_y}\tau_z\sigma_ys_0\nonumber\\
&-M\tau_0\sigma_zs_x+M\tau_z\sigma_0s_y,
\end{align}
where $F(A_0,\omega)=16\sqrt{2}tt_{a}J_1(A_0)J_1(\sqrt{2}A_0)/\omega$. Because of the emergence of the two 
light-induced terms, it is easy to check  
$[\bar{C}_{2z}\|\mathcal{T}]\mathcal{H}_{\rm eff}(\bm{k})[\bar{C}_{2z}\|\mathcal{T}]^{-1}\neq \mathcal{H}_{\rm eff}(-\bm{k})$,
$[\bar{E}\|\mathcal{T}C_{2z}]\mathcal{H}_{\rm eff}(\bm{k})[\bar{E}\|\mathcal{T}C_{2z}]^{-1}\neq\mathcal{H}_{\rm eff}(\bm{k})$
and $[\bar{E}\|\mathcal{T}|\bm{\tau}_{d}]\mathcal{H}_{\rm eff}(\bm{k})[\bar{E}\|\mathcal{T}|\bm{\tau}_{d}]^{-1}\neq\mathcal{H}_{\rm eff}(-\bm{k})$,
indicating the breaking of these three symmetries. Similarly, one can check that all other symmetries are preserved.

\section{III.  Analytical method for determining the spin-splitting texture}

To determine the spin-splitting texture in the momentum space, we have introduced the following quantity\cite{Hayami2020AM1,Hayami2020AM2},
\begin{align}
{\rm Tr}[e^{-\beta \mathcal{H}_{\rm eff}(\bm{k})}s_\mu]=\sum_s\frac{(-\beta)^s}{s!}g^\mu_s(\bm{k}),
\end{align}
where $\mu=0,x,y,z$, and $\beta$ is the inverse temperature. 

In our coplanar AFM model, we focus on the out-of-plane spin polarization component. The related coefficient $g^z_s(\bm{k})$ is given by
\begin{align}
g^z_s(\bm{k})={\rm Tr}[(\mathcal{H}_{\rm eff}(\bm{k}))^ss_z]=\frac{1}{2^{s-1}}{\rm Tr}[\{\mathcal{H}_{\rm eff}(\bm{k}),\{\mathcal{H}_{\rm eff}(\bm{k}),\dots\}\}s_z],\label{gz}
\end{align}
where $\mathcal{H}_{\rm eff}(\bm{k})$ repeats the pattern $s$ times and $\{\cdot,\cdot\}$ denotes the anticommutator defined by $\{A,B\}=AB+BA$.
Since the effective Hamiltonian commutes with itself, the product can be written as 
\begin{align}
\mathcal{H}_{\rm eff}(\bm{k})^2=\frac{1}{2}\{\mathcal{H}_{\rm eff}(\bm{k}),\mathcal{H}_{\rm eff}(\bm{k})\},
\end{align}
which introduces a factor $\frac{1}{2}$. With this relation, the product of $s$ Hamiltonians can be expressed in terms of $s-1$ anticommutator, yielding an overall factor of $\frac{1}{2^{s-1}}$. 

To have a nonzero $g^z_s(\bm{k})$, 
it is obvious that equal numbers of $s_x$ and $s_y$ terms in $\mathcal{H}_{\rm eff}(\bm{k})$ should be selected, indicating $s\ge 2$ is a necessary 
condition. 

\begin{figure}[t]
\centering
\includegraphics[width=0.65\textwidth]{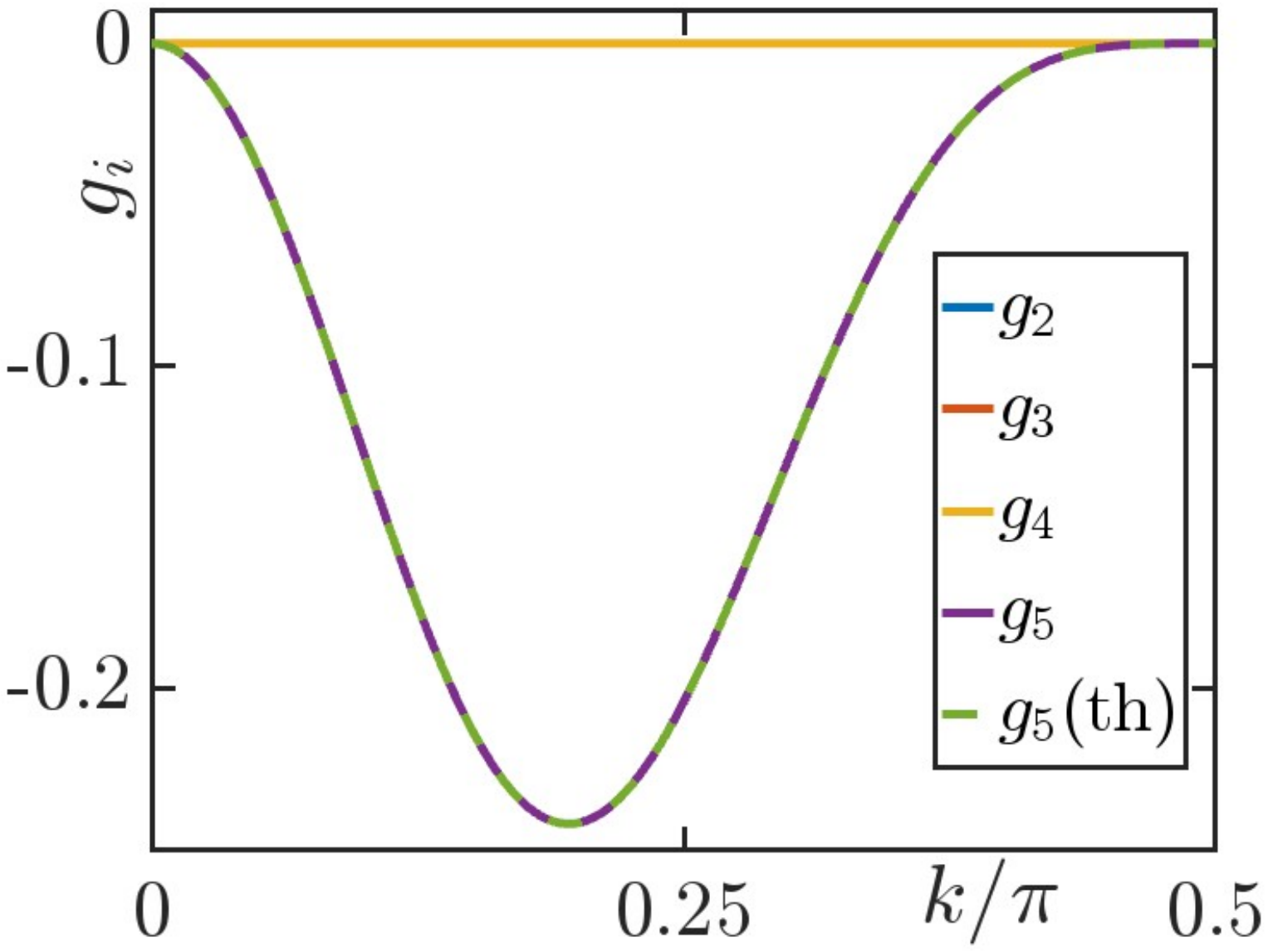}
\caption{Coefficients $g_i(\bm{k})$ along the line of $k_x=k_y=k$. Here, $g_5$(th) denotes the analytical result from Eq. (\ref{g5}), 
while $g_i$ with $1\leq i\leq5$ are determined by numerically calculating ${\rm Tr}[(\mathcal{H}_{\rm eff}(\bm{k}))^ss_z]$. 
The coefficients $g_{1}$, $g_{2}$, $g_{3}$, and $g_{4}$  are all zeros. Parameters: $t=0.4$, $t_s=0.7$, $t_a=0.3$, $M=0.5$, $A_0=0.6$, and $\omega=10$.
}\label{figs2}
\end{figure}

We find that the lowest-order contribution comes from the fifth order. The coefficient $g^z_5(\bm{k})$
can be easily determined by using Mathematica, with the result given by 
\begin{align}
g^z_5(\bm{k})=-64J_0(A_0)J_0(\sqrt{2}A_0)tt_sF(A_0,\omega)M^2\sin{2k_x}\sin{2k_y}(2+\cos{2k_x}+\cos{2k_y}).\label{g5}
\end{align}
Its momentum dependence can also be analytically derived by 
using the fact that  $s_x$ and $s_y$ terms must appear in equal times. Concretely, 
\begin{align}
g^z_5(\bm{k})=&{\rm Tr}[(\mathcal{H}_{\rm eff}(\bm{k}))^5s_z]\nonumber\\
=&\frac{1}{2^{4}}{\rm Tr}[\{\mathcal{H}_{\rm eff}(\bm{k}),\{\{\mathcal{H}_{\rm eff}(\bm{k}),\mathcal{H}_{\rm eff}(\bm{k})\},\{\mathcal{H}_{\rm eff}(\bm{k}),\mathcal{H}_{\rm eff}(\bm{k})\}\}\}s_z]\nonumber\\
\propto&{\rm Tr}[\{\mathcal{H}_{\rm eff}(\bm{k}),\{\{\mathcal{H}_{\rm eff}(\bm{k}),-M\tau_0\sigma_zs_x\},\{\mathcal{H}_{\rm eff}(\bm{k}),M\tau_z\sigma_0s_y\}\}\}s_z].
\end{align}
By applying the commutation and anticommutation relation of the terms in the Hamiltonian, the nonzero terms are given by
\begin{align}
&{\rm Tr}[\{\mathcal{H}_{\rm eff}(\bm{k}),\{\{-F(A_0,\omega)\sin{k_x}\sin{k_y}\cos{k_x}\tau_y\sigma_zs_0,-M\tau_0\sigma_zs_x\},
\{2J_0(A_0)t\cos{k_x}\tau_0\sigma_xs_0,M\tau_z\sigma_0s_y\}\}\}s_z]\nonumber\\
&+{\rm Tr}[\{\mathcal{H}_{\rm eff}(\bm{k}),\{\{2J_0(A_0)t\cos{k_y}\tau_x\sigma_0s_0,-M\tau_0\sigma_zs_x\},
\{F(A_0,\omega)\sin{k_x}\sin{k_y}\cos{k_y}\tau_z\sigma_ys_0,M\tau_z\sigma_0s_y\}\}\}s_z]\nonumber\\
=&4{\rm Tr}[\{\mathcal{H}_{\rm eff}(\bm{k}),\{F(A_0,\omega)M\sin{k_x}\sin{k_y}\cos{k_x}\tau_y\sigma_0s_x,2J_0(A_0)tM\cos{k_x}\tau_z\sigma_xs_y\}\}s_z]\nonumber\\
&+4{\rm Tr}[\{\mathcal{H}_{\rm eff}(\bm{k}),\{-2J_0(A_0)tM\cos{k_y}\tau_x\sigma_zs_x,F(A_0,\omega)M\sin{k_x}\sin{k_y}\cos{k_y}\tau_0\sigma_ys_y\}\}s_z]\nonumber\\
=&-8{\rm Tr}[\{\mathcal{H}_{\rm eff}(\bm{k}),2J_0(A_0)tF(A_0,\omega)M^2\sin{k_x}\sin{k_y}\cos^2{k_x}\tau_x\sigma_xs_z\}\}s_z]\nonumber\\
&-8{\rm Tr}[\{\mathcal{H}_{\rm eff}(\bm{k}),2J_0(A_0)tF(A_0,\omega)M^2\sin{k_x}\sin{k_y}\cos^2{k_y}\tau_x\sigma_xs_z\}\}s_z]\nonumber\\
=&-8{\rm Tr}[\{4J_0(\sqrt{2}A_0)t_s\cos{k_x}\cos{k_y}\tau_x\sigma_xs_0,2J_0(A_0)tF(A_0,\omega)M^2\sin{k_x}\sin{k_y}\cos^2{k_x}\tau_x\sigma_xs_z\}s_z]\nonumber\\
&-8{\rm Tr}[\{4J_0(\sqrt{2}A_0)t_s\cos{k_x}\cos{k_y}\tau_x\sigma_xs_0,2J_0(A_0)tF(A_0,\omega)M^2\sin{k_x}\sin{k_y}\cos^2{k_y}\tau_x\sigma_xs_z\}s_z]\nonumber\\
=&-16\times8\times8J_0(A_0)J_0(\sqrt{2}A_0)tt_sF(A_0,\omega)M^2\sin{k_x}\sin{k_y}\cos{k_x}\cos{k_y}(\cos^2{k_x}+\cos^2{k_y})\nonumber\\
=&-128J_0(A_0)J_0(\sqrt{2}A_0)tt_sF(A_0,\omega)M^2\sin{2k_x}\sin{2k_y}(2+\cos{2k_x}+\cos{2k_y}).
\end{align}
The momentum dependence is perfectly consistent with the result obtained  via Mathematica, but the overall expression differs 
by a factor, which is expected since there are many other arrangements of terms that lead to the same result. 
In Fig.~\ref{figs2}, we further show the consistency between the results obtained by analytical and 
numerical calculations.  

\section{IV. Low-frequency Floquet Hamiltonian and band structure}

In the main text, we derive the effective Floquet Hamiltonian under the high-frequency off-resonant condition, which requires the driving frequency to exceed the bandwidth. When this condition is not satisfied, the high-frequency expansion becomes unreliable. However, the emergence of $d$-wave spin splitting is symmetry-dictated and thus insensitive to the frequency. In fact, since the light-induced term scales inversely with frequency, the spin splitting is enhanced at lower frequencies. In this section, we introduce the full Floquet Hamiltonian, applicable for arbitrary frequency. The Floquet band structure within the first quasienergy Brillouin zone is obtained with a proper truncation of Floquet sectors.

\begin{figure}[h]
\centering
\includegraphics[width=0.9\textwidth]{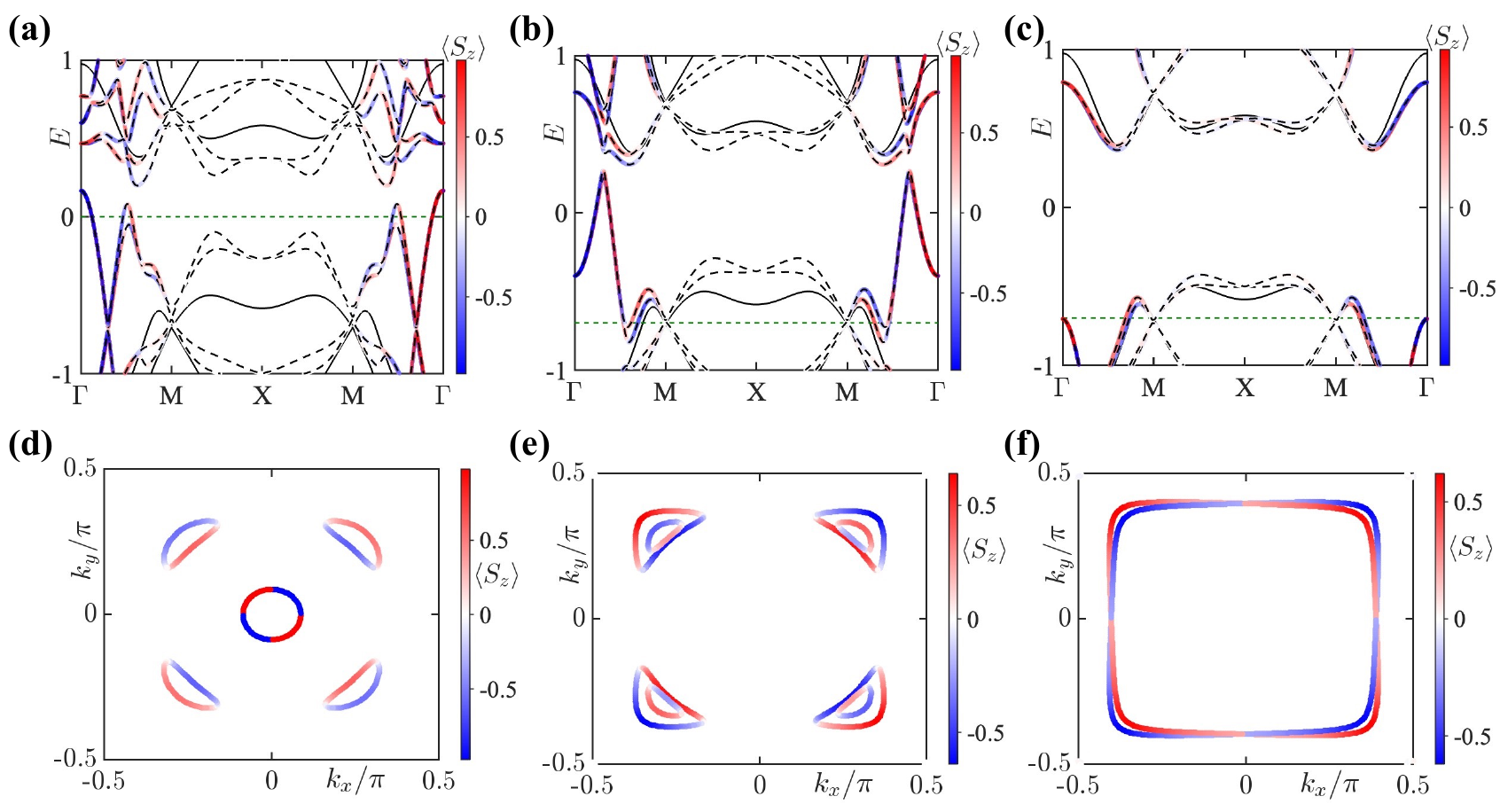}
\caption{(Color online) (a)-(c) Floquet Band structure and spin polarization for $\omega=2,3,5$, respectively. Green dashed lines indicate the equal-energy surfaces used to illustrate the spin-splitting patterns, located at $E=0,-0.7,-0.7$, respectively. 
(d)-(f) Spin polarizations on these equal-energy surfaces, corresponding to (a)–(c), respectively. 
Other parameters are $t=0.4, t_s=0.7, t_a=0.3, A_0=0.6$, and $M_x=M_y=0.5$.}\label{figs3}
\end{figure}

Based on the Fourier-transformed Floquet Hamiltonian and states, the Floquet time-evolution equation becomes\cite{Eckardt2015Highf}
\begin{align}
\sum_n [\mathcal{H}_{m-n}(\bm{k})+\delta_{mn}n\omega]|u_n\rangle= E|u_m\rangle,
\end{align}
where $\mathcal{H}(\bm{k})_{m-n}=\frac{1}{T}\int_0^Tdt e^{-i(m-n)\omega t}\mathcal{H}(\bm{k},t)$. 
In matrix form, this is an eigenvalue problem involving an infinite-dimensional matrix. 
However, because the eigenvalues are periodic modulo $\omega$, one need only determine the band structure 
within the first quasienergy Brillouin zone, which can be accurately obtained with a proper 
truncation of Floquet sectors. Following convention, we choose this zone as $[-\omega/2,\omega/2]$. 
To determine the band structure within the first quasienergy Brillouin zone accurately, we truncate the Floquet Hamiltonian at the $\mathcal{H}_{\pm4}$ components, 
yielding 
\begin{align}
\begin{pmatrix}
\mathcal{H}_0+2\omega&\mathcal{H}_1&\mathcal{H}_2&\mathcal{H}_3&\mathcal{H}_4\\
\mathcal{H}_{-1}&\mathcal{H}_0+\omega&\mathcal{H}_1&\mathcal{H}_2&\mathcal{H}_3\\
\mathcal{H}_{-2}&\mathcal{H}_{-1}&\mathcal{H}_0&\mathcal{H}_1&\mathcal{H}_2\\
\mathcal{H}_{-3}&\mathcal{H}_{-2}&\mathcal{H}_{-1}&\mathcal{H}_0-\omega&\mathcal{H}_1\\
\mathcal{H}_{-4}&\mathcal{H}_{-3}&\mathcal{H}_{-2}&\mathcal{H}_{-1}&\mathcal{H}_0-2\omega
\end{pmatrix}\left(
               \begin{array}{c}
                 |u_{2}\rangle \\
                 |u_{1}\rangle \\
                 |u_{0}\rangle \\
                 |u_{-1}\rangle \\
                 |u_{-2}\rangle \\
               \end{array}
             \right)=E\left(
               \begin{array}{c}
                 |u_{2}\rangle \\
                 |u_{1}\rangle \\
                 |u_{0}\rangle \\
                 |u_{-1}\rangle \\
                 |u_{-2}\rangle \\
               \end{array}
             \right)
.
\end{align}
After diagonalizing this Floquet Hamiltonian, we obtain the band structures and spin polarization textures, as shown in Fig.~\ref{figs3}. 
The results confirm that the $d$-wave spin splitting not only survives but becomes more prominent in the low-frequency regime.

\begin{figure}[h]
\centering
\includegraphics[width=0.65\textwidth]{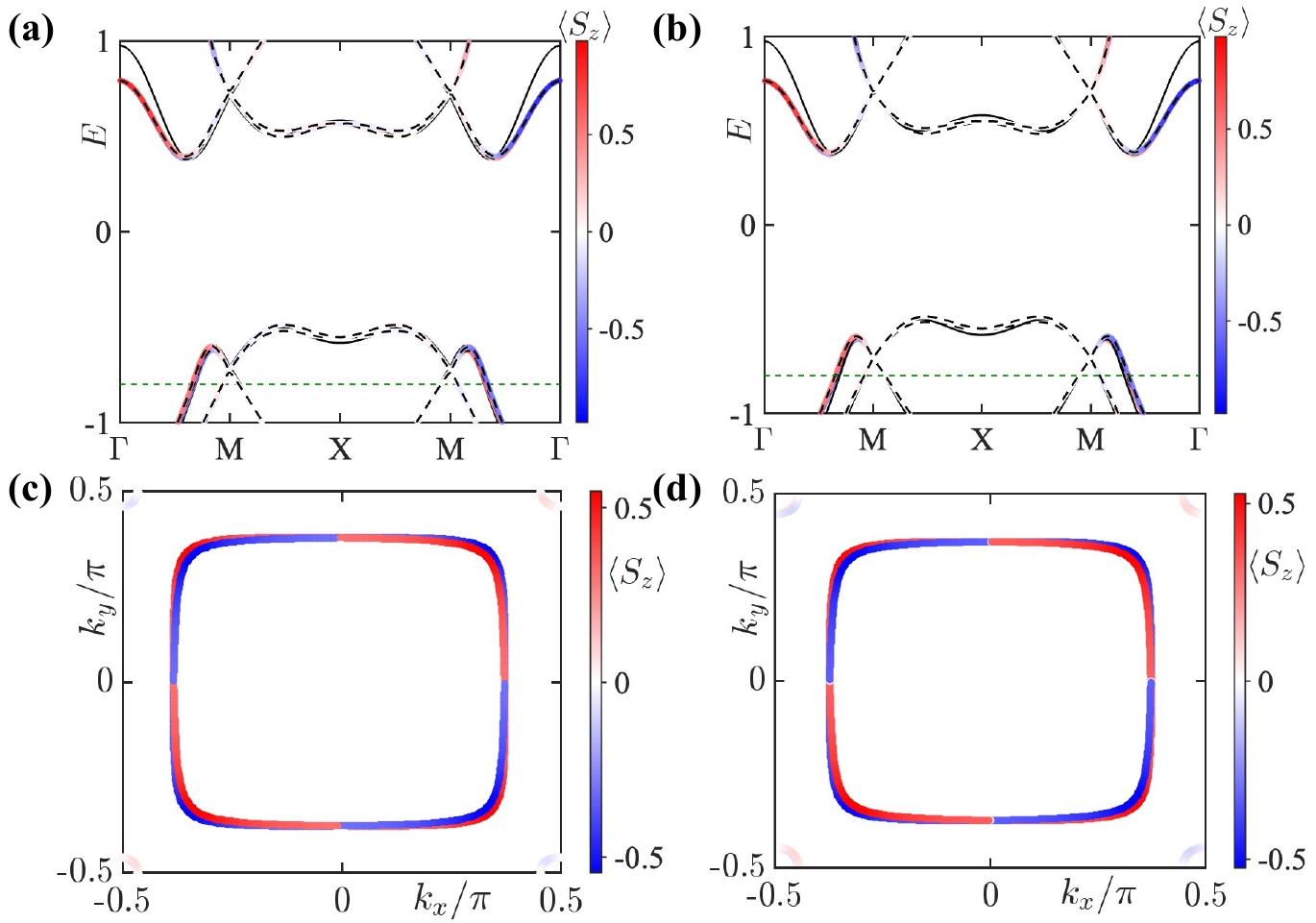}
\caption{(Color online) Floquet Band structures and spin polarizations from (a) the truncated full Floquet Hamiltonian and (b) the high-frequency effective Hamiltonian at $\omega=10$. Green dash lines indicate the equal-energy surfaces ($E=-0.8$)  used to illustrate the spin-splitting patterns. (c) and (d) Spin polarizations on these equal-energy surfaces, corresponding to (a) and (b), respectively. Parameters are $t=0.4,t_s=0.7,t_a=0.3,A_0=0.6$, and $M_x=M_y=0.5$.}\label{figs4}
\end{figure}

By comparing the band structures from the truncated full Floquet Hamiltonian and the high-frequency effective Hamiltonian, we find significant discrepancies at $\omega=5$, as expected for this deeply resonant regime. However, the two approaches converge rapidly with increasing frequency: as shown in Fig.~\ref{figs4}, the difference is already negligible at $\omega=10$. This validates the effective Hamiltonian as a useful tool for revealing the underlying symmetries and mechanisms of the $d$-wave spin-splitting texture.

\section{V. Generalization to other coplanar AFM structures}

In the main text, we construct a specific bilayer coplanar AFM structure as a proof of principle. 
Given the fundamental role of symmetry, our symmetry-based mechanism is apparently not limited to a special structure. 
To show its generality, here we introduce three distinct extensions.

\begin{figure}[h]
\centering
\includegraphics[width=0.65\textwidth]{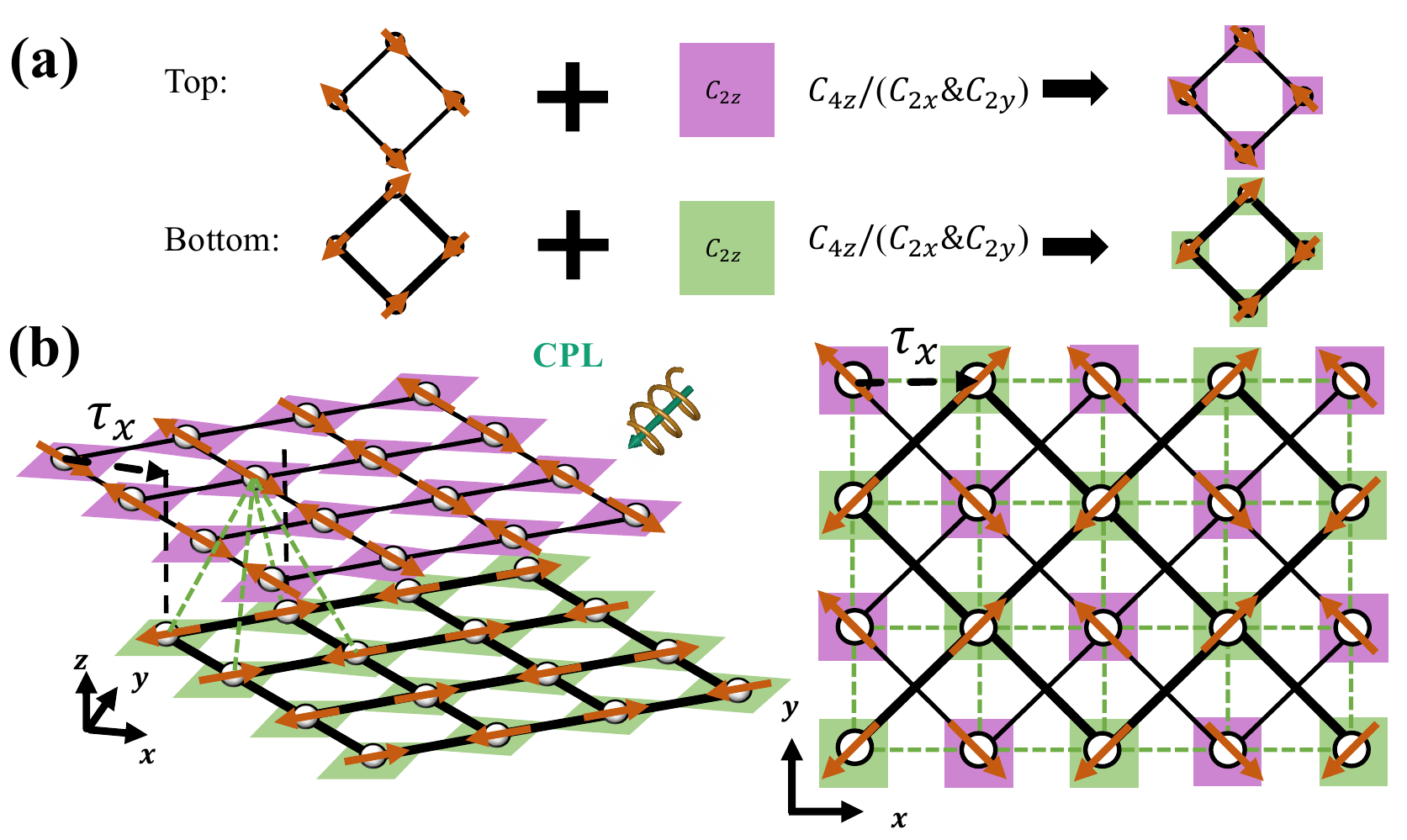}
\caption{(Color online) Schematics illustrating the generalization via additional sublattice configurations. (a) Unit cells of the top and bottom AFM layers. The mechanism can be generalized by introducing sublattice configurations that preserve $C_{2z}$ symmetry, which gives rise to unidirectionally polarized even-parity out-of-plane spin polarization. Furthermore, the sublattice configuration is required to respect either $C_{4z}$ symmetry or both $C_{2x}$ and $C_{2y}$ symmetries to eliminate the $s$-wave component and realize $d$-wave spin splitting. The purple and green squares denote symmetry-compatible sublattice configurations added to the top and bottom layers, respectively. (b) Schematic of the bilayer structure and its top view.}\label{figs5}
\end{figure}

1. \textbf{Generalization by enriching the sublattice configuration.}
Following the symmetry-guided route, we find that our mechanism can be directly generalized to other lattice geometries by enriching the sublattice configuration. Specifically, as shown in Fig.~\ref{figs5}, one approach is to introduce a $C_{2z}$-preserving sublattice configuration into the top/bottom-layer lattice. After this generalization, one can verify that the key symmetries discussed in the main text remain preserved: $[\bar{C}_{2z}\|\mathcal{T}]$, $[\bar{E}\|\mathcal{T}C_{2z}]$, $[\bar{E}\|\mathcal{T}|\bm{\tau}_d]$, $[C_{2z}\|C_{2z}]$, and $[C_{2z}\|E|\bm{\tau}_d]$. These symmetries enforce a spin-degenerate band structure before applying CPL. Under CPL, some symmetries are broken, while the remaining  $[C_{2z}\|C_{2z}]$ and $[C_{2z}\|E|\bm{\tau}_d]$ enforce $\langle s_{x,y}(\bm{k})\rangle=0$ and $\langle s_z(\bm{k})\rangle=\langle s_z(-\bm{k})\rangle$. To guarantee $d$-wave spin splitting, more specific symmetry constraints are required when constructing the sublattice configurations. Below, we provide three explicit generalizations.

\begin{figure}[h]
\centering
\includegraphics[width=0.65\textwidth]{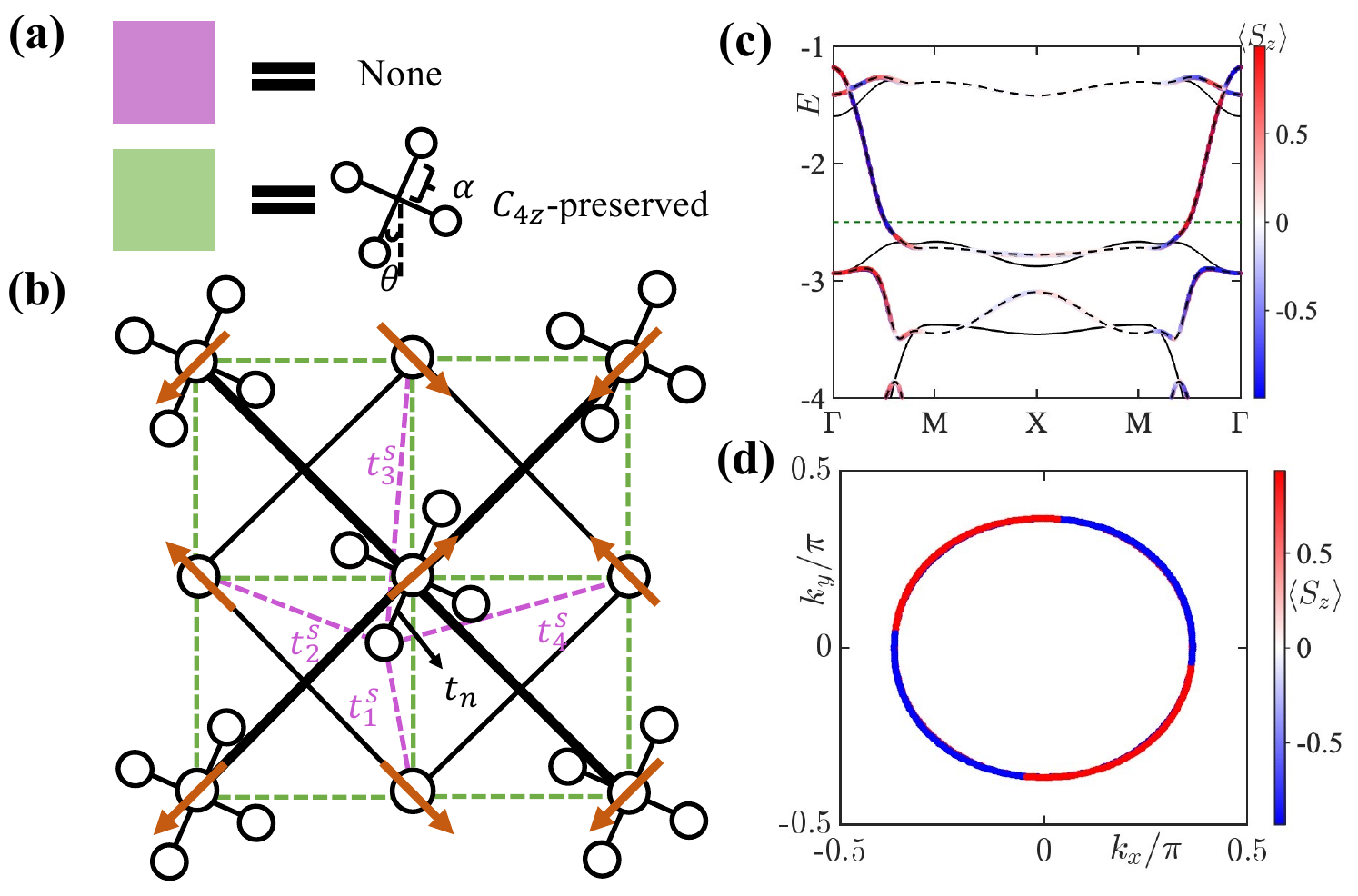}
\caption{(Color online) Schematics illustrating the generalization via additional $C_{4z}$-preserving sublattice structures. (a) No sublattice structure is added to the top layer, whereas a $C_{4z}$-preserving structure, characterized by bond length $\alpha$ and  rotation angle $\theta$,
is added to the bottom layer.  (b) Top view of the bilayer coplanar AFM. Intralayer hoppings between the original and added sublattices are indicated by solid black lines labeled $t_n$; interlayer hoppings are indicated by dashed purple lines labeled
 $t^s_1$, $t^s_2$, $t^s_3$ and $t^s_4$ (assuming $t^s_1>t^s_2>t>t^s_4>t^s_3$). (c) Band structures of the static system (solid black lines, spin-degenerate) and the CPL-driven system (dashed red and blue lines, spin-split). The dashed green line marks the equal-energy surface at $\epsilon=-2.5$. (d) Corresponding spin-splitting texture on the equal-energy surface shown in (c), exhibiting a rotated $d$-wave spin splitting. Other parameters are set to $\{t,t_1,t_2,A_0,\omega,M_x,M_y,\alpha,\theta,t_n,t^s_1,t^s_2,t^s_3,t^s_4\}=\{0.4,0.4,1,0.6,8,0.5,0.5,0.3,15^\circ,1.5,0.6,0.5,0.2,0.3\}$.}\label{figs6}
\end{figure}

(i) $C_{4z}$-preserving sublattice configuration. When the sublattice configuration additionally respects $C_{4z}$ symmetry, the coplanar AFM model respects $[C_{2(x-y)}\|C_{4z}|\bm{\tau}_x]$, which constrains the spin polarization as $\langle s_z(k_x,k_y)\rangle=-\langle s_z(k_y,-k_x)\rangle$. As shown in Fig.~\ref{figs6}, when a $C_{4z}$-preserving sublattice configuration is introduced into the bottom layer, the spin splitting on the equal-energy surface also exhibits the $d$-wave symmetry. Intriguingly, since this sublattice configuration breaks $C_{2x}$ and $C_{2y}$, the key symmetries $[\bar{E}\|\mathcal{T}C_{2x}M_z|\bm{\tau}_x]$ and $[\bar{E}\|\mathcal{T}C_{2y}M_z|\bm{\tau}_y]$ of the original bilayer model are broken. Consequently, the nodal lines---where spin degeneracy occurs---are no longer located at $k_x=0$ or $k_y=0$,  producing a rotated $d$-wave spin-splitting texture.

\begin{figure}[h]
\centering
\includegraphics[width=0.65\textwidth]{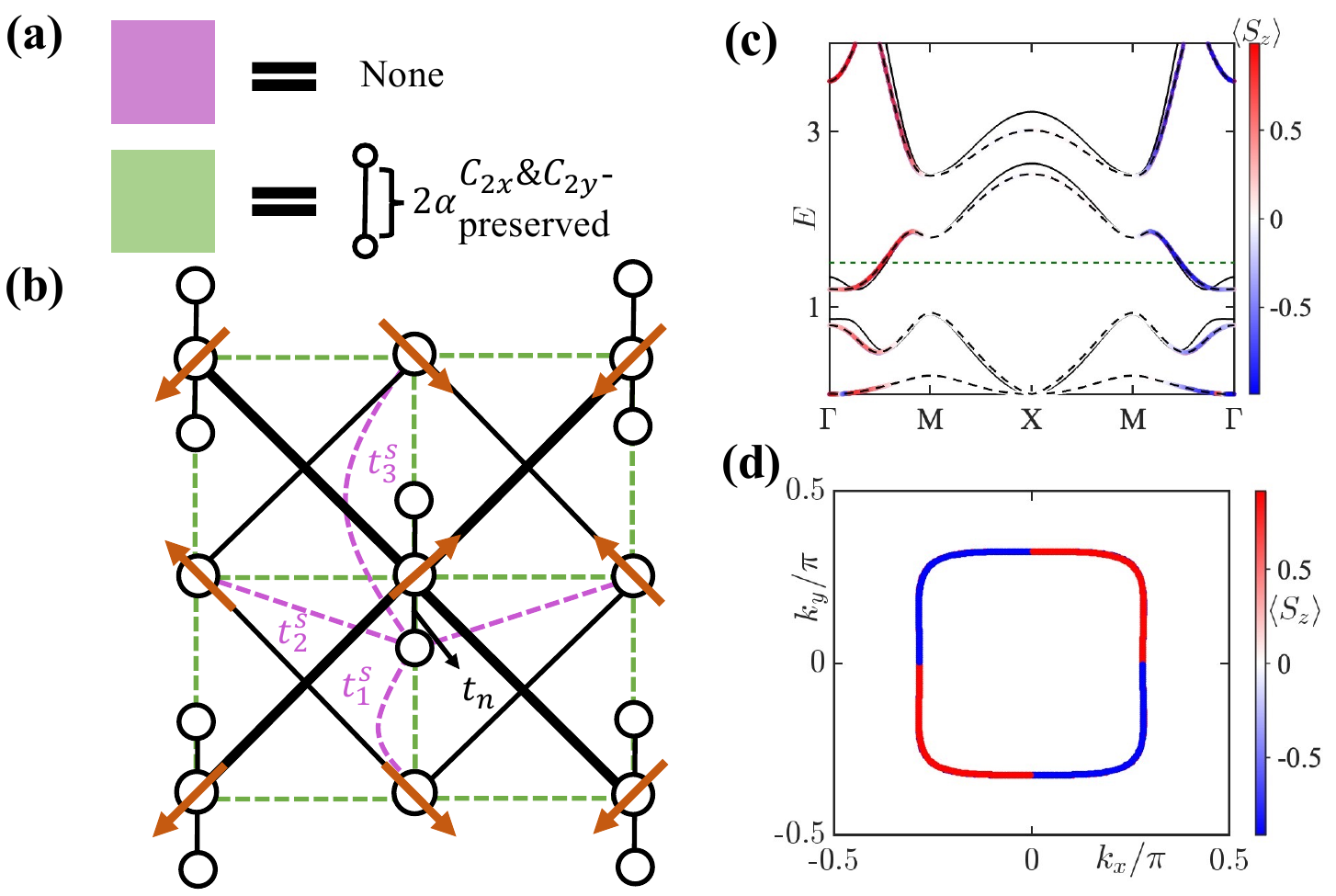}
\caption{(Color online) Schematics illustrating the generalization via additional $C_{2x}$- and $C_{2y}$-preserving sublattice structures. (a) No sublattice structure is added to the top layer, whereas a $C_{2x}$- and $C_{2y}$-preserving structure is added to the bottom layer. The bond 
length between the original sublattice and the added sublattices in the bottom layer is  $\alpha$. (b) Top view of the bilayer coplanar AFM. Intralayer hoppings between the original and added sublattices are indicated by solid black lines labeled $t_n$; 
 interlayer hoppings are indicated by dashed purple lines labeled $t^s_1$, $t^s_2$ and $t^s_3$ (assuming $t^s_1>t>t^s_2>t^s_3$). (c) Band structures of the static system (solid black lines, spin-degenerate) and the CPL-driven system (dashed red and blue lines, spin-split). The dashed green line marks the equal-energy surface at $\epsilon=1.5$. (d) Corresponding spin-splitting texture on the equal-energy surface shown in (c), indicating a $d$-wave spin splitting with two nodal lines at $k_x=0$ and $k_y=0$. Other parameters are set to $\{t,t_1,t_2,A_0,\omega,M_x,M_y,\alpha,t_n,t^s_1,t^s_2,t^s_3\}=\{0.4,0.4,1,0.6,8,0.5,0.5,0.2,1.5,0.6,0.35,0.3\}$.}\label{figs7}
\end{figure}

(ii) $C_{2x}$- and $C_{2y}$-preserving sublattice configuration. When the sublattice configuration additionally respects $C_{2x}$ and $C_{2y}$ symmetries, the coplanar AFM model respects symmetries $[\bar{E}\|\mathcal{T}C_{2x}M_z|\bm{\tau}_x]$ and $[\bar{E}\|\mathcal{T}C_{2y}M_z|\bm{\tau}_y]$, which constrain the spin polarization as $\langle s_z(k_x,k_y)\rangle=-\langle s_z(-k_x,k_y)\rangle$ and $\langle s_z(k_x,k_y)\rangle=-\langle s_z(k_x,-k_y)\rangle$, thereby forming two nodal lines at $k_x=0$ and $k_y=0$. As shown in Fig.~\ref{figs7}, when a $C_{2x}$- and $C_{2y}$-preserving sublattice configuration is introduced into the bottom layer,  the spin splitting 
on the equal-energy surface also exhibits $d$-wave characteristics, protected by the two effective mirror symmetries. 
Since the sublattice structure breaks $C_{4z}$ symmetry, the spin-splitting texture is more likely to be rectangular rather than square, owing to the breaking of $[C_{2(x-y)}\|C_{4z}|\bm{\tau}_x]$.

\begin{figure}[h]
\centering
\includegraphics[width=0.65\textwidth]{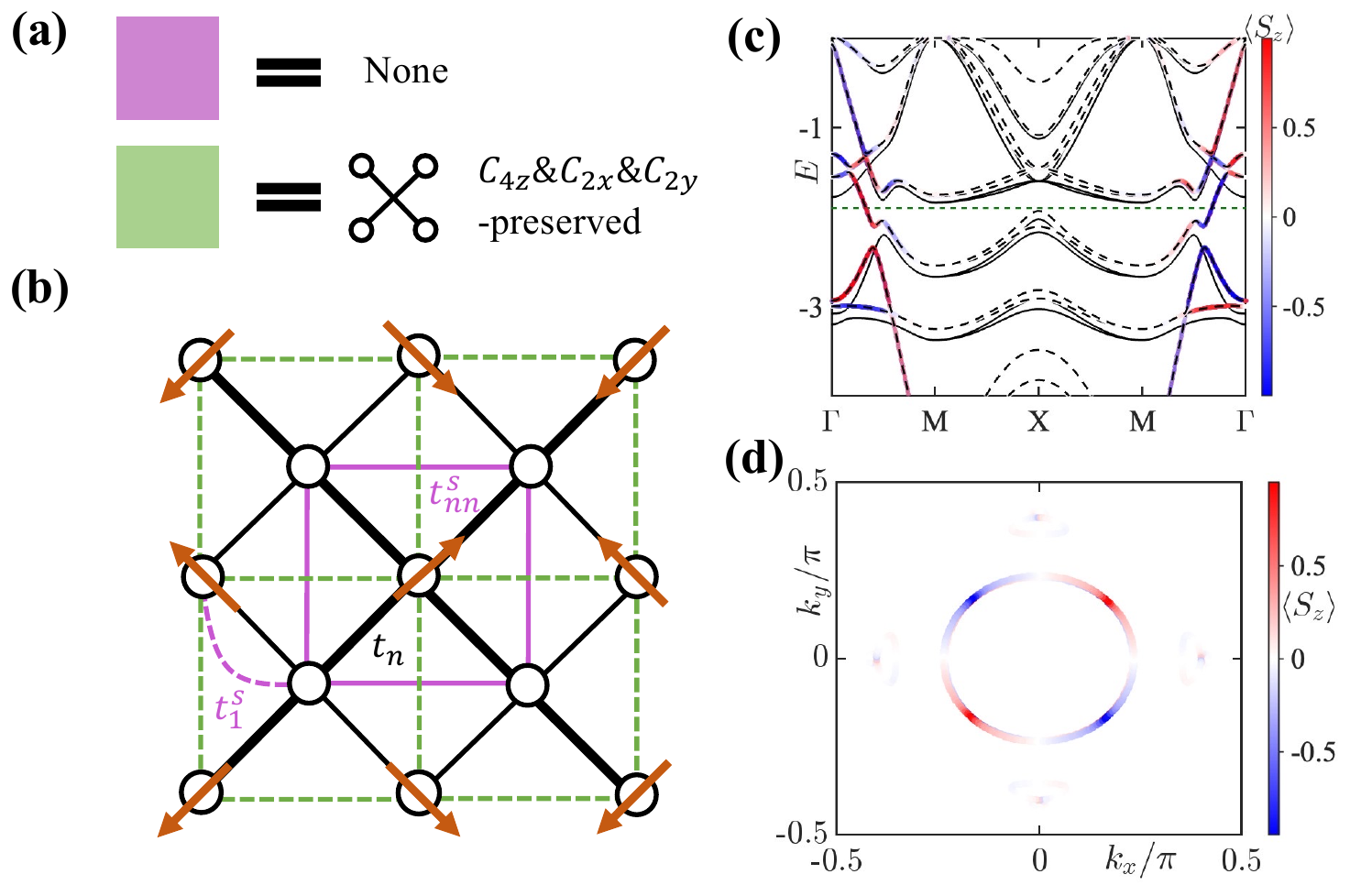}
\caption{(Color online) Schematics illustrating the generalization via 
additional $C_{4z}$-, $C_{2x}$- and $C_{2y}$-preserving structures. (a) No sublattice structure is added to the top layer, whereas a $C_{4z}$-, $C_{2x}$- and $C_{2y}$-preserving structure is added to the bottom layer, thereby forming a 
Lieb-like  lattice on the bottom layer. (b) Top view of the bilayer coplanar AFM. Nearest-neighbor (next-nearest-neighbor) intralayer hoppings 
involving the added sublattices are indicated by solid black (purple) lines labeled $t_n$ ($t^s_{nn}$); Nearest-neighbor interlayer hoppings are indicated by dashed purple lines labeled $t^s_1$. (c) Band structures of the static system (solid black lines, spin-degenerate) and the CPL-driven system (dashed red and blue lines, spin-split). The dashed green line marks the equal-energy surface at $\epsilon=-1.9$. (d) Corresponding spin-splitting texture on the equal-energy surface shown in (c). Other parameters are set to $\{t,t_1,A_0,\omega,M_x,M_y,t_n,t^s_1,t^s_{nn}\}=\{0.4,0.4,0.6,8,0.5,0.5,1.5,0.6,0.8\}$.}\label{figs8}
\end{figure}

(iii) $C_{4z}$-, $C_{2x}$- and $C_{2y}$-preserving sublattice configuration. When the sublattice configuration additionally respects $C_{4z}$, $C_{2x}$ and $C_{2y}$ symmetries, the coplanar AFM model respects symmetries $[C_{2(x-y)}\|C_{4z}|\bm{\tau}_x]$, $[\bar{E}\|\mathcal{T}C_{2x}M_z|\bm{\tau}_x]$ and $[\bar{E}\|\mathcal{T}C_{2y}M_z|\bm{\tau}_y]$, which constrains the spin polarization as $\langle s_z(k_x,k_y)\rangle=-\langle s_z(k_y,-k_x)\rangle$, $\langle s_z(k_x,k_y)\rangle=-\langle s_z(-k_x,k_y)\rangle$ and $\langle s_z(k_x,k_y)\rangle=-\langle s_z(k_x,-k_y)\rangle$. As shown in Fig.~\ref{figs8}, when a sublattice configuration respecting all of these symmetries is incorporated into the bottom layer, the resulting spin splitting on the constant-energy surfaces exhibits a $d_{xy}$-wave texture.

2. \textbf{Generalization by canting the magnetic order}. In our manuscript, we consider a 
coplanar AFM with an all-out magnetic order. However, it is the symmetries---not the 
specific magnetic order---that play the essential role. The emergence of $d$-wave spin splitting 
does not rely on a particular magnetic configuration. In the main text, we discussed its robustness 
against canting; here we provide an expanded analysis. As shown below, distinct types of canting can yield the 
same or different spin-splitting textures, depending on how the key symmetries are modified.

\begin{figure}[h]
\centering
\includegraphics[width=0.7\textwidth]{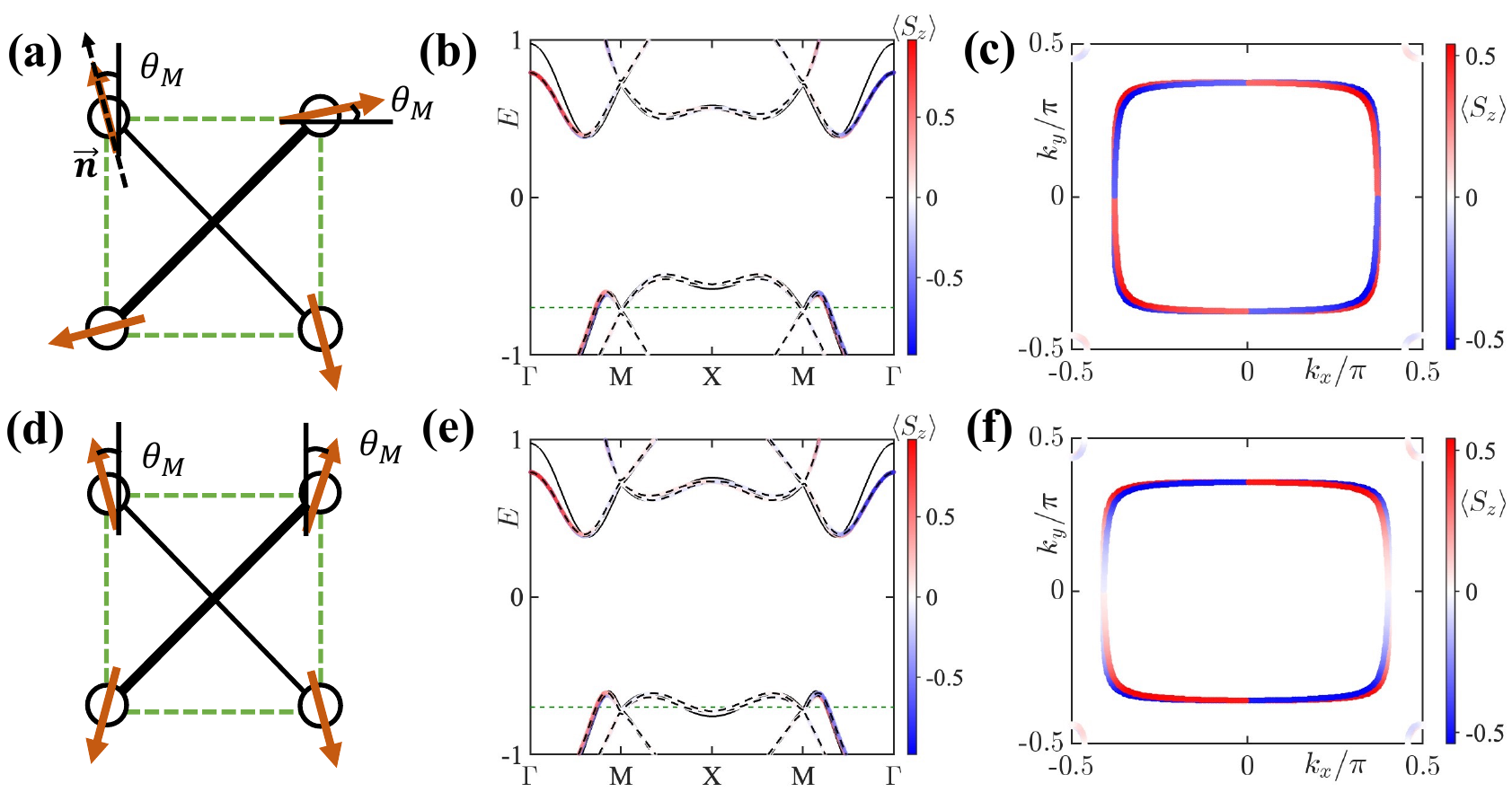}
\caption{(Color online) Schematics illustrating the generalization with different magnetic-order configuration. (a) $[C_{2\vec{\bm{n}}}\|C_{4z}|\bm{\tau}_x]$-preserving magnetic-order configuration. The symmetry operator $C_{2\vec{\bm{n}}}$ represents a twofold rotation along vector $\vec{\bm{n}}$ (dashed arrow) in spin space; $\theta_{M}$ describes the canting angle. (b) Band structures of the static system (solid black lines, spin-degenerate) and the CPL-driven system (dashed red and blue lines, spin-split). The dashed green line marks the equal-energy surface at $\epsilon=-0.7$. (c) Corresponding spin-splitting texture on the equal-energy surface shown in (b). (d) $[\bar{E}\|\mathcal{T}C_{2x}M_z|\bm{\tau}_x]$- and $[\bar{E}\|\mathcal{T}C_{2y}M_z|\bm{\tau}_y]$-preserving magnetic-order configuration.  (e) Band structures of the static system (solid black lines, spin-degenerate) and the CPL-driven system (dashed red and blue lines, spin-split). The dashed green line marks the equal-energy surface at $\epsilon=-0.7$. (f) Corresponding spin-splitting texture on the equal-energy surface shown in (e). Other parameters are set to $\{t,t_s,t_a,A_0,\omega,|M|,\theta_M\}=\{0.4,0.7,0.3,0.6,10,\sqrt{2}/2,30^\circ\}$.}\label{figs9}
\end{figure}

(i) $[C_{2(x-y)}\|C_{4z}|\bm{\tau}_x]$-breaking but $[C_{2\vec{\bm{n}}}\|C_{4z}|\bm{\tau}_x]$-preserving magnetic-order configuration. For the canted configuration shown in Fig.~\ref{figs9}(a), the original symmetry $[C_{2(x-y)}\|C_{4z}|\bm{\tau}_x]$ is broken, while another similar symmetry, $[C_{2\vec{\bm{n}}}\|C_{4z}|\bm{\tau}_x]$, can be identified. This symmetry can still protect the $d$-wave spin splitting, as shown in Figs.~\ref{figs9}(b) and (c). This explicit example demonstrates that our mechanism  is not restricted to a specific magnetic-order configuration. As long as the magnetic-order directions in the top and bottom layers are orthogonal, one can identify an alternative symmetry $[C_{2\vec{\bm{n}}}\|C_{4z}|\bm{\tau}_x]$ that enforces a $d$-wave texture.

(ii) $[C_{2(x-y)}\|C_{4z}|\bm{\tau}_x]$-breaking but $[\bar{E}\|\mathcal{T}C_{2x}M_z|\bm{\tau}_x]$- and $[\bar{E}\|\mathcal{T}C_{2y}M_z|\bm{\tau}_y]$-preserving magnetic-order configuration. For the canted magnetic-order configuration shown in Fig.~\ref{figs9}(d), the original symmetry $[C_{2(x-y)}\|C_{4z}|\bm{\tau}_x]$ is broken, and the constraint $\langle s_z(k_x,k_y)\rangle=-\langle s_z(k_y,-k_x)\rangle$ no longer holds. However, as discussed in the main text, this type of canting still preserves the key symmetries $[\bar{E}\|\mathcal{T}C_{2x}M_z|\bm{\tau}_x]$ and $[\bar{E}\|\mathcal{T}C_{2y}M_z|\bm{\tau}_y]$, which protect the existence of two nodal lines at $k_x=0$ and $k_y=0$, thereby giving rise to a $d$-wave spin splitting as shown in Figs.~\ref{figs9}(e) and (f). We emphasize 
that, due to this symmetry difference, the equal-energy surface for this case is typically rectangular, distinct from the square equal-energy surface in case (i).

3. \textbf{Generalization by replacing key symmetries}. Because distinct symmetries can impose 
equivalent constraints on the band structure and spin polarization, the symmetry combination enforcing 
unidirectional spin polarization and $d$-wave spin splitting is not unique. Below, we discuss how to generalize these key symmetries. 

(i) $[C_{2z}\|E|\bm{\tau}_d]$: This symmetry constrains $\langle s_{x,y}(\bm{k})\rangle=-\langle s_{x,y}(\bm{k})\rangle=0$, 
thereby ensuring unidirectional spin polarization. This symmetry can be extended to symmetries described by the general form $[C_{2z}\|U_{g_1}]$, where $U_{g_1}$ represents a lattice operation preserving $\bm{k}$ (i.e., $U_{g_1}\bm{k}\rightarrow\bm{k}$), such as a 
fractional translation operator $\boldsymbol{\tau}$ or the mirror operator $\mathcal{M}_z$ in a two-dimensional system.

(ii) $[C_{2z}\|C_{2z}]$: This symmetry  ensures even-parity out-of-plane spin polarization by enforcing $\langle s_{z}(\bm{k})\rangle=\langle s_{z}(-\bm{k})\rangle$. This symmetry can be extended to symmetries described by the general form $[U_z(\theta)\|U_{g_2}]$, where $U_z(\theta)$ represents an arbitrary rotation about the $z$-axis in spin space and $U_{g_2}$ represents a lattice operation that inverts $\bm{k}$ (i.e.,  $U_{g_2}\bm{k}\rightarrow-\bm{k}$), such as inversion $\mathcal{P}$ or twofold rotation  $C_{2z}$.

(iii) $[C_{2(x-y)}\|C_{4z}|\bm{\tau}_x]$: This symmetry  ensures $d$-wave out-of-plane spin polarization by enforcing $\langle s_{z}(k_x,k_y)\rangle=-\langle s_{z}(k_y,-k_x)\rangle$. This symmetry can be generalized to $[C_{2z_\perp}\|U_{g_3}]$, where $C_{2z_\perp}$ represents a twofold rotation operator about an axis perpendicular to the $z$-axis.  A specific example is $[C_{2\vec{\bm{n}}}\|C_{4z}|\bm{\tau}_x]$ discussed above. Here, $U_{g_3}$ represents a lattice operation that rotates $\bm{k}$ by $90^\circ$, such as  $C_{4z}$.

(iv) $[\bar{E}\|\mathcal{T}C_{2x}M_z|\bm{\tau}_{x}]$ and $[\bar{E}\|\mathcal{T}C_{2y}M_z|\bm{\tau}_{y}]$: 
These two effective mirror symmetries ensure the existence of nodal lines at $k_{x}=0$ and $k_y=0$ by enforcing $\langle s_{z}(k_x,k_y)\rangle=-\langle s_{z}(-k_x,k_y)\rangle$ and $\langle s_{z}(k_x,k_y)\rangle=-\langle s_{z}(k_x,-k_y)\rangle$, respectively. These symmetries can be generalized to $[\bar{E}U_z(\theta)\|\mathcal{T}C_{2z_\perp}]$, where the two corresponding axes $z_{\perp,1}$ and $z_{\perp,2}$ should be orthogonal. Here, the form $[\bar{E}\|\mathcal{T}C_{2z_\perp}]$ is required because it should remain preserved under Floquet engineering.

These extensions demonstrate the generality of our mechanism: the symmetries are essential, while the specific interlayer stacking, magnetic order, and lattice geometry are secondary. As long as the model or material preserves the key symmetries discussed above, 
the emergence of $d$-wave spin splitting under CPL is enforced.

\section{VI. Time-, spin- and angle-resolved photoemission photocurrent}

To observe the light-induced spin splitting in the coplanar AFM, pump-probe experiments with tr-ARPES provide a viable detection approach. Since electron dynamics is much faster than phonon dynamics that relax the system, the pump-induced quench effects on 
the bands are observable on short time scales. Furthermore, the spin polarization of the light-driven bands can be resolved through spin-resolved detection. In this section, we simulate the time- and spin-resolved energy spectra of our coplanar AFM model in a pump-probe setup.  We assume that the pump laser vector potential takes the Gaussian form
\begin{align}
 \bm{A}_{\rm pump}(t)=e^{-\frac{(t-t_0)^2}{2T_{\rm pump}^2}}\bm{A}(t),
\end{align}
where $\bm{A}(t)=A_0(\cos{\omega t},\sin{\omega t})$, with $T_{\rm pump}$ the pump laser width and $t_0$ its amplitude-peak time. 
Without loss of generality,  we set $t_0=0$ for simplification. On the other hand, the probe pulse profile is defined as 
\begin{align}
s(t,t_d)=e^{-\frac{(t-t_d)^2}{2T_{\rm probe}^2}},
\end{align}
where $T_{\rm probe}$ is the probe pulse width and $t_d$ its amplitude-peak time. The delay between pump and probe peaks is $\Delta t=t_{d}-t_{0}$;
with  $t_{0}=0$, this reduces to $t_{d}$. The time-resolved photocurrent (tr-ARPES intensity) at momenta $\bm{k}$, binding energy $\epsilon$, and pump-probe delay time $t_{d}$ is obtained as
\begin{align}
I(\bm{k},\epsilon,t_{d})={\rm Im}\int dt_1\int dt_2 s(t_1,t_{d})s(t_2,t_{d})e^{i\epsilon (t_1-t_2)} \text{Tr}[G^<(\bm{k},t_1,t_2)].
\end{align}
The lesser Green's function is defined as 
\begin{align}
G^<(\bm{k},t_1,t_2)&=U(\bm{k},t_1,0)G^<(\bm{k},0,0)U(\bm{k},t_2,0)^\dagger,\\
G^<(\bm{k},0,0)&=i\sum_\alpha |\psi_{k,\alpha}\rangle\langle\psi_{k,\alpha}|f(E_{k\alpha}),
\end{align}
where $f(E_{k\alpha})=\frac{1}{e^{(E_{k\alpha}-\mu)/T}+1}$ represents the Fermi-Dirac distribution of the initial equilibrium band occupation, and $U(\bm{k},t,t')=\hat{T}\exp{(-i\int_{t'}^t\mathcal{H}(\bm{k},\bar{t})d\bar{t})}$ represents the time-evolution operator. It  is numerically 
calculated through 
\begin{align}
 U(\bm{k},t,t')\approx \exp{[-i\mathcal{H}(\bm{k},t-\delta t)\delta t]}\exp{[-i\mathcal{H}(\bm{k},t-2\delta t)\delta t]}\cdots \exp{[-i\mathcal{H}(\bm{k},t'+\delta t)\delta t]}\exp{[-i\mathcal{H}(\bm{k},t')\delta t]}.
\end{align}
The special $d$-wave spin polarization can be revealed by the spin-resolved photocurrent. Its current intensity is defined as
\begin{align}
I_s(\bm{k},\epsilon,t_{d})={\rm Im}\int dt_1\int dt_2 s(t_1,t_{d})s(t_2,t_{d})e^{i\epsilon (t_1-t_2)}{\rm Tr}[s_zG^<(\bm{k},t_1,t_2)],
\end{align}
where $s_i$ represent the Pauli matrices that act on spin degree of freedom. 

\begin{figure}[h]
\centering
\includegraphics[width=0.8\textwidth]{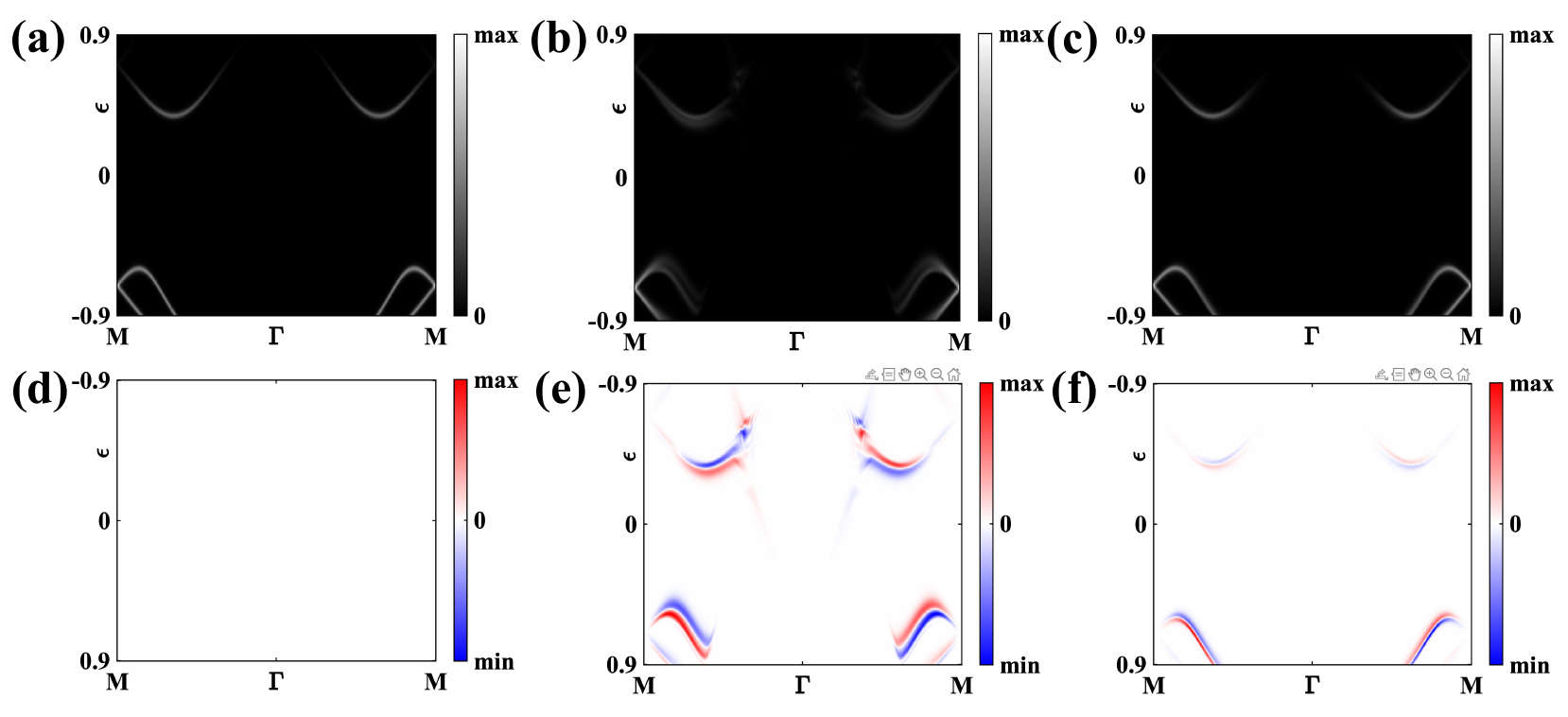}
\caption{(Color online) Simulated  time-resolved photocurrent spectra (a)-(c) and  spin-resolved photocurrent spectra (d)-(f) under different driving conditions along the high-symmetry path $M-\Gamma-M$. No driving is applied in (a) and (d). CPL with $A_0=0.6$ is applied in (b), (c), (e), and (f), with frequency $\omega=3$ for (b) and (e) and $\omega=10$ for (c) and (f).  Other parameters used in the calculation are $T_{\rm probe}=80$, $T_{\rm pump}=100$, $t_d=20$, $\mu=0.5$, $T=0.1$, $t=0.4$, $t_s=0.7$, $t_a=0.3$, and $M=0.5$.}\label{figs10}
\end{figure}

In the main text, we have shown the time- and spin-resolved photocurrent spectrum for CPL with $\omega=2$. Here we present two additional cases at higher frequencies. For comparison, Figs.~\ref{figs10}(a) and \ref{figs10}(d) show the Bloch bands and spin polarization without driving.  
When $\omega=3$ is applied, spin splitting is clearly observed [Fig.~\ref{figs10}(b)], with the spin polarization displaying $d$-wave symmetry [Fig.~\ref{figs10}(e)]. At $\omega=10$, the spin splitting weakens [Fig.~\ref{figs10}(c)], consistent with the effective Hamiltonian analysis; nevertheless, the $d$-wave spin polarization remains visible [Fig.~\ref{figs10}(f)].
We emphasize that the photocurrent spectra are obtained from a pump-probe setup. Although these spectra 
do not exactly match the energy spectra of the Floquet Hamiltonian, they accurately capture the light-induced 
$d$-wave spin splitting.

\section{VII. Drude spin conductivity}

The spin-split Fermi surface with its $d$-wave texture naturally leads to spin- and angle-dependent transport phenomena. Here we focus on the spin current generated by an electric field applied along a general direction, as characterized by the Drude spin conductivity. For simplicity, we investigate the high-frequency off-resonant regime, 
where non-equilibrium effects are weak and the distribution function can be approximated by the equilibrium Fermi-Dirac distribution~\cite{ChenFDrude2022,ZhuDrudesc2026}. Specifically, we consider a setup in which the model is connected to electrodes that maintain a stable chemical potential. In the high-frequency regime, the Floquet system can be effectively described by an effective Hamiltonian. When the system reaches equilibrium, we assume that the distribution is determined by the chemical potential of the electrodes. In this situation, we treat the transport problem using the Floquet effective Hamiltonian with a fixed chemical potential, which now be described with linear response theory.

In the linear response regime, the Drude spin-conductivity tensor reads~\cite{Xiao2010review,Yan2017PRLMn3Ir}
\begin{align}
\sigma^z_{ij}=-\sum_n\tau\int\frac{d^2\bm{k}}{(2\pi)^2}v^z_{n,i}(\bm{k})v_{n,j}(\bm{k})\left(\frac{\partial f(\epsilon)}{\partial \epsilon}\right)_{\epsilon=\epsilon_n},
\end{align}
where the velocity operator component is $v_{n,i}(\bm{k})=\langle u^n_{\bm{k}}|\frac{\partial \mathcal{H}_{\rm eff}(\bm{k})}{\partial k_{i}}|u^n_{\bm{k}}\rangle$, and the spin current operator is  $v^z_{n,i}(\bm{k})=\langle u^n_{\bm{k}}|\frac{1}{2}\{s_z,\frac{\partial \mathcal{H}_{\rm eff}(\bm{k})}{\partial k_{i}}\}|u^n_{\bm{k}}\rangle$, with 
$n$ being the band index and $|u^n_{\bm{k}}\rangle$ the wavefunction. 
Here, $f(\epsilon)$ describes the Fermi-Dirac distribution function and $\tau$ represents 
the relaxation time derived from the Boltzmann equation.

\begin{figure}[h]
\centering
\includegraphics[width=0.65\textwidth]{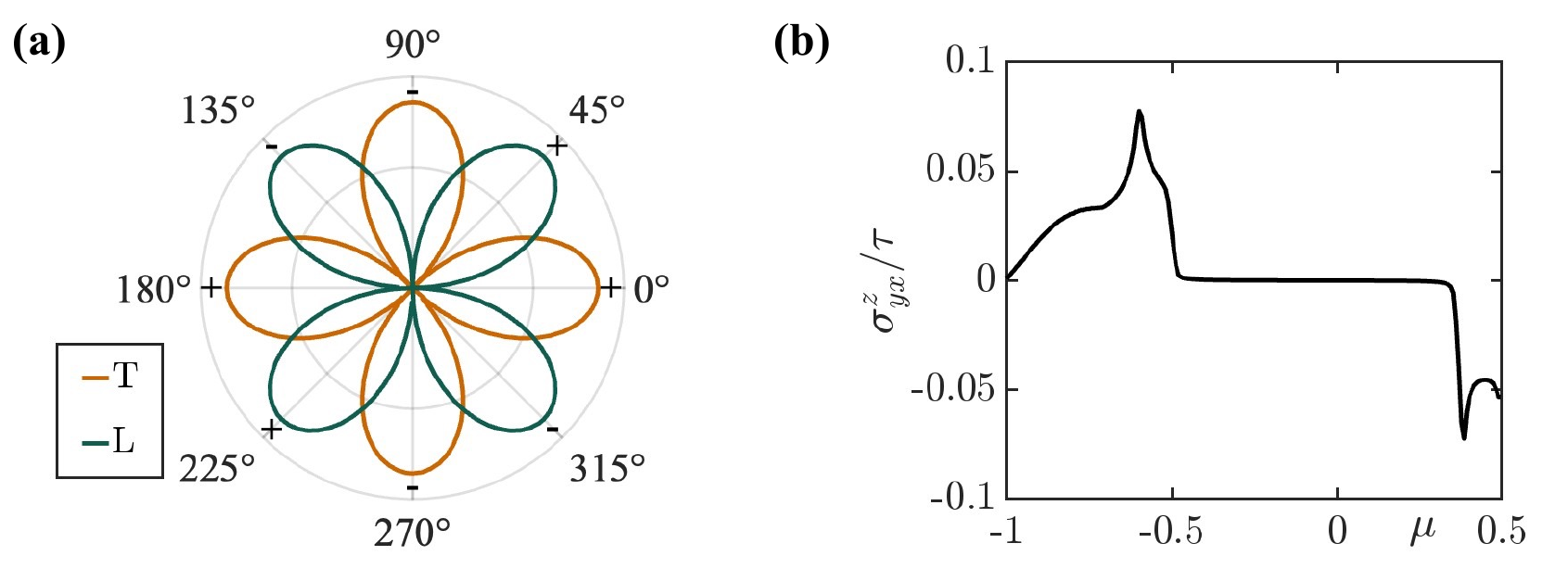}
\caption{(a) Angular dependence of the Drude spin conductivity at $\mu=-0.8$. 
The polar angle is defined between the applied electric field and the 
$x$-axis. Positive (negative) transverse (T) and longitudinal (L) spin conductivities 
are plotted with solid (dashed) green and orange lines, respectively.  
(b)  Transverse spin conductivity as a function of the chemical potential $\mu$. 
Parameters are $t=0.4$, $t_s=0.7$, $t_a=0.3$, $M_x=M_y=0.5$, $A_0=0.6$, and $\omega=10$ ($e=\hbar=1$ is set).
}\label{figs11}
\end{figure}

Figure \ref{figs11}(a) shows the angular dependence of the Drude spin conductivity $\sigma^z_{ij}$, 
revealing a clear $d$-wave pattern that directly reflects the underlying momentum-dependent spin splitting. 
While identical to the pattern in a  $C_{4z}\mathcal{T}$ protected $d$-wave altermagnet~\cite{Ma2021AM}, 
here it is enforced by the $[C_{2(x-y)}\|C_{4z}|\bm{\tau}_{x}]$ symmetry. Furthermore, the symmetries
 $[\bar{E}\|\mathcal{T}C_{2x}\mathcal{M}_{z}|\bm{\tau}_{x}]$ and 
  $[\bar{E}\|\mathcal{T}C_{2y}\mathcal{M}_{z}|\bm{\tau}_{y}]$ impose two key constraints: (i) the longitudinal spin current 
 vanishes when the electric field is aligned with the $x$ or $y$ axes,
giving $\sigma_{xx}^{z}(\theta=\frac{m}{2}\pi)=0$ ($m=0,1,2,3$);
(ii) the transverse spin current vanishes when the field is along $x=y$ or $x=-y$, yielding
$\sigma_{yx}^{z}(\theta=\pm\pi/4,\pm 3\pi/4)=0$. These symmetry constrains   
 produce the characteristic $d$-wave clover-like angular dependence. 
 Such an angular dependence suggests that the spin current can 
 be tuned continuously and switched on or off by varying the direction of the electric field, 
 a potentially valuable property in spintronic devices.

Figure \ref{figs11}(b) shows the transverse spin conductivity $\sigma_{yx}^{z}$
as a function of chemical potential $\mu$. Since the Drude spin conductivity arises 
from the spin-split Fermi surface, it vanishes inside the band gap. Above and below the gap, 
$\sigma_{yx}^{z}$ has opposite signs, reflecting the opposite spin-splitting textures of the 
conduction and valance bands.  

\section{VIII. Feasibility of the experimental realization and potential challenges}

Our work theoretically establishes a symmetry-guided route and proposes a specific bilayer 
coplanar AFM model for realizing light-induced $d$-wave spin splitting. we discuss 
the feasibility of experimental realization and identify potential challenges.

(1) Materials candidates: Van der Waals materials are ideal platforms for our proposal, given their highly flexible stacking capabilities. The stacking of antiferromagnetic van der Waals materials has been extensively explored in recent studies of unconventional magnetism, including light-induced odd-parity spin splitting in collinear AFMs~\cite{Huang2025AM7}, the realization of altermagnetism in twisted bilayer systems~\cite{LiuTwistAM2024}, and type-IV two-dimensional collinear magnets~\cite{BaiTypeIVAFM2025}. In these contexts, bilayer van der Waals materials have been predicted as ideal platforms for constructing special two-dimensional antiferromagnetic models. Our proposal requires only that the Hamiltonian satisfy certain spin-crystal symmetries and that each layer be a collinear AFM---rather than requiring exotic magnetic order. Stacking van der Waals materials thus represents an experimentally feasible route to realizing our proposal.

(2) Stability of the magnetic order under driving: The magnetic effect of light is much weaker than its electric effect, and thus the magnetic effect itself can hardly alter the ground-state magnetic configuration determined by the spin Hamiltonian, such as the Heisenberg Hamiltonian. A more direct way to assess light's influence on magnetic order is to derive the effective spin-exchange strength from a minimal Hubbard model \cite{Torre2021RMPheating,Mentink2015}. Under Floquet engineering, the effective exchange interaction changes from $J\approx 4t_h^2/U$ to $J\approx \sum_m4t_h^2[J_m(F)]^2/(U-m\hbar\omega)$, where $t_h$ and $U$ denote the hopping amplitude and local Coulomb repulsion, respectively, and $J_m(F)$ represents the $m$th Bessel function of the driving strength. Accordingly, a phase transition occurs only when the sign of the interaction changes, which requires $U<m\hbar\omega$ and, in particular, $m\hbar\omega\approx U$ because $J_m(F)$ decays rapidly as $m$ increases for small $F$. Therefore, the phase transition occurs only in specific driving regimes. More importantly, for weak driving, 
the renormalized exchange simplifies to $J\approx \sum_m[J_m(F)]^2(4t_h^2/U)$ in the low-frequency regime
($U\gg \hbar\omega$) or $J\approx 4t_h^2[J_0(F)]^2/U$ in the high-frequency regime ($U\ll \hbar\omega$);
in both cases, the interactions are merely rescaled by overall factors. One can therefore expect the magnetic order and its associated symmetries to remain stable, provided the driving stays below a threshold.

(3) Experimental tools: As theoretically demonstrated in Sec.~IV, the light-induced spin splitting is 
directly detectable via time- and angle-resolved photoemission spectroscopy (tr-ARPES)~\cite{Boschini2024RMPTrARPES,Zhou2023TrARPES}. 
It is noteworthy that the opening of a band gap between Floquet replica bands---a theoretical prediction dating back to 2009---has recently been observed in pump-probe experiments~\cite{Wang2026Floquet}. This marks significant progress in measuring optically driven band renormalization. The detection of our predicted light-induced spin splitting can be achieved through a similar approach. However, confirming the $d$-wave symmetry of the spin splitting requires spin-resolved tr-ARPES~\cite{Boschini2024RMPTrARPES}. Although this remains a highly complex technique demanding long integration times, detecting transient spin dynamics at a few momentum points along specific high-symmetry lines is well within the capability of current state-of-the-art spin-resolved tr-ARPES. The $d$-wave spin splitting can therefore be verified by measuring spin dynamics at momenta related by a 90$^{\circ}$ rotation. 

(4) Heating: Heating is a ubiquitous effect in Floquet systems that influences transport behavior through the distribution function. For isolated lattice systems, it is well established that a Floquet system eventually enters a featureless infinite-temperature state due to continuous energy absorption from the drive.  Nevertheless, there exist regimes in which heating effects are weak. For instance, in an insulator or semiconductor, when the driving frequency is either larger than the bandwidth or smaller than the band gap, off-resonant absorption is absent and the heating effect is weak. When the driving frequency is tuned to a spectral gap or to the high-frequency regime, recent studies have shown that heating can, in principle, be postponed to exponentially long times \cite{Torre2021RMPheating,Mori2016heating}. Since our mechanism is determined by the presence of key symmetries rather than by a specific frequency regime, the light-induced even-parity spin splitting is detectable through the Drude spin conductivity when heating is sufficiently suppressed by tuning the frequency. We emphasize that for pump-probe experiments, heating is not an issue even in the resonant regime~\cite{Zhou2023TrARPES}. This is due to the marked difference in time scales: the pump and probe laser pulses are typically of order 100 fs, whereas the electron-phonon relaxation time ranges from several hundred fs to several ps. The measurement is completed before the pumped electrons transfer their energy to phonons. 

(5) Transport of the steady state: When absorbed energy is dissipated to a thermal bath, the driven system can reach a steady state where driving and dissipation balance. Although the steady-state distribution typically deviates from the equilibrium Fermi-Dirac form, its symmetry may match that of the Floquet Hamiltonian. Consider our model: The key symmetry producing the $d$-wave clover-like spin conductivity is 
$[C_{2(x-y)}\|C_{4z}|\bm{\tau}_{x}]$. While the Floquet Hamiltonian respects this symmetry, the steady-state distribution need not, particularly in the resonant regime. The reason is: Although the energy spectrum respects the symmetry, the optical absorption rate can differ between momenta related by a $90^{\circ}$ 
rotation. Consequently, even with identical electron-phonon relaxation, the resulting steady-state distribution may break the symmetry. This issue is avoided in the off-resonant regime, where direct optical transitions are prohibited for all momenta. The symmetry is then preserved, or at most very weakly broken (through rare multiphoton processes). Therefore, provided the frequency is off-resonant, the $d$-wave clover-like spin conductivity should be detectable even when the driven system reaches a steady state with a nonthermal distribution~\cite{DehghaniDisspation,Li2025AM7}.

In summary, from material candidates to experimental observation, our proposal is well within the scope of current experimental capability. The requisite bilayer coplanar AFMs can be assembled from readily available van der Waals materials with flexible stacking techniques. The light-induced $d$-wave spin splitting is detectable via established pump-probe and spin-resolved tr-ARPES methodologies, with heating effects mitigated by the intrinsic separation of electronic and phononic time scales. The off-resonant driving regime permits steady-state transport measurements as an alternative detection route. These considerations collectively establish our proposal as an experimentally feasible route to realizing even-parity unidirectionally polarized spin splitting in coplanar AFMs.

}

\end{document}